\documentclass[reprint,aps]{revtex4}
\usepackage{amsmath}
\usepackage{graphicx}
\usepackage{dcolumn}
\usepackage{bm}

\usepackage{amssymb}
\usepackage{url}
\usepackage{caption,subcaption}
\usepackage[colorlinks=true,allcolors=blue]{hyperref}

\begin{document}

\title{Perturbative approaches in  relativistic kinetic theory and the emergence of first-order hydrodynamics}
\author{Gabriel S. Rocha}
\email{gabrielsr@id.uff.br}
\author{Gabriel S. Denicol}
\email{gsdenicol@id.uff.br}
\affiliation{Instituto de F\'{\i}sica, Universidade Federal Fluminense, Niter\'{o}i, Rio de Janeiro,
Brazil}
\author{Jorge Noronha}
\email{jn0508@illinois.edu}
\affiliation{Illinois Center for Advanced Studies of the Universe \& Department of Physics, University of Illinois Urbana-Champaign, Urbana, IL 61801, USA}

\begin{abstract}
Hydrodynamics can be formulated in terms of a perturbative series in derivatives of the temperature, chemical potential, and flow velocity around an equilibrium state. Different formulations for this series have been proposed over the years, which consequently led to the development of various hydrodynamic theories. In this work, we discuss the relativistic generalizations of the perturbative expansions put forward by Chapman and Enskog, and Hilbert, using general matching conditions in kinetic theory. This allows us to describe, in a comprehensive way, how different out-of-equilibrium definitions for the hydrodynamic fields affect the development of the hydrodynamic perturbative series. We provide a perturbative method for systematically deriving the hydrodynamic formulation recently proposed by Bemfica, Disconzi, Noronha, and Kovtun (BDNK) from relativistic kinetic theory. The various transport coefficients that appear in BDNK (at first-order) are explicitly computed using a new formulation of the relaxation time approximation for the Boltzmann equation. Assuming Bjorken flow, we also determine the hydrodynamic attractors of BDNK theory and compare the overall hydrodynamic evolution obtained using this formulation with that generated by the Israel-Stewart equations of motion and also kinetic theory.  
\end{abstract}

\maketitle

\section{Introduction}

Hydrodynamics is an effective theory constructed to describe the near-equilibrium dynamics of a many-body system in the regime where the time and length scales associated with microscopic interactions are much smaller than those defined by the variation of conserved quantities. In the context of kinetic theory, hydrodynamics is expected to emerge when the evolution of the system can be reasonably well described in terms of the few moments of the single-particle distribution function directly associated with conservation laws \cite{grad1958principles}. Understanding how this truncation in the number of degrees of freedom takes place, and how that process depends on the strength of the interactions, define a very active area of research especially in the relativistic regime \cite{Florkowski:2017olj,romatschke2019relativistic,Denicol:2021}. 

Throughout the years, these questions have been investigated in kinetic theory using mainly three approaches. The oldest method, pioneered by D.~Hilbert more than a century ago \cite{hilbert1912begrundung,cercignani:90mathematical}, was the first to recognize that hydrodynamic behavior could be systematically investigated via a perturbative expansion. However, even though the lowest order truncation of the Hilbert series led to the Euler equations, it was soon realized that this method did not lead to the celebrated non-relativistic Navier-Stokes equations \cite{cercignani:90mathematical}. The derivation of the latter from kinetic theory was achieved only later by S.~Chapman \cite{chapman1916vi,chapman1990mathematical} and D.~Enskog \cite{enskog1917kinetische}, using a nontrivial modification of Hilbert's ideas. Their method has become the standard way to derive hydrodynamics from the Boltzmann equation \cite{cercignani:90mathematical,struchtrup2005macroscopic}.  Grad \cite{grad:1949kinetic} formulated the third way within which such a reduction of degrees of freedom can take place in kinetic theory. Different than Hilbert or Chapman and Enskog, in Grad's approach the Boltzmann equation is converted into an infinite set of coupled equations of motion for the moments of the distribution function. Hydrodynamics then emerges from a truncation of such an infinite set of equations in terms of the lowest order moments, whose evolution equations are given by the conservation laws. The generalization of the moments method to relativistic kinetic systems can be found, e.g., in \cite{israel1979jm,deGroot:80relativistic,cercignani:02relativistic,Denicol:2012es,Denicol:2018rbw,rocha21-transient,Denicol:2021}.

Even though the Chapman-Enskog approach can be formulated in a relativistic covariant manner \cite{deGroot:80relativistic,cercignani:02relativistic,Denicol:2021}, fundamental issues appear after truncation --- the corresponding relativistic generalization of Navier-Stokes equations derived from this approach are incompatible with relativistic causality \cite{pichon:65etude,hiscock:85generic} and small perturbations around global equilibrium are unstable \cite{hiscock1983stability}. A solution to these problems was put forward by M\"uller \cite{muller1967paradoxon} and Israel and Stewart \cite{israel1979annals,israel1979jm}. In this approach, instead of using the  constitutive relations characteristic of the Navier-Stokes formulation, the dissipative currents are promoted to independent variables that, in turn, obey additional equations of motion that describe how those variables relax towards their Navier-Stokes values. Then, causality and stability constrain the possible values of relaxation times and coupling constants between the dissipative currents \cite{hiscock1983stability,denicol2008stability,Pu:2009fj,biswas2020causality,brito2020linear,Bemfica:2020xym}. Recently, it has been rigorously proven that stability in a causal theory is a Lorentz invariant statement \cite{Bemfica:2020zjp,Gavassino:2021owo} and conditions under which thermodynamic stability imply causality in relativistic fluids have been derived \cite{Gavassino:2021kjm}. The Israel-Stewart (IS) formulation can be derived from the Boltzmann equation by truncating an infinite tower of equations obeyed by irreducible moments of the single-particle distribution function generalizing Grad's \cite{grad:1949kinetic} moments method \cite{israel1979jm,Denicol:2012es,Denicol:2018rbw,rocha21-transient} or by power-counting schemes \cite{denicol2012derivation,Denicol:2019iyh}. 

A new solution to the causality and stability problems was recently proposed by Bemfica, Disconzi, Noronha, and Kovtun (BDNK) \cite{bemfica:18causality,kovtun:19first,bemfica2019nonlinear,Hoult:2020eho,Bemfica:2020zjp}. This formulation, which is deeply rooted on effective field theory arguments, when truncated to first-order in derivatives also leads to constitutive relations for the dissipative currents as in  Navier-Stokes theory, but with the important difference that it  includes time-like derivatives in these relations (i.e., time derivatives of the hydrodynamic fields are present in the constitutive relations even in the local rest frame of the fluid). Conditions that ensure causality, strong hyperbolicity, and well-posedness of  solutions of the equations of motion, which are valid even in the full nonlinear regime and when the fluid is coupled to Einstein's equations, were derived in  \cite{bemfica:18causality,bemfica2019nonlinear,DisconziBemficaRodriguezShaoSobolevConformal,DisconziBemficaGraber,Bemfica:2020zjp} and stability was proven to hold \cite{bemfica:18causality,kovtun:19first,bemfica2019nonlinear,Hoult:2020eho,Bemfica:2020zjp}. BDNK theory also motivates the investigation of alternative definitions of the hydrodynamic variables out of equilibrium, the so-called hydrodynamic frames \cite{Kovtun:12}, due to the fact that causality in this framework requires that some of the coefficients associated with energy diffusion and the non-equilibrium correction to energy density must be nonzero in the first-order theory. This framework has also been explored numerically in Refs.~\cite{Pandya:2021ief,Bantilan:2022ech,Pandya:2022pif}, where it was compared to other hydrodynamic formulations.      

Matching conditions define the local equilibrium state, which serve as starting point of hydrodynamic expansions. Consequently, they define the temperature $T$, chemical potential $\mu$, and fluid 4-velocity $u^{\mu}$ of non-equilibrium systems. In kinetic theory, these definitions are implemented by constraints in the moments of the single-particle distribution function. Even though they have a significant effect on the truncation procedures leading to different formulations of hydrodynamics, the interplay between these definitions and hydrodynamic attractors \cite{Heller:2015dha} are far from understood. In fact, Ref.\ \cite{Heller:2015dha} proposed that  hydrodynamics should be seen as an attractor which determines the late-time behavior of many-body systems and, thus, dictates their approach towards equilibrium. This has been confirmed in Bjorken \cite{bjorken1983highly} and Gubser \cite{Gubser:2010ze} flow backgrounds (see, e.g.,  Refs.~\cite{Heller:2015dha,Strickland:2017kux,Denicol:2018pak} and \cite{Denicol:2019lio, Behtash:2017wqg}), and this topic   remains under active investigation (for reviews, see \cite{Florkowski:2017olj,Romatschke:2017ejr,Berges:2020fwq,Soloviev:2021lhs}). 

In this work, we provide a systematic perturbative procedure for the derivation of BDNK theory starting from the relativistic Boltzmann equation. The various transport coefficients that appear in BDNK (at first-order) are explicitly determined, for the first time, using the new formulation of the relaxation time approximation for the Boltzmann equation proposed in \cite{rocha:21}. For completeness, we also provide a comprehensive review of the relativistic Hilbert series and formulate Chapman-Enskog theory for general matching conditions. 
Focusing on Bjorken flow, we  determine the hydrodynamic attractors of BDNK theory and compare the evolution obtained using this approach with that obtained using 
Israel-Stewart equations of motion and also kinetic theory (in the relaxation time approximation).

This paper is organized as follows. In Sec.~\ref{sec:fl-dyn-vars} we review the formulation of hydrodynamic equations using general matching conditions. Next, in Sec.~\ref{sec:BEq-fl-dyn}, we show how this general procedure is implemented within kinetic theory.  In Sec.~\ref{sec:pert-expn}, we discuss the various perturbative procedures used to derive hydrodynamic equations of motion from the Boltzmann equation. We implement the well-known Chapman-Enskog expansion, used to derive relativistic Navier-Stokes theory, using general matching conditions and provide formulas that can be used to determine its transport coefficients. We then proceed to discuss the Hilbert expansion, which leads to another perturbative hydrodynamic formulation. Afterwards, we introduce the perturbative procedure suitable for the derivation of BDNK formalism in kinetic theory. In Sec.~\ref{sec:RTA}, we use the new relaxation time approximation for the relativistic Boltzmann equation proposed in Ref.~\cite{rocha:21} to explicitly compute the transport coefficients present in first-order BDNK, and also those present in the Hilbert series, in the case of an ultrarelativistic gas. In Sec.~\ref{sec:comparison} we initiate our discussion of Bjorken flow. We first discuss the behavior of Hilbert's equations of motion and compare their exact solution in Bjorken with the solution of  relativistic Navier-Stokes theory. For completeness, we also outline the corresponding Israel-Stewart equations of motion in Bjorken flow under general matching conditions, following \cite{rocha21-transient}. We present numerical and also exact solutions of BDNK theory for different matching prescriptions and investigate the corresponding hydrodynamic attractor. Comparisons between solutions of BDNK, Israel-Stewart, and kinetic theory (obtained by solving a system of moment equations) are also made.  Sec. \ref{sec:concl} summarizes the main text and states our conclusions and future plans. Appendix \ref{sec:MIS-gen-match} summarizes the 19 moments truncation procedure used to obtain the IS equations of motion under general matching conditions, originally performed in Ref.~\cite{rocha21-transient}. Appendix \ref{sec:zero-mode-CE} shows the procedure necessary to derive the Chapman-Enskog constitutive relations when the basis contains the zero modes of the collision term. Appendix \ref{sec:choice-basis} gives further details on the choice of basis used to compute transport coefficients. Finally, Appendix \ref{sec:BEq-EoM} gives the details concerning the moment equations in Bjorken flow. 

\emph{Notation:} We use a mostly \emph{minus} 
metric signature and natural units, $\hbar=c=k_B=1$.

\section{Fluid dynamical variables}
\label{sec:fl-dyn-vars}

The main dynamical equations in hydrodynamics come from the local conservation of energy, momentum, and net charge,
\begin{equation}
\begin{aligned}
\label{eq:consv-eqns}
    \partial_{\mu}N^{\mu} = 0,\\
    \partial_{\mu}T^{\mu \nu} = 0,
\end{aligned}
\end{equation}
where $N^{\mu}$ is the net-charge 4-current and $T^{\mu \nu}$ is the energy-momentum tensor. Without any loss of generality, these tensors can be decomposed in terms of a time-like normalized 4-vector $u^{\mu}$, $u^{\mu}u_{\mu} = 1$, in the following way
\begin{equation}
\begin{aligned}
\label{eq:decompos-numu-tmunu}
   N^{\mu} &= n u^{\mu} + \nu^{\mu},  \\
    T^{\mu \nu} &= \varepsilon u^{\mu} u^{\nu} - P \Delta^{\mu \nu} + h^{\mu} u^{\nu} + h^{\nu} u^{\mu} + \pi^{\mu \nu}.
\end{aligned}
\end{equation}
The 4-vector $u^{\mu}$ is identified as the fluid 4-velocity and will be formally defined after the imposition of the so-called matching conditions -- a procedure that will be discussed later. Above, we also defined the projection operator $\Delta^{\mu \nu} = g^{\mu \nu} - u^{\mu} u^{\nu}$. Each term introduced in the decomposition done above can be expressed in terms of projections and contractions of the conserved currents, 
\begin{equation}
\begin{aligned}
\label{eq:definitions}
    n &\equiv u_{\mu}N^{\mu} , \, \, \varepsilon \equiv u_{\mu}u_{\nu}T^{\mu\nu}, \, \, P \equiv -\frac{1}{3}\Delta_{\mu\nu}T^{\mu\nu},  \\
    \nu^{\mu} &\equiv \Delta^{\mu}_{\nu} N^{\nu}, \, \, h^{\mu} \equiv \Delta^{\mu}_{\nu} u_{\lambda} T^{\nu\lambda}, \, \, \pi^{\mu\nu} \equiv \Delta^{\mu\nu}_{\alpha\beta} T^{\alpha\beta},
\end{aligned}
\end{equation}
which are identified as the total particle density, the total energy density, the total pressure, the particle diffusion 4-current, the energy diffusion 4-current, and the shear-stress tensor, respectively. We note that we have introduced above the doubly-symmetric traceless tensor $\Delta^{\mu \nu \alpha \beta} = \left( \Delta^{\mu \alpha } \Delta^{\nu \beta} + \Delta^{\nu \alpha } \Delta^{\mu \beta} \right)/2 - \Delta^{\mu \nu} \Delta^{\alpha \beta}/3$.

Next, we introduce a reference local equilibrium state \cite{israel1979annals} and separate the particle density, energy density, and pressure into equilibrium and non-equilibrium parts. In general, 
\begin{equation}
\begin{aligned}
\label{eq:definitions2}
    n &\equiv n_0(\alpha,\beta) + \delta n, \\
    \varepsilon &\equiv \varepsilon_0(\alpha,\beta) + \delta \varepsilon, \\ 
    P &\equiv P_0(\alpha,\beta) + \Pi,
\end{aligned}
\end{equation}
where $\alpha$ and $\beta$ are the thermal potential and inverse temperature of this (fictitious) reference equilibrium state. The densities $n_0$, $\varepsilon_0$ and $P_0$ are the local equilibrium net-charge density, energy density, and pressure, respectively, and are determined in terms of the temperature and thermal potential using an equation of state. The variables $\alpha$ and $\beta$ are determined by the matching conditions. The variables $\delta n$, $\delta \varepsilon$, and $\Pi$ represent the corresponding non-equilibrium corrections for the net-charge density, energy density, and pressure, respectively. 

The most widely employed matching condition imposes that the particle and energy densities are completely determined by $\alpha$ and $\beta$ alone as if the fluid were in local equilibrium. That is, in this case we have the constraints
\begin{equation}
\begin{aligned}
\label{eq:decompos-numu-tmunu2}
   n &\equiv n_0(\alpha,\beta) 
   \Longleftrightarrow \delta n \equiv 0, \\
   \varepsilon &\equiv \varepsilon_0(\alpha,\beta) 
   \Longleftrightarrow \delta \varepsilon \equiv 0 .
\end{aligned}
\end{equation}
Then, the temperature and thermal potential are determined by inverting the thermodynamic functions $n_0(\alpha,\beta)$ and $\varepsilon_0(\alpha,\beta)$. In order to complete the matching procedure, we must also define the local rest frame of the fluid, i.e., we have to define the 4-velocity introduced in the tensor decomposition of the conserved currents. In this case, two different constraints are widely employed in the field. The first is the so-called Landau matching condition (or Landau picture) which stipulates that no energy diffusion should occur in the rest frame of the fluid \cite{landau:59fluid},
\begin{equation}
\label{eq:Landau}
T^{\mu}_{\ \nu} u^{\nu} \equiv \varepsilon u^\mu \Longleftrightarrow h^\mu \equiv 0. 
\end{equation}
This condition is frequently used in fluid-dynamical simulations of ultrarelativistic heavy-ion collisions, see \cite{Schenke:2010rr,Noronha-Hostler:2013gga,Shen:2014vra,NunesdaSilva:2020bfs}.
The second is the Eckart matching condition (or Eckart picture) which imposes that no particle diffusion should occur in the rest frame of the fluid \cite{eckart:40fluid},
\begin{equation}
\label{eq:Eckart}
N^{\mu} \equiv n u^{\mu} \Longleftrightarrow \nu^\mu \equiv 0. 
\end{equation}
This condition is commonly used in astrophysics applications  \cite{Most:2021rhr,chabanov:21-general}. 

It is important to remark that different matching conditions, which differ from Landau or Eckart's, can in principle be chosen \cite{Tsumura:2006hnr,Tsumura:2007wu,Van:2007pw,Van:2011yn}. Indeed, this choice reflects the fact that there is no unique definition of temperature, chemical potential, and flow velocity in an out of equilibrium state. However, different choices of such matching conditions do affect the properties of the hydrodynamic equations of motion -- for instance, some choices can lead to well-defined behavior while for others causality may be lost.

In the following, we shall perform our calculations assuming a wide set of matching conditions, which will be defined in the next section in the context of kinetic theory. In this general case, the conserved currents have the form
\begin{equation}
\begin{aligned}
\label{eq:decompos-numu-tmunu3}
   N^{\mu} &= (n_{0} + \delta n) u^{\mu} + \nu^{\mu}  \\
    T^{\mu \nu} &= (\varepsilon_{0} + \delta \varepsilon )u^{\mu} u^{\nu} - (P_{0} + \Pi) \Delta^{\mu \nu} + h^{\mu} u^{\nu} + h^{\nu} u^{\mu} + \pi^{\mu \nu}.
\end{aligned}
\end{equation}
Substituting this expression into the conservation laws \eqref{eq:consv-eqns}, we obtain the following equations of motion (already projected into their components parallel and orthogonal to $u^\mu$), 
\begin{subequations}
 \label{eq:basic-hydro-EoM}
\begin{align}
 \label{eq:hydro-EoM-n0}
 Dn_{0}+D\delta n + (n_{0}+\delta n) \theta + \partial_{\mu} \nu^{\mu} &= 0, \\
\label{eq:hydro-EoM-eps}
 D\varepsilon_{0}+D\delta \varepsilon + (\varepsilon_{0}+\delta \varepsilon + P_{0} + \Pi) \theta - \pi^{\mu \nu} \sigma_{\mu \nu} + \partial_{\mu}h^{\mu} + u_{\mu} Dh^{\mu} &= 0, \\
\label{eq:hydro-EoM-umu}
(\varepsilon_{0} + \delta \varepsilon + P_{0} + \Pi)Du^{\mu} - \nabla^{\mu}(P_{0} + \Pi) + h^{\mu} \theta + h^{\alpha} \Delta^{\mu \nu} \partial_{\alpha}u_{\nu} +  \Delta^{\mu \nu} Dh_{\nu} + \Delta^{\mu \nu} \partial_{\alpha}\pi^{\alpha}_{ \ \nu} &= 0,
\end{align}
\end{subequations}
where we make use of the irreducible decomposition of the 4-derivative, $\partial_{\mu} = u_{\mu}D + \nabla_{\mu}$, with $D \equiv u^{\nu} \partial_{\nu}$ being the comoving time derivative and $\nabla_{\mu} = \Delta_{\mu}^{\ \nu} \partial_{\nu}$ the 4-gradient, while $\theta \equiv \partial_{\mu}u^{\mu}$ is the scalar expansion rate and $\sigma^{\mu\nu} \equiv \Delta^{\mu\nu}_{\alpha\beta}\partial_{\alpha}u_{\beta} $ is the shear tensor.

Different approaches can be used to supplement the equations of motion  \eqref{eq:hydro-EoM-n0}-\eqref{eq:hydro-EoM-umu} with information about viscous effects. One may provide constitutive relations satisfied by the non-equilibrium fields and  express them in terms of derivatives of the fluid-dynamical variables that already appear in equilibrium. This is the case of the relativistic formulation of Navier-Stokes (NS) theory \cite{landau:59fluid} and of the recently proposed BDNK theory \cite{bemfica:18causality,kovtun:19first,bemfica2019nonlinear,Hoult:2020eho,Bemfica:2020zjp} -- both these frameworks will be discussed in detail in the remainder of this paper in the context of kinetic theory. As mentioned in the previous section, another way to describe viscous effects consists in postulating that the non-equilibrium fields are independent dynamical variables which obey their own equations of motion (this can be achieved using entropy arguments or the moments method in kinetic theory). That is the case of the Israel-Stewart  \cite{israel1979annals} and the Denicol-Niemi-Molnar-Rischke (DNMR) \cite{denicol2012derivation} formulations of relativistic hydrodynamics, where the non-equilibrium variables relax asymptotically to Navier-Stokes-like constitutive relations. In this regard, a generalized version of the moments method \cite{Denicol:2012es} has been recently proposed to consider the case of general matching conditions in Ref.~\cite{rocha21-transient}. Second-order theories have also been derived under general matching conditions using entropy arguments in \cite{noronha2021transient}. One of the goals of this paper is to compare the fluid-dynamical solutions that emerge from all these different formalisms and better understand their differences and domain of applicability, in particular when uncommon matching conditions are employed.

\section{Boltzmann equation and fluid-dynamical variables}
\label{sec:BEq-fl-dyn}

The relativistic Boltzmann equation is a non-linear integro-differential equation for the single-particle momentum distribution function $f(\textbf{x},\textbf{p}) \equiv f_{\textbf{p}}$ \cite{deGroot:80relativistic}. Assuming that we have a one-component gas composed of classical particles that only interact through binary collisions, the relativistic Boltzmann equation becomes,
\begin{equation}
\label{eq:EdBoltzmann}
p^{\mu }\partial_{\mu }f_{\mathbf{p}} = \int dQ \ dQ^{\prime} \ dP^{\prime} W_{pp' \leftrightarrow qq'} ( f_{\mathbf{q}}f_{\mathbf{q}'} 
-
f_{\mathbf{p}}f_{\mathbf{p}'}  )   \equiv C\left[ f_{\mathbf{p}}\right],  
\end{equation}
where $p^{\mu }=( \sqrt{m^2+\vert\textbf{p}\vert^2},\textbf{p})$ is the 4-momentum of the particle and we introduced the transition rate $W_{pp' \leftrightarrow qq'}$ and the phase space integration measure $dP = d^{3}p/[(2\pi)^{3} p^{0}]$. 

The particle 4-current and energy-momentum tensor can be expressed as integrals of $f_{\mathbf{p}}$. Multiplying Eq.~\eqref{eq:EdBoltzmann} by $1$ or $p^{\nu}$ and integrating it in momentum-space, one obtains the conservation laws \eqref{eq:consv-eqns} and identifies
\begin{equation}
\label{eq:currents_kin}
\begin{aligned}
N^{\mu} &= \int dP \, p^{\mu} f_{\textbf{p}} , \\
T^{\mu\nu} &= \int dP \, p^{\mu}p^{\nu} f_{\textbf{p}} .
\end{aligned}
\end{equation}
Hence, from the tensor decomposition \eqref{eq:decompos-numu-tmunu3}, it follows that,
\begin{equation}
\begin{aligned}
\label{eq:def_kinetic}
    n &\equiv \int dP \, E_{\textbf{p}} f_{\textbf{p}} , \, \, \varepsilon \equiv \int dP \, E_{\textbf{p}}^2 f_{\textbf{p}}, \, \, P \equiv -\frac{1}{3} \int dP \, \Delta_{\mu \nu}p^{\mu}p^{\nu} f_{\textbf{p}},  \\
    \nu^{\mu} &\equiv \int dP \, p^{\langle \mu \rangle} f_{\textbf{p}}, \, \, h^{\mu} \equiv \int dP \, E_{\textbf{p}} p^{\langle \mu \rangle} f_{\textbf{p}}, \, \, \pi^{\mu\nu} \equiv \int dP \, p^{\langle \mu}p^{\nu \rangle} f_{\textbf{p}} \\
    \delta n &\equiv \int dP \, E_{\textbf{p}} \delta f_{\textbf{p}}, \, \, \delta \varepsilon \equiv \int dP \, E_{\textbf{p}}^2 \delta f_{\textbf{p}}, \, \,  
    \Pi \equiv -\frac{1}{3} \int dP \, \Delta_{\mu \nu}p^{\mu}p^{\nu}\delta f_{\textbf{p}}, 
\end{aligned}
\end{equation}
where $\delta f_{\textbf{p}} \equiv f_{\textbf{p}}-f_{0\textbf{p}}$ is the non-equilibrium component of the distribution function, with $f_{0\textbf{p}}$ being the distribution function in local equilibrium. Above, we  introduced the $\ell$-th rank projection operator  $p^{\langle \mu_{1}} \cdots p^{\mu_{\ell} \rangle} \equiv \Delta^{\mu_{1} \cdots \mu_{\ell}}_{\nu_{1} \cdots \nu_{\ell}} p^{\nu_{1}} \cdots p^{\nu_{\ell}}$, given in terms of the $2\ell$-rank projection tensor $\Delta^{\mu_{1} \cdots \mu_{\ell}}_{\nu_{1} \cdots \nu_{\ell}}$. The latter is constructed from the $\Delta^{\mu \nu}$ projectors in such a way that it is completely symmetric with respect to permutations in any of the  indices $\mu_{1} \cdots \mu_{\ell}$ and $\nu_{1} \cdots \nu_{\ell}$, separately, and also traceless within each subset of indices \cite{deGroot:80relativistic}. For classical particles, the local equilibrium distribution function is given by 
\begin{equation}
\label{eq:f_0}
    f_{0\textbf{p}} \equiv \exp{ \left(\alpha -\beta E_{\mathbf{p}} \right)}. 
\end{equation}
It is also convenient to introduce the deviation function
\begin{equation}
\label{eq:df}
    \phi_{\textbf{p}} \equiv \frac{\delta f_{\textbf{p}}}{f_{0\textbf{p}}},    
\end{equation}
which will be widely employed in the remainder of this paper.

As already mentioned, the local thermodynamic variables $\alpha$, $\beta$, and $u^{\mu}$ are defined by matching conditions. In kinetic theory, these definitions act as constraints on the deviation function $\phi_{\bf p}$. For instance, the Landau matching conditions, Eqs.~\eqref{eq:decompos-numu-tmunu2} and \eqref{eq:Landau}, lead to the following constraints
\begin{equation}
\begin{aligned}
\label{eq:Landau_kinetic}
    \langle E_{\textbf{p}} \phi_{\textbf{p}} \rangle_{0} = 0, \, \, 
    \langle E_{\textbf{p}}^{2} \phi_{\textbf{p}} \rangle_{0} = 0, \, \,
    \left\langle E_{\textbf{p}} p^{\langle \mu \rangle} \phi_{\textbf{p}} \right\rangle_{0} =0,\\
\end{aligned}
\end{equation}
where we make use of the notation $\langle \cdots \rangle_0 = \int dP (\cdots) f_{0\textbf{p}}$ to denote integrals with respect to the equilibrium distribution. The Eckart prescription, Eqs.~\eqref{eq:decompos-numu-tmunu2} and \eqref{eq:Eckart}, differs from the Landau one only in the vector constraint, which is replaced by
\begin{equation}
\begin{aligned}
\label{eq:Eckart_kinetic}
  \left\langle  p^{\langle \mu \rangle} \phi_{\textbf{p}} \right\rangle_{0} =0.
\end{aligned}
\end{equation}

Recent developments have created the demand for more general definitions of the reference equilibrium state \cite{bemfica:18causality,bemfica2019nonlinear,Bemfica:2020zjp}. And indeed, in kinetic theory, these constrains can be readily generalized with the usage of arbitrary moments of the single-particle distribution function. These can be written in general as
\begin{equation}
\begin{aligned}
\label{eq:matching_kinetic1}
    \left\langle \, g_{\textbf{p}}  \phi_{\textbf{p}} \right\rangle_{0} = 0, \ 
    \left\langle \, h_{\textbf{p}}  \phi_{\textbf{p}} \right\rangle_{0} = 0, \
    \left\langle \, q_{\textbf{p}} p^{\langle \mu \rangle} \phi_{\textbf{p}} \right\rangle_{0} =0, 
\end{aligned}
\end{equation}
where $g_{\textbf{p}}$ and $h_{\textbf{p}}$ are linearly independent functions and $q_{\textbf{p}}$ is a generic function of the microscopic energy. In this work, we shall use
\begin{equation}
\begin{aligned}
\label{eq:matching_kinetic2}
    g_{\textbf{p}} = E_{\textbf{p}}^q, \ 
    h_{\textbf{p}} = E_{\textbf{p}}^s, \ 
    q_{\textbf{p}} = E_{\textbf{p}}^z, \ 
\end{aligned}
\end{equation}
which reduce to Landau matching conditions when $(q,s,z) =(1,2,0)$ while Eckart matching is recovered when $(q,s,z) = (1,2,1)$. We note that, from the point of view of the Boltzmann equation, the choice of matching conditions is arbitrary.


\section{Perturbative Expansions}
\label{sec:pert-expn}

The long wavelength, long time behavior of a microscopic theory, the hydrodynamic limit, can be implemented in kinetic theory as a truncation of a perturbative series \cite{deGroot:80relativistic,cercignani:02relativistic,cercignani:90mathematical}. The perturbative parameter is the ratio between  typical microscopic and macroscopic scales, e.g. the mean free path and the length scale at which the hydrodynamic fields vary significantly, respectively. 
The first implementation of such an expansion was developed by Hilbert in the non-relativistic case \cite{hilbert1912begrundung,hilbert1989grundzuge,cercignani:90mathematical}. The lowest order truncation of the 
Hilbert expansion leads to the Euler equations, which provided the first microscopic derivation of a fluid-dynamical theory. Nevertheless, its higher order truncations failed to reproduce any reliable dissipative fluid-dynamical theory \cite{cercignani:90mathematical}. Afterwards, an improved perturbative series expansion was put forward by Chapman and Enskog \cite{chapman1916vi,chapman1990mathematical,enskog1917kinetische}, independently. Chapman and Enskog's approach was broadly favored, since it led to Navier-Stokes theory -- the most widely employed fluid-dynamical theory in the non-relativistic regime. 

In this section, we discuss the relativistic generalizations of the Chapman-Enskog and Hilbert series. We then present a novel perturbative scheme that can be used to systematically derive the BDNK equations from kinetic theory. We present microscopic expressions for the transport coefficients appearing in all fluid-dynamical theories emerging from each perturbative scheme, for arbitrary matching conditions.

\subsection{Chapman-Enskog expansion}
\label{sec:Chap-Ensk-1}

The most widespread formalism used in the derivation of relativistic fluid dynamics from kinetic theory is the Chapman-Enskog expansion \cite{deGroot:80relativistic}. In the relativistic regime, this formalism leads to equations of motion for the macroscopic quantities which violate causality and are linearly unstable around global equilibrium \cite{pichon:65etude,hiscock1983stability,hiscock:85generic,Pu:2009fj,bemfica:18causality,Gavassino:2020ubn,brito2020linear,Gavassino:2021kjm} and, for this reason, such theory cannot be applied to determine the spacetime evolution of relativistic fluids. Nevertheless, Chapman-Enskog theory illustrates the basic aspects of the derivation of fluid dynamics from kinetic theory and shall be discussed in this section assuming general matching conditions.

In this formalism, a perturbative solution of the Boltzmann equation is constructed in terms of an expansion in powers of gradients of the fluid-dynamical fields. In practice, one converts the original Boltzmann equation into the following perturbative problem, introducing the book-keeping (dimensionless) parameter $\epsilon$,
\begin{equation}
\label{eq:chap-ensk}
\begin{aligned}
\epsilon E_{\mathbf{p}} Df_{\mathbf{p}} + \epsilon p^\mu \nabla_\mu f_{\mathbf{p}}   = C[f_{\mathbf{p}}].
\end{aligned}
\end{equation}
Conservation of particle number (in binary collisions), energy, and momentum in microscopic collisions guarantee that the first two moments of the collision term vanish \cite{deGroot:80relativistic,cercignani:02relativistic},
\begin{equation}
\label{eq:non-lin-C-prop}
\begin{aligned}
\int dP \,C[f_{\mathbf{p}}]=0,
\int dP \,p^{\mu}C[f_{\mathbf{p}}]=0.
\end{aligned}
\end{equation}
This leads to the general conservation laws \eqref{eq:consv-eqns}, which are actually non-perturbative in the parameter $\epsilon$,
\begin{equation}
\label{eq:non-pert-csv}
\begin{aligned}
\partial_{\mu} \int dP \, p^{\mu} f_{\bf p} = 0, \
\partial_{\nu} \int dP \, p^{\mu} p^{\nu} f_{\bf p}=0.
\end{aligned}
\end{equation}

We then consider the following asymptotic solution for $f_{\mathbf{p}}$, 
\begin{equation}
\label{eq:chapman-expn-SPDF}
f_{\mathbf{p}} = \sum_{i=0}^{\infty} \epsilon^{i} f^{(i)}_{\mathbf{p}}.    
\end{equation}
The Boltzmann equation is solved order by order in $\epsilon$ \cite{cercignani:02relativistic,deGroot:80relativistic}, leading to  equations that can be solved to obtain the expansion coefficients $f^{(i)}_{\mathbf{p}}$. A solution of the original equation is then recovered by setting $\epsilon=1$. Since $\epsilon$ is inserted multiplying a gradient of $f_{\mathbf{p}}$, it effectively serves as a book-keeping parameter to count powers or/and order of gradients. As we shall demonstrate in the following, the zeroth-order solution of this series will lead to ideal fluid dynamics and the first order solution to Navier-Stokes theory. In practice, it is extremely complicated to obtain any solution beyond first-order. 

To zeroth-order in $\epsilon$ one obtains the following nonlinear equation for $f^{(0)}_{\mathbf{p}}$,
\begin{equation}
\label{eq:chap-ensk0}
\begin{aligned}
0 = C[f^{(0)}_{\mathbf{p}}].
\end{aligned}
\end{equation}
The solution to this equation is well-known and one readily identifies $f^{(0)}_{\mathbf{p}}$ as the local equilibrium distribution function. In the classical limit, this amounts to 
\begin{equation}
\label{eq:f0}
\begin{aligned}
 f^{(0)}_{\mathbf{p}}  = \exp{(\alpha-\beta u_\mu p^\mu)} \equiv f_{0\bf{p}},
\end{aligned}
\end{equation}
with $\alpha$ and $\beta$ being Lorentz scalars and $u^\mu$ a unitary time-like 4-vector ($u^\mu u_\mu=1$). These quantities, are at this point, arbitrary functions of spacetime coordinates. We naturally identify $\beta$ as the inverse temperature, $\alpha$ as the thermal potential, and $u^\mu$ as the 4-velocity. As already stated, these quantities are formally defined by the matching conditions introduced in Eq.~\eqref{eq:matching_kinetic1}.

The algebraic equations obtained from the higher-order terms in $\epsilon$ are obtained by expanding the collision term and the comoving time derivative of $f_{\mathbf{p}}$ in powers of $\epsilon$,
\begin{equation}
\label{eq:chap-ensk-DandC}
\begin{aligned}
Df_{\mathbf{p}}&= D^{(0)}f_{\mathbf{p}} + \epsilon D^{(1)}f_{\mathbf{p}} + \epsilon^2 D^{(2)}f_{\mathbf{p}} + \ldots, 
\\
C[f] &= \epsilon C^{(1)} + \epsilon^2 C^{(2)} + \ldots, 
\end{aligned}
\end{equation}
where $D^{(n)}f_{\mathbf{p}}$ and $C^{(n)}$ denote the $n$-th order contribution in $\epsilon$ of the comoving derivative of $f_{\mathbf{p}}$ and the collision term, respectively. The expansion of the collision term is simply obtained by replacing the expansion of the single-particle distribution into the expression for the collision term. We have already used that the zeroth-order contribution of the collision term is zero, see Eq.\ \eqref{eq:chap-ensk0}. The expansion of the comoving derivative is more convoluted and shall be explained later. Then, 
\begin{equation}
\label{eq:chap-enskn}
\begin{aligned}
 E_{\mathbf{p}} D^{(n-1)}f_{\mathbf{p}} +  p^\mu \nabla_\mu f^{n-1}_{\mathbf{p}}   = C^{(n)}, \, \, \, n \geq 1.
\end{aligned}
\end{equation}

The first order solution satisfies
\begin{equation}
\label{eq:chap-ensk1}
\begin{aligned}
E_{\mathbf{p}} D^{(0)}f_{\mathbf{p}} +  p^\mu \nabla_\mu 
 f_{0\bf{p}}  = f_{0\bf{p}} \hat{L}\phi_{\mathbf{p}},
\end{aligned}
\end{equation}
where $\phi_\mathbf{p} \equiv f^{(1)}_{\mathbf{p}}/f_{0\bf{p}}$ defines the first correction to the local equilibrium distribution and $\hat{L}$ is the linearized collision operator 
\begin{equation}
\begin{aligned}
&  \hat{L}\phi_{\mathbf{p}} \equiv
\int dQ \ dQ^{\prime} \ dP^{\prime} \tilde{\mathcal{W}}_{pp'  \leftrightarrow qq'} f_{\mathbf{p}'}^{\rm{eq}} (\phi_{\mathbf{p}} + \phi_{\mathbf{p}'}  - \phi_{\mathbf{q}} - \phi_{\mathbf{q}'} ).
\end{aligned}    
\end{equation}
The first order solution $\phi_{\bf p}^{(1)}$ is obtained by inverting the linear operator $\hat{L}$. Before doing so, let us discuss how to compute $D^{(0)}f_{\bf p}$. Naively, one would identify this quantity as $D f_{0\bf{p}}$ (as was done by Hilbert, see the next subsection). However, Chapman and Enskog argued that the conservation laws introduce higher order contributions in $\epsilon$ into this derivative. This can be seen using that
\begin{equation}
\label{eq:df/dt}
\begin{aligned}
D f_{0\bf{p}} = \left( D\alpha - E_{\mathbf{p}} D\beta  -  \beta Du_{\mu} p^{\langle \mu \rangle}\right) f_{0\bf{p}}.
\end{aligned}    
\end{equation}
Then, with the conservation laws \eqref{eq:basic-hydro-EoM}, we re-express this term as  
\begin{equation}
\begin{aligned}
D f_{0\bf{p}} 
=
\left[ \mathcal{A}_{\bf p} \theta   -  \frac{\beta}{\varepsilon_{0} + P_{0}} p_{\langle \mu \rangle} \nabla^{\mu}P_{0}\right] f_{0\bf{p}}
+
\mathcal{O}(\delta n, \delta \varepsilon, \Pi, \nu^{\mu}, h^{\mu}, \pi^{\mu \nu}), 
\end{aligned}    
\end{equation}
where we introduced the  function
\begin{equation}
\label{eq:def-Ap}
\begin{aligned}
& \mathcal{A}_{\bf p} =%
\frac{- I_{3,0}I_{1,0} + I_{2,0}(I_{2,0} + I_{2,1})
-
E_{\bf p} 
I_{1,0}I_{2,1}}{I_{3,0} I_{1,0} - I_{2,0}^{2}}, 
\end{aligned}    
\end{equation}
which is written in terms of the thermodynamic integrals
\begin{equation}
\begin{aligned}
& I_{n,q} = \frac{1}{(2q+1)!!} \int dP \left( -\Delta^{\lambda \sigma}p_{\lambda} p_{\sigma} \right)^{q} E_{\mathbf{p}}^{n-2q} f_{0\bf{p}}.
\end{aligned}    
\end{equation}
Note that the dissipative currents $\delta n, \delta \varepsilon, \Pi, \nu^{\mu}, h^{\mu}, \pi^{\mu \nu}$ vanish in equilibrium and, thus, they are at least of $\mathcal{O}(\epsilon)$. $D^{(0)}f$ is the zeroth-order contribution of $Df_{\bf p}$ and, hence, we can identify it as the zeroth-order contribution in $\epsilon$ of $Df_{0\bf{p}}$,
\begin{equation}
\label{eq:time-like-D-subst}
\begin{aligned}
D^{(0)} f_{\bf p}
=
\left[ \mathcal{A}_{\bf p} \theta   -  \frac{\beta}{\varepsilon_{0} + P_{0}} p_{\langle \mu \rangle}\nabla^{\mu}P_{0}\right] f_{0\bf{p}}.
\end{aligned}    
\end{equation}
Thus, the equation above is a constraint that must be enforced when determining the first-order solution of the Chapman-Enskog expansion. In practice, it guarantees that any time-like derivative of a fluid-dynamical field must always be replaced by space-like derivatives of these fields. It is this feature that will make the Chapman-Enskog series an expansion solely in powers of \textit{space-like} gradients (i.e., there are no time derivatives in the local rest frame of the fluid). Afterwards, we use the identity 
\begin{equation}
\begin{aligned}
\label{eq:Gibbs-duhem}
\nabla_{\mu} \beta = \frac{1}{\varepsilon_{0}+ P_{0}} \left( n_{0} \nabla_{\mu} \alpha  - \beta \nabla_{\mu} P_{0} \right),
\end{aligned}
\end{equation}
which can be derived directly from Gibbs-Duhem relation, and Eq.~\eqref{eq:chap-ensk1} becomes
\begin{equation}
\label{eq:eq-int-tensors-CE}
\begin{aligned}
&
\left( \mathcal{A}_{\bf p} -  \frac{\beta}{3} \Delta^{\lambda \sigma}p_{\lambda} p_{\sigma} \right) \theta + \left(1 - \frac{n_{0} E_{\mathbf{p}}}{\varepsilon_{0}+ P_{0}} \right) p^{\langle \mu \rangle} \nabla_{\mu} \alpha 
-
\beta p^{\langle \mu}p^{\nu \rangle} \sigma_{\mu \nu} 
=  \hat{L}\phi_{\mathbf{p}}.
\end{aligned}    
\end{equation}

The linear operator $\hat{L}$ satisfies several fundamental properties that are relevant to our discussion. First, it is self-adjoint, 
\begin{equation}
\label{eq:selfadjoint}
\begin{aligned}
\int dP f_{0\bf{p}} A_{\mathbf{p}}  \hat{L}B_{\mathbf{p}} = \int dP f_{0\bf{p}} B_{\mathbf{p}} \hat{L}A_{\mathbf{p}},
\end{aligned}
\end{equation}
with $A_{\mathbf{p}}$ and $B_{\mathbf{p}}$ being arbitrary functions of momentum (modulo some assumptions to ensure convergence). Furthermore, this operator has 5 degenerate eigenfunctions (the five microscopic quantities that are conserved in binary elastic collisions) with a vanishing eigenvalue,
\begin{equation}
\label{eq:zeromodes}
\begin{aligned}
\hat{L}1 = 0, \, \, \,  \hat{L}p^\mu=0.
\end{aligned}
\end{equation}
The self-consistency aspect of this approach may be demonstrated by multiplying Eq.\ \eqref{eq:eq-int-tensors-CE} by $1$ or $p^\nu$ and verifying if these compatibility conditions are indeed satisfied. Using properties \eqref{eq:selfadjoint} and \eqref{eq:zeromodes} of the linear collision operator, one finds the equations
\begin{equation}
\label{eq:Moments0}
\begin{aligned}
&
\int dP \, \left[ \left( \mathcal{A}_{\bf p} -  \frac{\beta}{3} \Delta^{\mu \nu}p_{\mu} p_{\nu} \right) \theta + \left(1 - \frac{n_{0} E_{\mathbf{p}}}{\varepsilon_{0}+ P_{0}} \right) p^{\langle \mu \rangle} \nabla_{\mu} \alpha 
-
\beta p^{\langle \mu}p^{\nu \rangle} \sigma_{\mu \nu} 
\right] = 0,\\
&
\int dP \, p^{\lambda}\left[ \left( \mathcal{A}_{\bf p} -  \frac{\beta}{3} \Delta^{\mu \nu}p_{\mu} p_{\nu} \right) \theta + \left(1 - \frac{n_{0} E_{\mathbf{p}}}{\varepsilon_{0}+ P_{0}} \right) p^{\langle \mu \rangle} \nabla_{\mu} \alpha 
-
\beta p^{\langle \mu}p^{\nu \rangle} \sigma_{\mu \nu} 
 \right] = 0.
\end{aligned}
\end{equation}
These conditions are automatically satisfied without imposing any further constraints to the solution since
\begin{equation}
\label{eq:orthg-srcs}
\begin{aligned}
\int dP \, \left( \mathcal{A}_{\bf p} -  \frac{\beta}{3} \Delta^{\mu \nu}p_{\mu} p_{\nu} \right) 
= 0, 
\quad 
\int dP \, E_{\bf p} \left( \mathcal{A}_{\bf p} -  \frac{\beta}{3} \Delta^{\mu \nu}p_{\mu} p_{\nu} \right) 
= 0,
\quad
\int dP \Delta^{\mu \nu}p_{\mu} p_{\nu} \left(1 - \frac{n_{0} E_{\mathbf{p}}}{\varepsilon_{0}+ P_{0}} \right) 
 =
0,
\end{aligned}    
\end{equation}
and the tensors $1$, $p^{\langle \mu \rangle}$, and $p^{\langle \mu} p^{\nu \rangle}$ are elements of an orthogonal basis \cite{denicol2012derivation}. This demonstrates that the approximation is consistent with the fundamental properties of the linearized collision term. Note that if the time-like derivatives of the distribution function were not properly evaluated within the perturbative scheme, this would not be the case.

Equation \eqref{eq:eq-int-tensors-CE} is an inhomogeneous linear integral equation for $\phi_{\bf p}$. The general solution of this equation is written as
\begin{equation}
\label{eq:phi-hom-phi-part}
\begin{aligned}
& \phi_{\bf p} = \phi_{\bf p}^{\rm{hom}} + \phi_{\bf p}^{\rm{part}},
\end{aligned}    
\end{equation}
where $\phi_{\bf p}^{\rm{hom}}$ is the solution to the homogeneous equation, $\hat{L}\phi_{\bf{p}}=0$, and 
$\phi_{\bf p}^{\rm{part}}$ is a particular solution to the inhomogeneous equation. Given the zero-modes of the collision operator, the homogeneous component is
\begin{equation}
\label{eq:hom-solution-CE}
\phi_{\bf p}^{\rm{hom}} = a + b_{\mu} p^{\mu},    
\end{equation}
where $a$ and $b_{\mu}$ are arbitrary real-valued constants, which will be determined by imposing the matching conditions \eqref{eq:matching_kinetic1} and \eqref{eq:matching_kinetic2}. The arbitrariness in the choice of these constants is reflected on the fact that the choice of matching conditions is also arbitrary in kinetic theory.

Since $\hat{L}$ is a linear operator, the particular solution $\phi_{\bf p}^{\rm{part}}$ must have the general form 
\begin{equation}
\label{eq:part-sol-phi-CE}
\begin{aligned}
\phi_{\bf p}^{\rm{part}} = \mathcal{S}_{\bf p} \theta +  \mathcal{V}_{\bf p} p^{\langle \mu \rangle} \nabla_{\mu} \alpha 
+
\mathcal{T}_{\bf p} p^{\langle \mu}p^{\nu \rangle} \sigma_{\mu \nu}, 
\end{aligned}    
\end{equation}
where $\mathcal{S}_{\bf p}$, $\mathcal{V}_{\bf p}$, and $\mathcal{T}_{\bf p}$ are unknown functions of $E_{\bf{p}}$.
The next step is to replace the particular solution \eqref{eq:part-sol-phi-CE} into Eq.~\eqref{eq:eq-int-tensors-CE},
\begin{equation}
\label{eq:int-SVT}
\begin{aligned}
&
\left[\left( \mathcal{A}_{\bf p} -  \frac{\beta}{3} \Delta^{\lambda \sigma}p_{\lambda} p_{\sigma} \right) \theta + \left(1 - \frac{n_{0} E_{\mathbf{p}}}{\varepsilon_{0}+ P_{0}} \right) p^{\langle \mu \rangle} \nabla_{\mu} \alpha 
-
\beta p^{\langle \mu}p^{\nu \rangle} \sigma_{\mu \nu} \right] f_{0\bf{p}}
\\
&
=  
\theta f_{0\bf{p}}\hat{L}\left[\mathcal{S}_{\bf p}\right] +  \nabla_{\mu} \alpha
f_{0\bf{p}}\hat{L} \left[\mathcal{V}_{\bf p} p^{\langle \mu \rangle} \right] 
+
\sigma_{\mu \nu}
f_{0\bf{p}}\hat{L} \left[\mathcal{T}_{\bf p} p^{\langle \mu}p^{\nu \rangle} \right]  .
\end{aligned}    
\end{equation}
This results in coupled integral equations for $\mathcal{S}$, $\mathcal{V}$, and $\mathcal{T}$. We now expand these functions using a complete basis of functions of $E_{\bf p}$, $P_{n}^{(\ell)}$, $n, \ell=0,1, \cdots$,
\begin{equation}
\label{eq:SVT-sum-infty}
\begin{aligned}
&
\mathcal{S}_{\bf p} =  \sum_{n = 0}^{\infty} s_{n} P_{n}^{(0)}, \ \
\mathcal{V}_{\bf p} =  \sum_{n = 0}^{\infty} v_{n} P_{n}^{(1)}, \ \
\mathcal{T}_{\bf p} = \sum_{n = 0}^{\infty} t_{n} P_{n}^{(2)}.
\end{aligned}    
\end{equation}
Equation \eqref{eq:int-SVT} can be decoupled by multiplying it by the basis elements $P_{n}^{(\ell)}p^{\langle \mu_{1}}  \cdots p^{\mu_{\ell} \rangle}$ and then integrating over momentum. This leads to the following systems of equations,
\begin{subequations}
\label{eq:sys-SVT}
\begin{align}
&
\label{eq:sys-SVTa}
\sum_{n} S_{rn} s_{n} = 
\int dP P_{r}^{(0)}\left( \mathcal{A}_{\bf p} -  \frac{\beta}{3} \Delta^{\lambda \sigma}p_{\lambda} p_{\sigma} \right)f_{0\bf{p}} \equiv  A_{r},
, \\
&
\label{eq:sys-SVTb}
\sum_{n} V_{rn} v_{n} = \int dP \left(\Delta^{\mu \nu}p_{\mu} p_{\nu} \right) P_{r}^{(1)} \left(1 - \frac{n_{0} E_{\mathbf{p}}}{\varepsilon_{0}+ P_{0}} \right) f_{0\bf{p}} \equiv B_{r},\\
&
\label{eq:sys-SVTc}
\sum_{n} T_{rn} t_{n} =- \beta \int dP \left(\Delta^{\mu \nu}p_{\mu} p_{\nu} \right)^{2} P_{r}^{(2)}f_{0\bf{p}} \equiv
C_{r} ,
\end{align}    
\end{subequations}
where we defined the following integrals of the linearized collision term
\begin{equation}
\label{eq:mat-def-SVT-gn}
\begin{aligned}
& 
S_{rn} \equiv \int dP P_{r}^{(0)}\hat{L}\left[P_{n}^{(0)}\right] f_{0\bf{p}},\\
&
V_{rn} \equiv \int dP  P_{r}^{(1)} p^{\langle \mu \rangle} \hat{L} \left[P_{n}^{(1)} p_{\langle \mu \rangle} \right] f_{0\bf{p}},\\
&
T_{rn} \equiv \int dP 
P_{r}^{(2)}
p^{\langle \mu}p^{\nu \rangle}
\hat{L} \left[P_{n}^{(2)} p_{\langle \mu}p_{\nu \rangle} \right] f_{0\bf{p}}.
\end{aligned}    
\end{equation} 
Equations \eqref{eq:sys-SVT} can be schematically inverted as 
\begin{equation}
\label{eq:coeffs-svt}
\begin{aligned}
& 
s_{n} = \sum_{m} [S^{-1}]_{nm} A_{m},
\quad
v_{n} = \sum_{m} [V^{-1}]_{nm} B_{m},
\quad
t_{n} = \sum_{m} [T^{-1}]_{nm} C_{m}.  
\end{aligned}    
\end{equation}
We note that, if the basis contains parts of the homogeneous solution, the corresponding terms must be removed from the inversion procedure. We discuss an example of this procedure in Appendix \ref{sec:zero-mode-CE}. Nevertheless, we remark that this will not be the case for the basis employed in this work.    

The coefficients of the homogeneous solution are obtained from the matching conditions \eqref{eq:matching_kinetic1} and \eqref{eq:matching_kinetic2}, which when substituted in Eq.~\eqref{eq:phi-hom-phi-part} lead to the conditions
\begin{equation}
\label{eq:a-bu-bmu}
\begin{aligned}
& I_{q,0} a + I_{q+1,0} b_{\mu} u^{\mu} 
=
- \left\langle E_{\bf p}^{q} \mathcal{S}_{\bf p} \right\rangle \theta, 
\\
& I_{s,0} a + I_{s+1,0} b_{\mu} u^{\mu} 
=
- \left\langle E_{\bf p}^{s} \mathcal{S}_{\bf p} \right\rangle \theta, 
\\
& I_{z+2,1} b_{\langle \mu \rangle} 
=
\frac{1}{3} \left\langle \left(\Delta^{\mu \nu}p_{\mu} p_{\nu} \right) E_{\bf p}^{z} \mathcal{V}_{\bf p} \right\rangle_{0} \nabla_{\mu} \alpha,
\end{aligned}    
\end{equation}
where it was used that $p^{\mu} = E_{\bf p} u^{\mu} + p^{\langle \mu \rangle}$. These are solved with 

\begin{equation}
\label{eq:a,b-mu}
\begin{aligned}
&
a = \frac{I_{q+1,0} \langle E_{\bf p}^{s} \mathcal{S}_{\bf p} \rangle_{0} - \langle E_{\bf p}^{q} \mathcal{S}_{\bf p} \rangle_{0} I_{s+1,0}}{G_{s+1,q}}
\theta,
\\
&
b^{\mu} u_{\mu} = \frac{\langle E_{\bf p}^{q} \mathcal{S}_{\bf p}\rangle_{0} I_{s,0} - I_{q,0} \langle E_{\bf p}^{s} \mathcal{S}_{\bf p} \rangle_{0}}{G_{s+1,q}} \theta, \\
&
b_{\langle \mu \rangle} 
=
\frac{1}{3}\frac{\left\langle \left(\Delta^{\mu \nu}p_{\mu} p_{\nu} \right) E_{\bf p}^{z} \mathcal{V}_{\bf p} \right\rangle_{0}}{I_{z+2,1}} \nabla_{\mu} \alpha, 
\end{aligned}    
\end{equation}
where $G_{n,m} = I_{n,0} I_{m,0} - I_{n-1,0}I_{m+1,0}$.

Finally, combining the results displayed above, we obtain the solution for the first order Chapman-Enskog deviation function
\begin{equation}
\begin{aligned}
&
\phi_{\bf p} = 
\tilde{\mathcal{S}}_{\bf p} 
\theta 
+
\tilde{\mathcal{V}}_{\bf p} 
p^{\langle \mu \rangle} \nabla_{\mu} \alpha 
+
\mathcal{T}_{\bf p} p^{\langle \mu}p^{\nu \rangle} \sigma_{\mu \nu}, 
\end{aligned}    
\end{equation}
where we defined the scalar functions of $E_{\bf p}$ 
\begin{equation}
\label{eq:til-SVT-sol}
\begin{aligned}
& 
\tilde{\mathcal{S}_{\bf p}} = \sum_{n \geq 0} \sum_{m \geq 0} [\hat{S}^{-1}]_{nm} A_{m} \left( P_{n}^{(0)}
+
\frac{I_{q+1,0} \mathcal{I}_{sn}^{(0)} - \mathcal{I}_{qn}^{(0)} I_{s+1,0}}{G_{s+1,q}}
+
\frac{\mathcal{I}_{qn}^{(0)} I_{s,0} - I_{q,0} \mathcal{I}_{sn}^{(0)}}{G_{s+1,q}} E_{\bf p} 
\right),
\\
&
\tilde{\mathcal{V}}_{\bf p} 
=
\sum_{n \geq 0} \sum_{m \geq 0}  [\hat{V}^{-1}]_{nm} B_{m} \left( P_{n}^{(1)} 
-
\frac{\mathcal{I}_{zn}^{(1)}}{I_{z+2,1}} 
\right), \\
&
\mathcal{T}_{\bf p} =
\sum_{n \geq 0} \sum_{m \geq 0} [T^{-1}]_{nm} C_{m} P_{n}^{(2)},
\end{aligned}    
\end{equation}
and the following thermodynamic integral  
\begin{equation}
\begin{aligned}
& \mathcal{I}_{mn}^{(\ell)} = \frac{(-1)^{\ell}}{(2\ell +1)!!} \left\langle \left(\Delta^{\mu \nu}p_{\mu} p_{\nu} \right)^{\ell} E_{\bf p}^{m} P_{n}^{(\ell)} \right\rangle_{0}.
\end{aligned}
\end{equation}
This solution can then be used to obtain the constitutive relations satisfied by the non-equilibrium corrections under general matching conditions. Indeed, definitions \eqref{eq:def_kinetic} yield  
\begin{equation}
\label{eq:def-coeffs}
\begin{aligned}
\Pi & = - \zeta \theta,  \
\delta n =  - \xi \theta, \
 \delta \varepsilon = \chi \theta,\\
\nu^{\mu} & = \kappa \nabla^{\mu} \alpha, \
h^{\mu} = - \lambda \nabla^{\mu} \alpha, \\
\pi^{\mu \nu} & = 2 \eta \sigma^{\mu \nu},
\end{aligned}
\end{equation}
with transport coefficients given by 
\begin{equation}
\label{eq:coeffs-chap-ensk}
\begin{aligned}
&
\zeta = \sum_{n \geq 2} \sum_{m \geq 2}[\hat{S}^{-1}]_{nm} A_{m}
H_{n}^{(\zeta)}, 
\ \
\xi = \sum_{n \geq 2} 
 \sum_{m \geq 2} [\hat{S}^{-1}]_{nm} A_{m} H_{n}^{(\xi)},
\ \
\chi = - \sum_{n \geq 2} 
 \sum_{m \geq 2} [\hat{S}^{-1}]_{nm} A_{m}
H_{n}^{(\chi)},\\
&
\kappa = \sum_{n \geq 1} \sum_{m \geq 1}  [\hat{V}^{-1}]_{nm} B_{m} J_{n}^{(\kappa)}, 
\ \
\lambda = \sum_{n \geq 1} \sum_{m \geq 1}  [\hat{V}^{-1}]_{nm} B_{m} J_{n}^{(\lambda)},\\
&
\eta =  \sum_{n \geq 0} \sum_{m \geq 0} [T^{-1}]_{nm} C_{m} \mathcal{I}_{0n}^{(2)},
\end{aligned}    
\end{equation}
where
\begin{equation}
\label{eq:H,J-expn}
\begin{aligned}
& H_{n}^{(\zeta)}  =   - \frac{1}{3} 
\left(m^{2} \mathcal{I}_{0n}^{(0)} 
-
\mathcal{I}_{2n}^{(0)}
\right)
-
\frac{1}{3}
\frac{m^{2} G_{q+1,0} - G_{q+1,2}}{G_{s+1,q}}
\mathcal{I}_{sn}^{(0)}+
\frac{1}{3}
\frac{m^{2} G_{s+1,0} - G_{s+1,2}}{G_{s+1,q}}
\mathcal{I}_{qn}^{(0)},   \\
&
H_{n}^{(\xi)}  = \mathcal{I}_{1n}^{(0)}
+
\frac{G_{q+1,1}}{G_{s+1,q}}
\mathcal{I}_{sn}^{(0)}
-
\frac{G_{s+1,1}}{G_{s+1,q}}
\mathcal{I}_{qn}^{(0)}, 
\\
&
H_{n}^{(\chi)} = \mathcal{I}_{2n}^{(0)} 
+
\frac{G_{q+1,2}}{G_{s+1,q}}
\mathcal{I}_{sn}^{(0)}
-
\frac{G_{s+1,2}}{G_{s+1,q}}
\mathcal{I}_{qn}^{(0)}, \\
&
J_{n}^{(\kappa)}  =  - \mathcal{I}_{0n}^{(1)} 
+
\frac{I_{2,1}}{I_{z+2,1}}
\mathcal{I}_{zn}^{(1)},
\\
&
J_{n}^{(\lambda)}  =  \mathcal{I}_{1n}^{(1)} 
-
\frac{I_{3,1}}{I_{z+2,1}}
\mathcal{I}_{zn}^{(1)} .
\end{aligned}   
\end{equation}
The transport coefficients are in general quite involved functions of temperature and chemical potential. Some simplification can be made with the usage of phenomenological approximations of the collision term, such as the relaxation time approximation \cite{andersonRTA:74,rocha:21}. It is also relevant to point out that the choice of matching conditions affect greatly some of the transport coefficients, which explicitly depend on the parameters $q$, $s$, and $z$ necessary to define the matching conditions. Indeed, if we use the Landau prescription, $(q,s,z) = (1,2,1)$, we have $\xi = \chi = \lambda = 0$. If, instead, we use the Eckart prescription, then $\xi = \chi = \kappa = 0$. Alternatively, in a matching condition defined such that $q=0$ and $s=2$, we would have $\zeta = 0$. It should also be noted that, in the massless limit, due to the fact that $\mathcal{A}_{\bf p} = -\frac{\beta}{3} E_{\bf p}^{2} $ and $\Delta^{\mu \nu}p_{\mu} p_{\nu} = E_{\bf p}^{2}$, we have that $\mathcal{S}_{\bf p} = 0$, implying that all transport coefficients related to scalar non-equilibrium fields must vanish, i.e.,  $\zeta = \xi = \chi = 0$. Moreover, we note that the combinations 
\begin{equation}
\label{eq:matching-inv-NS}
\begin{aligned}
& \zeta +  \left(\frac{\partial P_{0}}{\partial n_{0}} \right)_{\varepsilon_{0}} \xi
+ \left(\frac{\partial P_{0}}{\partial \varepsilon_{0}} \right)_{n_{0}} \chi 
=
\sum_{n \geq 2} \sum_{m \geq 2}[\hat{S}^{-1}]_{nm} A_{m} \mathcal{H}_{n},
\\
&  \kappa + \frac{n_{0}}{\varepsilon_{0} + P_{0}} \lambda =  \sum_{n \geq 1} \sum_{m \geq 1}  [\hat{V}^{-1}]_{nm} B_{m}  \mathcal{J}_{n}, 
\end{aligned}    
\end{equation}
are matching-invariant. In the above equations,
\begin{equation}
\label{eq:match-inv-NS-defs}
\begin{aligned}
& \left(\frac{\partial P_{0}}{\partial n_{0}} \right)_{\varepsilon_{0}} =  \frac{I_{3,1} I_{2,0} - I_{2,1} I_{3,0}}{I_{2,0}^{2} - I_{1,0}I_{3,0}},
\quad 
\left(\frac{\partial P_{0}}{\partial \varepsilon_{0}} \right)_{n_{0}} 
=
\frac{I_{2,1} I_{2,0} - I_{3,1} I_{1,0}}{I_{2,0}^{2} - I_{1,0}I_{3,0}}, \\
& \mathcal{H}_{n} = -  \frac{1}{3} 
\left(m^{2} \mathcal{I}_{0n}^{(0)} 
-
\mathcal{I}_{2n}^{(0)}
\right)
+
\left(\frac{\partial P_{0}}{\partial n_{0}} \right)_{\varepsilon_{0}}
\mathcal{I}_{1n}^{(0)}
+
\left(\frac{\partial P_{0}}{\partial \varepsilon_{0}} \right)_{n_{0}}
\mathcal{I}_{2n}^{(0)},
\\
&
\mathcal{J}_{n} = - \mathcal{I}_{0n}^{(1)} + \frac{n_{0}}{\varepsilon_{0} + P_{0}}  \mathcal{I}_{1n}^{(1)}.\\
\end{aligned}    
\end{equation}
These expressions can be derived using  \eqref{eq:orthg-srcs} with the identification of the $I_{n,m}$ expressions with the thermodynamic derivatives above. 

The equations of motion are obtained replacing the first order solution for $f_{\bf p}$ in the exact conservation laws \eqref{eq:basic-hydro-EoM}, where the non-equilibrium corrections above, $\{\delta \varepsilon, \delta n,\Pi,\nu^\mu,h^\mu,\pi^{\mu\nu}\}$, are determined in terms of the hydrodynamic variables by the constitutive relations in Eq.\ \eqref{eq:def-coeffs}. Using the results in \cite{Bemfica:2020zjp}, it is straightforward to demonstrate that the resulting equations of motion are acausal for any choice of matching condition. Therefore, one can see that the choice of matching condition cannot render the hydrodynamic theory emerging from the first-order truncation of the Chapman-Enskog expansion causal and (linearly) stable.

\subsection{Hilbert expansion}
\label{sec:Hilb-expn}

In this section we discuss the other perturbative formalism used to derive a fluid-dynamical framework from kinetic theory: the Hilbert expansion \cite{hilbert1912begrundung,cercignani:90mathematical,grad1958principles}. This approach was originally developed by D.~Hilbert in the non-relativistic context, prior to the Chapman-Enskog theory. As previously mentioned, the Hilbert approach is not as widely employed as the former since it does not lead to Navier-Stokes theory, and has not been worked out in detail in the relativistic regime. Nevertheless, understanding the Hilbert expansion allows us to comprehend basic aspects and assumptions made in perturbative derivations of fluid dynamics and, for this reason, we find it useful to work out the details of this formalism (including its transport coefficients) in this section.

The starting point of the Hilbert expansion is identical to that of Chapman-Enskog theory, where one introduces a perturbative parameter into the Boltzmann equation, 
\begin{equation}
\label{eq:hilb-}
\begin{aligned}
 \epsilon p^\mu \partial_\mu f_{\mathbf{p}}   = C[f_{\mathbf{p}}].
\end{aligned}
\end{equation}
As shown in the last section, fundamental properties of the full collision integral \eqref{eq:non-lin-C-prop} lead to non-perturbative conservation laws (see Eq.~\eqref{eq:non-pert-csv}), which act as constraints on $f_{\mathbf{p}}$. Also similarly to Chapman-Enskog theory, we impose a perturbative solution for the single-particle distribution function, as in Eq.~\eqref{eq:chapman-expn-SPDF},
\begin{equation}
\label{eq:hilb-expn-SPDF}
f_{\mathbf{p}} = \sum_{i=0}^{\infty} \epsilon^{i} f^{(i)}_{\mathbf{p}}.    
\end{equation}
Then, solutions are found order by order iteratively. After this task is performed, the book-keeping parameter $\epsilon$ is set to one. 

The zeroth order solution is identical to the one found in Chapman-Enskog theory, and satisfies
\begin{equation}
\label{eq:hilb0}
\begin{aligned}
0 = C[f^{(0)}_{\mathbf{p}}],
\end{aligned}
\end{equation}
leading to the local equilibrium distribution function, already displayed in Eq.~\eqref{eq:f0}. Thus, the Hilbert expansion also recovers ideal fluid dynamics as its zeroth-order solution. 
The next order solutions will differ from those of Chapman-Enskog theory and are obtained from the equations
\begin{equation}
\label{eq:n-th-ord-hilbert}
\begin{aligned}
´p^{\mu}\partial_{\mu}\left( f_{0\bf{p}} \phi^{(n-1)}_{\bf p} \right)
=
f_{0\bf{p}}\Hat{L}[\phi^{(n)}_{\bf p}]
+
\sum_{j=1}^{n-1}J[f_{\bf p}^{(n-j)},f_{\bf p}^{(j)}], \ n \geq 1,
\end{aligned}    
\end{equation}
where we defined the bilinear form of the collision operator
\begin{equation}
\begin{aligned}
& J[f_{\bf p}, g_{\bf p}]
=
\int dQ \ dQ^{\prime} \ dP^{\prime} W_{pp' \leftrightarrow qq'} (f_{\mathbf{p}} g_{\mathbf{p}'} - g_{\mathbf{q}}f_{\mathbf{q}'} ), 
\end{aligned}    
\end{equation}
and introduced the notation $\phi^{(0)}_{\bf p} =1, \ \phi^{(n \geq 1)}_{\bf p} = f^{(n)}_{\bf p}/f_{0\bf{p}}$. We note that the equations above are different than the equations resulting from Chapman-Enskog theory. In the former, time-like and space-like derivatives of the single-particle distribution function are explicitly separated on the left-hand side of the equation. Such time-like derivatives of the distribution function are then expanded in $\epsilon$, leading to a rearrangement of the perturbative series,  see Eqs.~\eqref{eq:chap-ensk-DandC}, \eqref{eq:chap-enskn}, and \eqref{eq:chap-ensk1}. This expansion of the time-like derivatives is then determined systematically using the zero modes of the linear collision operator, as shown in \eqref{eq:time-like-D-subst}. Historically speaking, this approach was understood as a correction to the Hilbert series. As already mentioned, Chapman and Enskog's approach was broadly favored, since it led to Navier-Stokes theory and, thus,  provided the first microscopic derivation of this widely employed fluid-dynamical theory. In the remainder of this section we discuss the original framework proposed by Hilbert and its implications. 

The main feature of the Hilbert expansion is the emergence of an infinite set of conservation laws that must be solved independently order by order. This can be seen by multiplying Eq.~\eqref{eq:n-th-ord-hilbert} with 1 and $p^{\mu}$ and integrating it in momentum space, leading to 
\begin{equation}
\label{eq:hilb-eoms-f}
\begin{aligned}
& \int dP \left[p^{\mu}\partial_{\mu}\left( f_{0\bf{p}} \phi^{(n-1)}_{\bf p} \right)\right] = 0, 
\int dP \,p^{\alpha} \left[p^{\mu}\partial_{\mu}\left( f_{0\bf{p}} \phi^{(n-1)}_{\bf p} \right) \right] = 0, \quad n \geq 1
\end{aligned}    
\end{equation}
where we used properties \eqref{eq:selfadjoint} and \eqref{eq:zeromodes} of the collision operator, $\hat{L}$, and the following property of the bilinear collision term, $J[f,g]$, \cite{grad1958principles},
\begin{equation}
\label{eq:non-lin-C-prop-fg}
\begin{aligned}
\int dP \,J[f_{\mathbf{p}},g_{\mathbf{p}}]=0,
\int dP \,p^{\mu}J[f_{\mathbf{p}},g_{\mathbf{p}}]=0.
\end{aligned}
\end{equation}
This implies that the conservation laws of particle number, energy, and momentum obtained from \eqref{eq:hilb-} must be solved independently order by order in the perturbative parameter $\epsilon$, 
\begin{equation}
\label{eq:cons-tmunu-hilb}
\begin{aligned}
&
\partial_{\mu} N^{\mu}_{(k)} = 0, \quad 
\partial_{\mu} T^{\mu \nu}_{(k)} = 0, \quad k \geq 0 .
\end{aligned}
\end{equation}
It is then convenient to decompose this set of conserved currents in terms of the 4-velocity, as explained in Sec.~\ref{sec:fl-dyn-vars},
\begin{subequations}
\label{eq:decompos-curr-hilb}
\begin{align}
N^{\mu}_{(k)} & \equiv \int dP p^{\mu} f^{(k)}_{\bf p}  = n_{(k)} u^{\mu} + \nu^{\mu}_{(k)}  \\
    T^{\mu \nu}_{(k)} & \equiv \int dP p^{\mu}p^{\nu} f^{(k)}_{\bf p} = \varepsilon_{(k)} u^{\mu} u^{\nu} - \Pi_{(k)} \Delta^{\mu \nu} + h^{\mu}_{(k)} u^{\nu} + h^{\nu}_{(k)} u^{\mu} + \pi^{\mu \nu}_{(k)}, \quad k \geq 0,
\end{align}    
\end{subequations}
where $n_{(k)}$, $\varepsilon_{(k)}$, $\Pi_{(k)}$, $\nu^{\mu}_{(k)}$, $h^{\mu}_{(k)}$, and $\pi^{\mu \nu}_{(k)}$ denote, respectively, the $k$-th order contribution to particle density, energy density, bulk viscous pressure, particle diffusion 4-current, energy diffusion 4-current, and shear-stress tensor. Furthermore, at zeroth order, $n_{(0)} = n_{0}$, $\varepsilon_{(0)} = \varepsilon_{0}$, $\Pi_{(0)} = P_{0}$, $\nu^{\mu}_{(0)} = 0$, $h^{\mu}_{(0)} = 0$, and $\pi^{\mu \nu}_{(0)} = 0$. As we shall see later, this will be essential in determining the free parameters that appear in the homogeneous solutions for $\phi_{\bf{p}}^{(n)}$ in Eq.~\eqref{eq:n-th-ord-hilbert}. 

For the sake of completeness, we derive the fluid-dynamical equations stemming from the Hilbert expansion truncated at first order. First, we note 
that the five unknown fields contained in $f_{\bf{p}}^{(0)}$, i.e., the temperature, thermal potential, and 4-velocity, must be determined. In the Hilbert expansion this task is performed by deriving equations for these variables. Using Eq.~\eqref{eq:hilb-eoms-f} with $n=1$ or, equivalently, Eq.~\eqref{eq:cons-tmunu-hilb} with $k=0$, one finds the conservation laws,
\begin{subequations}
 \label{eq:euler-eqns}
\begin{align}
 \label{eq:hydro-EoM-n0-euler}
D n_{0} + n_{0} \theta &= 0, \\
\label{eq:hydro-EoM-eps-euler}
D\varepsilon_{0} + (\varepsilon_{0}+ P_{0}) \theta &= 0, \\
\label{eq:hydro-EoM-umu-euler}
(\varepsilon_{0} + P_{0})Du^{\mu} - \nabla^{\mu}P_{0}  &= 0,
\end{align}
\end{subequations}
where, as already explained in Sec.~\ref{sec:fl-dyn-vars}, $\varepsilon_{0}$, $P_{0}$, and $n_{0}$ are functions of $\alpha$ and $\beta$. We note that the conservation laws above are identical to those obeyed by an ideal fluid, even when the actual system described is out of equilibrium.

We now proceed to determine the first order correction. First, we take Eq.~\eqref{eq:n-th-ord-hilbert} for $n=1$, which reduces to
\begin{equation}
\label{eq:1st-ord-hilbert}
\begin{aligned}
´p^{\mu}\partial_{\mu}f_{0\bf{p}} 
=
f_{0\bf{p}}\Hat{L}[\phi^{(1)}_{\bf p}].
\end{aligned}    
\end{equation}
The left-hand side of Eq.~\eqref{eq:1st-ord-hilbert} can be irreducibly written as
\begin{equation}
\begin{aligned}
p^{\mu} \partial_{\mu} f_{0\bf{p}} = \left[ E_{\mathbf{p}} D\alpha - E_{\mathbf{p}}^{2} D\beta - \frac{\beta}{3} \Delta^{\lambda \sigma}p_{\lambda} p_{\sigma} \theta +   p^{\langle \mu \rangle} \nabla_{\mu}\alpha - E_{\mathbf{p}} p^{\langle \mu \rangle} \left( \beta Du_{\mu} + \nabla_{\mu} \beta \right) - \beta p^{\langle \mu}p^{\nu \rangle} \sigma_{\mu \nu} \right] f_{0\bf{p}}.
\end{aligned}    
\end{equation}
 Note that the time-like derivatives of temperature, thermal potential, and 4-velocity can be substituted by space-like ones analogously to what occurred in Chapman-Enskog theory. Nevertheless, here we have the fundamental difference that this substitution is exact, and not perturbative, since $\alpha$, $\beta$, and $u^\mu$ satisfy the ideal fluid-dynamical equations, see Eqs.~\eqref{eq:euler-eqns}. Hence, Eq.~\eqref{eq:1st-ord-hilbert} becomes 
\begin{equation}
\label{eq:int-SVT-hil}
\begin{aligned}
&
\left[\left( \mathcal{A}_{\bf p} -  \frac{\beta}{3} \Delta^{\lambda \sigma}p_{\lambda} p_{\sigma} \right) \theta + \left(1 - \frac{n_{0} E_{\mathbf{p}}}{\varepsilon_{0}+ P_{0}} \right) p^{\langle \mu \rangle} \nabla_{\mu} \alpha 
-
\beta p^{\langle \mu}p^{\nu \rangle} \sigma_{\mu \nu} \right] f_{0\bf{p}}
=  
 f_{0\bf{p}}\hat{L}\phi_{\bf p}^{(1)},
\end{aligned}    
\end{equation}
which is mathematically equivalent to Eq.\ \eqref{eq:int-SVT} obtained in Chapman-Enskog theory, with $\mathcal{A}_{\bf p}$ already being defined in Eq.~\eqref{eq:def-Ap}. This allows us to proceed with the same steps performed from Eq.~\eqref{eq:phi-hom-phi-part} to Eq.~\eqref{eq:coeffs-svt}, leading to the particular solution
\begin{equation}
\begin{aligned}
&
\phi_{\bf p}^{\rm{part}} = 
\mathcal{S}_{\bf p} 
\theta 
+
\mathcal{V}_{\bf p} 
p^{\langle \mu \rangle} \nabla_{\mu} \alpha 
+
\mathcal{T}_{\bf p} p^{\langle \mu}p^{\nu \rangle} \sigma_{\mu \nu},
\end{aligned}    
\end{equation}
where $\mathcal{S}$, $\mathcal{V}$, and $\mathcal{T}$ are given in Eqs.~\eqref{eq:SVT-sum-infty} and \eqref{eq:coeffs-svt}. 

As before, this solution must be complemented by a homogeneous solution, constructed from a linear combination of the zero-modes of the collision operator 
\begin{equation}
\label{eq:hom-solution-hilb}
\phi_{\bf p}^{\rm{hom}} = a + b_{\mu} p^{\mu}.    
\end{equation}
The five unknown fields that appear in the homogeneous solution will be determined in the same way as done previously for the temperature, thermal potential, and 4-velocity at zeroth order. We derive equations of motion for these quantities using Eq.~\eqref{eq:hilb-eoms-f} with $n=2$ or, equivalently, Eq.~\eqref{eq:cons-tmunu-hilb} with $k=1$. We note that this procedure is carried out order by order, always determining the free parameters of the homogeneous solution using the constraints from Eq.~\eqref{eq:cons-tmunu-hilb}. This guarantees that the energy-momentum tensor and the particle 4-current are always exactly conserved, even when truncated at a given order. We note that this is a crucial difference with respect to the traditional Chapman-Enskog theory, where the undetermined coefficients of the homogeneous solution, $a$ and $b^{\mu}$, are determined using matching conditions. 

Then, using decomposition \eqref{eq:decompos-curr-hilb} for $k=1$ and taking into account that the zeroth order currents obey the zeroth order conservation laws \eqref{eq:euler-eqns} separately, we have
\begin{subequations}
 \label{eq:hydro-EoMs-hilb}
\begin{align}
 \label{eq:hydro-EoM-n0-hilb}
D n_{(1)} + n_{(1)} \theta + \partial_{\mu} \nu^{\mu}_{(1)} &= 0, \\
\label{eq:hydro-EoM-eps-hilb}
D \varepsilon_{(1)} + (\varepsilon_{(1)} + \Pi_{(1)}) \theta - \pi^{\mu \nu}_{(1)} \sigma_{\mu \nu} + \partial_{\mu}h^{\mu}_{(1)} + u_{\mu} Dh^{\mu}_{(1)} &= 0, \\
\label{eq:hydro-EoM-umu-hilb}
(\varepsilon_{(1)} + \Pi_{(1)}) Du^{\mu} - \nabla^{\mu}\Pi_{(1)} + h^{\mu}_{(1)} \theta + h^{\alpha}_{(1)} \Delta^{\mu \nu} \partial_{\alpha}u_{\nu} +  \Delta^{\mu \nu} Dh_{(1)\nu} + \Delta^{\mu \nu} \partial_{\alpha}\pi^{\alpha}_{(1)\nu} &= 0.
\end{align}
\end{subequations}
These equations are complemented with the constitutive relation satisfied by the shear-stress tensor, 
\begin{equation}
\begin{aligned}
& \pi^{\mu \nu}_{(1)} = 2 \eta \sigma^{\mu \nu},
\end{aligned}    
\end{equation}
where the transport coefficient $\eta$ is identical to the one obtained in Chapman-Enskog theory, see Eq.~\eqref{eq:coeffs-chap-ensk}. Furthermore, the variables $n_{(1)}$, $\varepsilon_{(1)}$, $\Pi_{(1)}$, $\nu^{\mu}_{(1)}$, and $h^{\mu}_{(1)}$ can be expressed in terms of the fields $a$ and $b^\mu$ and gradients of $\alpha$ and $u^\mu$. Using the decomposition \eqref{eq:decompos-curr-hilb} and definitions \eqref{eq:def_kinetic}, we have
\begin{equation}
\label{eq:constrains-hilbs-ab}
\begin{aligned}
& n_{(1)} = a I_{1,0} + (b_{\mu} u^{\mu}) I_{2,0} - \xi_{H} \theta, \\
& \varepsilon_{(1)} = a I_{2,0} + (b_{\mu} u^{\mu}) I_{3,0} + \chi_{H} \theta, \\
& \Pi_{(1)} = a I_{2,1} + (b_{\mu} u^{\mu}) I_{3,1} + \zeta_{H} \theta, \\
& \nu^{\mu}_{(1)} = - I_{2,1} b^{\langle \mu \rangle}
+
\kappa_{H} \nabla^{\mu} \alpha,
\\ 
& h^{\mu}_{(1)} = - I_{3,1} b^{\langle \mu \rangle}
-
\lambda_{H} \nabla^{\mu} \alpha,
\end{aligned}    
\end{equation}
where we defined the following transport coefficients
\begin{equation}
\begin{aligned}
& 
\xi_{H} = - \left\langle E_{\bf p} \mathcal{S}_{\bf p} \right\rangle_{0}, 
\quad  
\chi_{H} = \left\langle E_{\bf p}^{2} \mathcal{S}_{\bf p} \right\rangle_{0},
\quad
\zeta_{H} = - \frac{1}{3} \left\langle \left( \Delta^{\alpha \beta} p_{\alpha} p_{\beta}\right) \mathcal{S}_{\bf p} \right\rangle_{0}, \\
&
\kappa_{H} = \frac{1}{3} \left\langle \left( \Delta^{\alpha \beta} p_{\alpha} p_{\beta}\right) \mathcal{V}_{\bf p} \right\rangle_{0},
\quad
\lambda_{H} = - \frac{1}{3} \left\langle \left( \Delta^{\alpha \beta} p_{\alpha} p_{\beta}\right) E_{\bf p} \mathcal{V}_{\bf p} \right\rangle_{0}.
\end{aligned}    
\end{equation}
 These transport coefficients depend on the temperature and thermal potential, which are determined by the zeroth order equations of motion \eqref{eq:euler-eqns}. We note that the homogeneous solution has five independent degrees of freedom while $n_{(1)}$, $\varepsilon_{(1)}$, $\Pi_{(1)}$, $\nu^{\mu}_{(1)}$, and $h^{\mu}_{(1)}$ define a total of nine degrees of freedom. Using \eqref{eq:constrains-hilbs-ab}, we can derive the constraints which relate the three scalar fields and the two vector fields, which are given by
\begin{subequations}
\label{eq:constrains-hilbs-fin}
\begin{align}
& 
\label{eq:constrains-hilbs-fin-sca}
G_{3,1} \Pi_{(1)} = (I_{2,1} I_{3,0} - I_{3,1} I_{2,0}) (n_{(1)} + \xi_{H} \theta) 
+
(I_{2,1} I_{2,0} - I_{3,1} I_{1,0}) 
(\varepsilon_{(1)} - \chi_{H} \theta)
+
\zeta_{H} \theta, \\
\label{eq:constrains-hilbs-fin-vec}
& \frac{1}{I_{2,1}} \nu^{\mu}_{(1)} - \frac{1}{I_{3,1}} h^{\mu}_{(1)}  
= 
\left(\frac{\kappa_{H}}{I_{2,1}} 
+
\frac{\lambda_{H}}{I_{3,1}} \right) \nabla^{\mu} \alpha .
\end{align}    
\end{subequations}
In the massless limit, $\zeta_{H} = \chi_{H} = \xi_{H} = 0$ and the first constraint reduces to $\Pi_{(1)} = (1/3)  \varepsilon_{(1)}$, which is consistent with the tracelessness property of the of the energy-momentum tensor in this case. As for the second constraint, it reduces to $\nu^{\mu} - (\beta/4)h^{\mu} = (n_{0}/12) \nabla^{\mu}\alpha$.

The system formed by the partial differential equations \eqref{eq:euler-eqns} and 
\eqref{eq:hydro-EoMs-hilb},  together with the constraints \eqref{eq:constrains-hilbs-fin} correspond to the fluid-dynamical equations that emerge from the first-order truncation of the Hilbert expansion. These equations are not of the form of Navier-Stokes theory and, for this reason, were readily abandoned for applications in the non-relativistic regime. In the relativistic regime they are not often employed as well even though they do not appear to display the same pathologies of  relativistic Navier-Stokes theory.

\subsection{New perturbative expansion}
\label{sec:Chap-Ensk-2}

In the previous sections we discussed two traditional perturbative frameworks that can be employed to derive fluid dynamical equations in relativistic kinetic theory. Nevertheless, both frameworks have fundamental flaws that must be addressed. As mentioned before, the Chapman-Enskog expansion leads to fluid-dynamical equations in the relativistic regime that are acausal and linearly unstable around global equilibrium. On the other hand, the Hilbert expansion leads to an infinite set of conservation laws, overestimating the number of conserved quantities in the fluid. Therefore, it cannot correctly describe the type of collective excitations that appear near equilibrium. In this section, we discuss another perturbative procedure that leads to relativistic fluid-dynamical equations that do not contain the above mentioned undesired and unphysical features: the BDNK equations \cite{bemfica:18causality,Kovtun:2019hdm,bemfica2019nonlinear,Hoult:2020eho,Bemfica:2020zjp}.

For the practical purposes discussed in this paper, the main difference between the BDNK equations and relativistic Navier-Stokes theory is that the former is built upon constitutive relations for the dissipative currents that do not only contain space-like derivatives of the fluid-dynamical variables. Alternatively, one can say that the Navier-Stokes formulation did not include all the possible terms that appear in a first-order formulation. Even though this may appear to be a minor difference, it has been proven that the addition of time-like derivatives of the fluid-dynamical variables to the constitutive relations can change the character of the equations of motion in such a way that causal and stable formulations of hydrodynamics computed at first-order in derivatives can be be obtained \cite{bemfica:18causality,Kovtun:2019hdm,bemfica2019nonlinear,Hoult:2020eho,Bemfica:2020zjp}, as long as a judicious choice for the definition of the hydrodynamic variables out of equilibrium are employed. 

The main reason the Chapman-Enskog expansion leads to Navier-Stokes theory and not to BDNK theory is the replacement of time-like derivatives by space-like ones that occurs when obtaining the perturbative solution for the time-like derivatives of the distribution function (cf.~Eq.~\eqref{eq:time-like-D-subst}, for instance). As already discussed, this replacement is essential to guarantee the validity of the compatibility conditions \eqref{eq:Moments0} in Chapman-Enskog theory.
As for the Hilbert expansion, the time-like derivatives are \emph{exactly} substituted by space-like ones due to the fact that the equations of motion include the Euler equations explicitly. In both cases, the zero modes of the linearized collision operator lead to conditions that force the replacement of time-like derivatives of the fluid-dynamical variables by space-like ones. 

In the following we construct a perturbative solution using moments of the Boltzmann equation and \textit{not} the Boltzmann equation itself. We first integrate the Boltzmann equation with the complete and irreducible basis
$P_{n}^{(\ell)}(\beta E_{\bf p}) p_{\langle \mu_{1}} \cdots  p_{\mu_{\ell} \rangle}$ used in the last sections (the functions $P_{n}^{(\ell)}$ are not necessarily orthogonal). The perturbative book-keeping parameter $\epsilon$ is then inserted on the left-hand side of all moment equations,
\begin{equation}
\label{eq:function-moments-gn}
\begin{aligned}
& \epsilon \int dP P_{n}^{(\ell)}(\beta E_{\bf p}) p_{\langle \mu_{1}} \cdots  p_{\mu_{\ell} \rangle} p^{\mu}\partial_{\mu} f_{\mathbf{p}} 
 =
 \int dP P_{n}^{(\ell)}(\beta E_{\bf p}) p_{\langle \mu_{1}} \cdots  p_{\mu_{\ell} \rangle}
 C[f_{\mathbf{p}}]. 
\end{aligned}    
\end{equation}
Naturally, if the basis elements correspond to $1,p^\mu$, we obtain the usual conservation laws already displayed in \eqref{eq:non-pert-csv}. These conservation laws will be treated non-perturbatively, as was the case of Chapman-Enskog theory. Thus, from now on, we shall only consider the remaining basis elements in our analysis.

Then, as usual, one assumes an asymptotic series solution for $f_{\mathbf{p}}$,
\begin{equation}
\label{eq:chapman-expn-SPDF2}
f_{\mathbf{p}} = \sum_{i=0}^{\infty} \epsilon^{i} f^{(i)}_{\mathbf{p}},    
\end{equation}
and Eq.~\eqref{eq:function-moments-gn} is solved order-by-order in the perturbative parameter. Indeed, at $\mathcal{O}(\epsilon^{0})$, we have
\begin{equation}
\label{eq:BDN-gn-mod-Ozero}
\begin{aligned}
& 0
 =
 \int dP P_{n}^{(\ell)}(\beta E_{\bf p}) p_{\langle \mu_{1}} \cdots  p_{\mu_{\ell} \rangle}
 C[f_{\mathbf{p}}^{(0)}]. 
\end{aligned}
\end{equation}
The fact that integrals over arbitrary basis elements all vanish implies that $C[f_{\mathbf{p}}^{(0)}] = 0$ and, thus, $f_{\mathbf{p}}^{(0)} = f_{0\bf{p}}$. Next, collecting all terms of first order in $\epsilon$, we obtain
\begin{equation}
\label{eq:BDN-gn-2-o1-}
\begin{aligned}
&
\int dP P_{n}^{(\ell)}(\beta E_{\bf p}) p_{\langle \mu_{1}} \cdots  p_{\mu_{\ell} \rangle}  p^{\mu}\partial_{\mu} f_{0\bf{p}} 
= 
\int dP P_{n}^{(\ell)}(\beta E_{\bf p}) p_{\langle \mu_{1}} \cdots  p_{\mu_{\ell} \rangle}f_{0\bf{p}}\hat{L}\phi^{(1)}_{\mathbf{p}}.
\end{aligned}
\end{equation}
Here, we emphasize that the zero modes of the linearized collision operator do not enter this set of equations, i.e., the basis elements $1$ and $p^{\mu}$ are not present in this equation. This implies that the compatibility conditions that require the exchange of time-like derivatives of $f_{\mathbf{p}}$ by space-like ones in Chapman-Enskog theory, see Eqs.~\eqref{eq:Moments0}, do not appear in this case. This is a consequence of performing the perturbative procedure on moments of the Boltzmann equation and not on the Boltzmann equation itself. The term inside each integral on the left-hand sides can be irreducibly written as
\begin{equation}
\begin{aligned}
p^{\mu} \partial_{\mu} f_{0\bf{p}} = \left[ E_{\mathbf{p}} D\alpha - E_{\mathbf{p}}^{2} D\beta - \frac{\beta}{3} \Delta^{\lambda \sigma}p_{\lambda} p_{\sigma} \theta +   p^{\langle \mu \rangle} \nabla_{\mu}\alpha - E_{\mathbf{p}} p^{\langle \mu \rangle} \left( \beta Du_{\mu} + \nabla_{\mu} \beta \right) - \beta p^{\langle \mu}p^{\nu \rangle} \sigma_{\mu \nu} \right] f_{0\mathbf{p}}.
\end{aligned}    
\end{equation}

Since $\hat{L}$ is a linear operator, Eq.~\eqref{eq:BDN-gn-2-o1-} implies that the solution for $\phi_{\bf p}$ 
can be expressed as the sum of a homogeneous and a particular solution,
\begin{equation}
\label{eq:phi-hom-phi-part2}
\begin{aligned}
& \phi_{\bf p} = \phi_{\bf p}^{\rm{hom}} + \phi_{\bf p}^{\rm{part}},
\end{aligned}    
\end{equation}
where the homogeneous has the usual form
\begin{equation}
\label{eq:phi-hom-cd}
\phi_{\bf p}^{\rm{hom}} = a + b_{\mu} p^{\mu}.    
\end{equation}
Since we do not have any self-consistency or compatibility conditions that impose the replacement of time-like derivatives of fluid-dynamical variables by space-like ones, the particular
solution has the general form
\begin{equation}
\label{eq:part-sol-phi-BDN}
\begin{aligned}
\phi_{\bf p}^{\rm{part}} = \mathcal{S}_{\bf p}^{(\alpha)} D\alpha 
+
\mathcal{S}_{\bf p}^{(\beta)} 
D\beta 
+
\mathcal{S}_{\bf p}^{(\theta)} \theta 
+
\mathcal{V}_{\bf p}^{(\alpha)} p^{\langle \mu \rangle} \nabla_{\mu} \alpha 
+
\mathcal{V}_{\bf p}^{(\beta)} p^{\langle \mu \rangle} (\nabla_{\mu} \beta + \beta D u_{\mu} )
+
\mathcal{T}_{\bf p} p^{\langle \mu}p^{\nu \rangle} \sigma_{\mu \nu}. 
\end{aligned}    
\end{equation}
The following steps are essentially the same as those applied in Chapman-Enskog and Hilbert procedures, and involve the inversion of the linearized collision operator (in the subspace excluding its zero modes). We assume that the functions $\mathcal{S}$, $\mathcal{V}$, and $\mathcal{T}$ can be expanded in the complete basis $P_{n}^{(\ell)}$,
\begin{equation}
\label{eq:SVT-gn-expn}
\begin{aligned}
\mathcal{S}_{\bf p}^{(\alpha, \beta , \theta)} = \sum_{n \geq 0} s_{n}^{(\alpha, \beta , \theta)} P_{n}^{(0)},\qquad 
\ \ 
\mathcal{V}_{\bf p}^{(\alpha, \beta)} = \sum_{n\geq 0} v_{n}^{(\alpha, \beta)} P_{n}^{(1)},\qquad
 \ \ 
\mathcal{T}_{\bf p} = \sum_{n\geq 0} t_{n} P_{n}^{(2)},
\end{aligned}    
\end{equation}
which leads to the following system of linear equations,
\begin{equation}
\label{eq:sys-SVT2-BDN-gn}
\begin{aligned}
&
\sum_{n} S_{rn} s_{n}^{(\alpha, \beta, \theta)} = A_{r}^{(\alpha, \beta, \theta)}, \\
&
\sum_{n} V_{rn} v_{n}^{(\alpha, \beta)} = B_{r}^{(\alpha, \beta)}, \\
&
\sum_{n} T_{rn} t_{n} = C_{r}, 
\end{aligned}    
\end{equation}
to be solved for the coefficients $s_{n}^{(\alpha, \beta, \theta)}$, $v_{n}^{(\alpha, \beta)}$, and $t_{n}$. The matrices $S$, $V$, and $T$ were already defined in  \eqref{eq:mat-def-SVT-gn}. We further define the thermodynamic integrals,
\begin{equation}
\label{eq:source-t-int-BDN}
\begin{aligned}
&
A_{r}^{(\alpha)} =  \mathcal{I}_{1r}^{(0)}, \ \
A_{r}^{(\beta)} = - \mathcal{I}_{2r}^{(0)},
\ \
A_{r}^{(\theta)} = \frac{\beta}{3} \left(  \mathcal{I}_{2r}^{(0)} - m^{2} \mathcal{I}_{0r}^{(0)} \right) , 
\\
&
B_{r}^{(\alpha)} = - 3 \mathcal{I}_{0r}^{(1)}\ \
B_{r}^{(\beta)} =  3 \mathcal{I}_{1r}^{(1)}, \\
&
C_{r} = - 15 \beta \mathcal{I}_{0r}^{(2)}.
\end{aligned}    
\end{equation}
Equations \eqref{eq:sys-SVT2-BDN-gn} can be schematically inverted as
\begin{equation}
\label{eq:coeffs-svt-CE2}
\begin{aligned}
& 
s_{n}^{(\alpha, \beta, \theta)} = \sum_{m} [S^{-1}]_{nm} A_{m}^{(\alpha, \beta, \theta)}
\quad
v_{n}^{(\alpha, \beta)} = \sum_{m} [V^{-1}]_{nm} B_{m}^{(\alpha, \beta)}
\quad
t_{n} = \sum_{m} [T^{-1}]_{nm} C_{m}.  
\end{aligned}    
\end{equation}

Then, we proceed to obtain the homogeneous solution $\phi^{\rm{hom}}_{\bf p}$. This is made by substituting Eqs.~\eqref{eq:phi-hom-phi-part2} and \eqref{eq:phi-hom-cd} in the general matching conditions \eqref{eq:matching_kinetic1} and \eqref{eq:matching_kinetic2}, in complete analogy with Eq.~\eqref{eq:a-bu-bmu} in the Chapman-Enskog expansion. In the present case, this procedure yields
\begin{equation}
\begin{aligned}
& I_{q,0} a + I_{q+1,0} b_{\mu} u^{\mu} 
=
- \left\langle E_{\bf p}^{q} \mathcal{S}^{(\alpha)}_{\bf p} \right\rangle D\alpha 
- \left\langle E_{\bf p}^{q} \mathcal{S}^{(\beta)}_{\bf p} \right\rangle D\beta  
- \left\langle E_{\bf p}^{q} \mathcal{S}^{(\theta)}_{\bf p} \right\rangle \theta 
\\
& I_{s,0} a + I_{s+1,0} b_{\mu} u^{\mu} 
=
- \left\langle E_{\bf p}^{s} \mathcal{S}^{(\alpha)}_{\bf p} \right\rangle D\alpha 
- \left\langle E_{\bf p}^{s} \mathcal{S}^{(\beta)}_{\bf p} \right\rangle D\beta  
- \left\langle E_{\bf p}^{s} \mathcal{S}^{(\theta)}_{\bf p} \right\rangle \theta 
\\
& I_{z+2,1} b_{\langle \mu \rangle} 
=
\frac{1}{3} \left\langle \left(\Delta^{\mu \nu}p_{\mu} p_{\nu} \right) E_{\bf p}^{z} \mathcal{V}_{\bf p}^{(\alpha)} \right\rangle_{0} \nabla_{\mu} \alpha
+
\frac{1}{3} \left\langle \left(\Delta^{\mu \nu}p_{\mu} p_{\nu} \right) E_{\bf p}^{z} \mathcal{V}_{\bf p}^{(\beta)} \right\rangle_{0} \left( \nabla_{\mu} \beta + \beta D u_{\mu}\right),
\end{aligned}    
\end{equation}
where we used again that $p^{\mu} = E_{\bf p} u^{\mu} + p^{\langle \mu \rangle}$. The equations above can be solved for $a$, $b^{\mu} u_{\mu}$, and $b_{\langle \lambda \rangle}$, which gives 
\begin{equation}
\label{eq:hom-sol-c-dmu}
\begin{aligned}
&
a = 
\frac{I_{q+1,0} \langle E_{\bf p}^{s} \mathcal{S}_{\bf p}^{(\alpha)} \rangle_{0} - \langle E_{\bf p}^{q} \mathcal{S}_{\bf p}^{(\alpha)} \rangle_{0} I_{s+1,0}}{G_{s+1,q}}
D\alpha
+
\frac{I_{q+1,0} \langle E_{\bf p}^{s} \mathcal{S}_{\bf p}^{(\beta)} \rangle_{0} - \langle E_{\bf p}^{q} \mathcal{S}_{\bf p}^{(\beta)} \rangle_{0} I_{s+1,0}}{G_{s+1,q}}
D\beta 
+
\frac{I_{q+1,0} \langle E_{\bf p}^{s} \mathcal{S}_{\bf p}^{(\theta)} \rangle_{0} - \langle E_{\bf p}^{q} \mathcal{S}_{\bf p}^{(\theta)} \rangle_{0} I_{s+1,0}}{G_{s+1,q}}
\theta,
\\
&
b^{\mu} u_{\mu} = 
\frac{\langle E_{\bf p}^{q} \mathcal{S}_{\bf p}^{(\alpha)}\rangle_{0} I_{s,0} - I_{q,0} \langle E_{\bf p}^{s} \mathcal{S}_{\bf p}^{(\alpha)} \rangle_{0}}{G_{s+1,q}} D\alpha
+
\frac{\langle E_{\bf p}^{q} \mathcal{S}_{\bf p}^{(\beta)}\rangle_{0} I_{s,0} - I_{q,0} \langle E_{\bf p}^{s} \mathcal{S}_{\bf p}^{(\beta)} \rangle_{0}}{G_{s+1,q}} D\beta 
+
\frac{\langle E_{\bf p}^{q} \mathcal{S}_{\bf p}^{(\theta)}\rangle_{0} I_{s,0} - I_{q,0} \langle E_{\bf p}^{s} \mathcal{S}_{\bf p}^{(\theta)} \rangle_{0}}{G_{s+1,q}} \theta, \\
&
b_{\langle \lambda \rangle} 
=
\frac{1}{3}\frac{\left\langle \left(\Delta^{\mu \nu}p_{\mu} p_{\nu} \right) E_{\bf p}^{z} \mathcal{V}_{\bf p}^{(\alpha)} \right\rangle_{0}}{I_{z+2,1}} \nabla_{\lambda} \alpha
+
\frac{1}{3}\frac{\left\langle \left(\Delta^{\mu \nu}p_{\mu} p_{\nu} \right) E_{\bf p}^{z} \mathcal{V}_{\bf p}^{(\beta)} \right\rangle_{0}}{I_{z+2,1}} \left(\nabla_{\lambda} \beta + \beta D u_{\lambda} \right).
\end{aligned}    
\end{equation}

Finally, combining the homogeneous solution found above with the particular solution derived in Eq.~\eqref{eq:coeffs-svt-CE2}, we obtain the complete first-order solution of the modified perturbative procedure introduced in this section. The solution can be expressed as,
\begin{equation}
\label{eq:part-sol-phi-BDN-final}
\begin{aligned}
\phi_{\bf p} = \tilde{\mathcal{S}}_{\bf p}^{(\alpha)} D\alpha 
+
\tilde{\mathcal{S}}_{\bf p}^{(\beta)} 
D\beta 
+
\tilde{\mathcal{S}}_{\bf p}^{(\theta)} \theta 
+
\tilde{\mathcal{V}}_{\bf p}^{(\alpha)} p^{\langle \mu \rangle} \nabla_{\mu} \alpha 
+
\tilde{\mathcal{V}}_{\bf p}^{(\beta)} p^{\langle \mu \rangle} (\nabla_{\mu} \beta + \beta D u_{\mu} )
+
\mathcal{T}_{\bf p} p^{\langle \mu}p^{\nu \rangle} \sigma_{\mu \nu}, 
\end{aligned}    
\end{equation}
where we define the momentum-dependent functions
\begin{equation}
\begin{aligned}
& 
\tilde{\mathcal{S}}_{\bf p}^{(\alpha, \beta, \theta)} = 
\sum_{n} \sum_{m} [S^{-1}]_{nm} A_{m}^{(\alpha, \beta, \theta)} \left( P_{n}^{(0)}
+
\frac{I_{q+1,0} \mathcal{I}_{sn}^{(0)} - \mathcal{I}_{qn}^{(0)} I_{s+1,0}}{G_{s+1,q}}
+
\frac{\mathcal{I}_{qn}^{(0)} I_{s,0} - I_{q,0} \mathcal{I}_{sn}^{(0)}}{G_{s+1,q}} E_{\bf p} 
\right),
\\
&
\tilde{\mathcal{V}}_{\bf p}^{(\alpha, \beta)} 
=
\sum_{n} \sum_{m}  [V^{-1}]_{nm} B_{m}^{(\alpha, \beta)} \left( P_{n}^{(1)} 
-
\frac{\mathcal{I}_{zn}^{(1)}}{I_{z+2,1}} 
\right),  \\
&
\mathcal{T}_{\bf p} =
\sum_{n} \sum_{m} [T^{-1}]_{nm} C_{m} P_{n}^{(2)}.
\end{aligned}    
\end{equation}
Replacing the solution in \eqref{eq:part-sol-phi-BDN-final} into the definition of the dissipative currents \eqref{eq:def_kinetic}, we obtain the following constitutive relations
\begin{equation}
\label{eq:def-coeffs-BDN-gn}
\begin{aligned}
& \Pi 
= \zeta^{(\alpha)} D\alpha - \zeta^{(\beta)} \frac{D\beta }{\beta} - \zeta^{(\theta)} \theta,
\ \
 \delta n 
= \xi^{(\alpha)} D\alpha - \xi^{(\beta)} \frac{D\beta }{\beta} - \xi^{(\theta)} \theta,
\ \
 \delta \varepsilon 
= \chi^{(\alpha)} D\alpha - \chi^{(\beta)}  \frac{D\beta }{\beta} - \chi^{(\theta)} \theta,\\
& \nu^{\mu} 
= \kappa^{(\alpha)} \nabla^{\mu} \alpha - \kappa^{(\beta)} \left( \frac{1}{\beta} \nabla^{\mu} \beta +   D u^{\mu}\right), \ \
 h^{\mu} 
= \lambda^{(\alpha)} \nabla^{\mu} \alpha -  \lambda^{(\beta)} \left( \frac{1}{\beta} \nabla^{\mu} \beta +   D u^{\mu}\right),
\\
&
\pi^{\mu \nu} = 2 \eta \sigma^{\mu \nu},
\end{aligned}
\end{equation}
where the microscopic expressions for the fourteen transport parameters introduced above are given by 
\begin{equation}
\begin{aligned}
& \zeta^{(\alpha)} =  \sum_{n,m} [S^{-1}]_{nm} A^{(\alpha)}_{m} H^{(\zeta)}_{n}, 
\
\zeta^{(\beta)} = -\beta \sum_{n,m} [S^{-1}]_{nm} A^{(\beta)}_{m} H^{(\zeta)}_{n}, 
\
\zeta^{(\theta)} = -\sum_{n,m} [S^{-1}]_{nm} A^{(\theta)}_{m} H^{(\zeta)}_{n},
\\
& \xi^{(\alpha)} = \sum_{n,m} [S^{-1}]_{nm} A^{(\alpha)}_{m} H^{(\xi)}_{n}, 
\
\xi^{(\beta)} = -\beta \sum_{n,m} [S^{-1}]_{nm} A^{(\beta)}_{m} H^{(\xi)}_{n}, 
\
\xi^{(\theta)} = -\sum_{n,m} [S^{-1}]_{nm} A^{(\theta)}_{m} H^{(\xi)}_{n},\\
& \chi^{(\alpha)} = \sum_{n,m} [S^{-1}]_{nm} A^{(\alpha)}_{m} H_{n}^{(\chi)},   
\
\chi^{(\beta)} = -\beta \sum_{n,m} [S^{-1}]_{nm} A^{(\beta)}_{m} H_{n}^{(\chi)},  
\
\chi^{(\theta)} = -\sum_{n,m} [S^{-1}]_{nm} A^{(\theta)}_{m} H_{n}^{(\chi)},
\\
& \kappa^{(\alpha)} = \sum_{n,m} [V^{-1}]_{nm} B^{(\alpha)}_{m} J_{n}^{(\kappa)}, 
\
\kappa^{(\beta)} = - \beta \sum_{n,m} [V^{-1}]_{nm} B^{(\beta)}_{m} J_{n}^{(\kappa)}, 
\\
& \lambda^{(\alpha)} = \sum_{n,m} [V^{-1}]_{nm} B^{(\alpha)}_{m} J_{n}^{(\lambda)}, 
\
 \lambda^{(\beta)} = - \beta \sum_{n,m} [V^{-1}]_{nm} B^{(\beta)}_{m} J_{n}^{(\lambda)},
\\
& \eta = \sum_{n,m} [T^{-1}]_{nm} C_{m}  \mathcal{I}_{0n}^{(2)}.
\end{aligned}    
\end{equation}
The functions $H^{(\zeta,\xi,\chi)}, J^{(\kappa,\lambda)}$ were already defined in Eqs.~\eqref{eq:H,J-expn}.
We further notice that the shear coefficient has the same expression as in Chapman-Enskog and Hilbert expansions. Furthermore, in the massless limit, since $\delta T^{\mu}_{\ \mu} = \delta \varepsilon - 3 \Pi = 0$, we have that $3\zeta^{(\alpha)} = \chi^{(\alpha)}$, $3\zeta^{(\beta)} = \chi^{(\beta)}$, and $3\zeta^{(\theta)} = \chi^{(\theta)}$. Also in this limit, since $\Delta^{\lambda \sigma}p_{\lambda} p_{\sigma} = -E_{\bf p}^{2}$, we have that $3 \xi^{(\theta)} = \xi^{(\beta)}$ and $3 \chi^{(\theta)} =  \chi^{(\beta)}$, even though they are in general not zero. This is in contrast to what happened in the traditional Chapman-Enskog expansion where $\zeta$, $\xi$, and $\chi$ vanish identically in the $m \rightarrow 0$ limit.  
The constitutive relations \eqref{eq:def-coeffs-BDN-gn}, combined with the conservation laws \eqref{eq:consv-eqns}, lead to the BDNK equations \cite{bemfica:18causality,kovtun:19first,bemfica2019nonlinear,Bemfica:2020zjp,Hoult:2020eho}. 

We also note that, as in Navier-Stokes theory (see Eq.~\eqref{eq:coeffs-chap-ensk}), the majority of the coefficients are strongly dependent on the parameters $q$ and $s$ that specify the matching conditions employed. In fact, for Landau matching conditions, $(q,s,z) = (1,2,1)$, we have $\xi^{(\alpha, \beta, \theta)} = \chi^{(\alpha, \beta, \theta)} = \lambda^{(\alpha, \beta)} = 0$, while for Eckart matching conditions, $(q,s,z) = (1,2,0)$, one finds $\xi^{(\alpha, \beta, \theta)} = \chi^{(\alpha, \beta, \theta)} = \lambda^{(\alpha, \beta)} = 0$. Furthermore, for matching conditions that respect $(q,s)=(0,2)$ we have that $\zeta^{(\alpha, \beta, \theta)} = 0$.

The approach presented in this section provides a systematic way to derive the BDNK equations from kinetic theory at nonzero chemical potential. Early work in this direction was presented in Refs.\ \cite{bemfica:18causality,bemfica2019nonlinear}, but the latter did not employ an irreducible basis nor gave explicit expressions for all the transport parameters that are valid at zero and nonzero chemical potential. 

Furthermore, we point out that Navier-Stokes theory can be obtained from the BDNK equations by replacing the time-like derivatives of $\beta$, $\alpha$, and $u^\mu$ in the constitutive relations \eqref{eq:def-coeffs-BDN-gn} using a first-order truncation of the  conservation laws. Performing this substitution, we find the relation between the transport coefficients appearing in BDNK theory and those of Navier-Stokes theory. The result is,
\begin{subequations}
\label{eq:rel-BDNK-NS}
\begin{align}
& \zeta = \left(\frac{\partial P_{0}}{\partial n_{0}} \right)_{\varepsilon_{0}} \beta \zeta^{(\alpha)} 
+ \left(\frac{\partial P_{0}}{\partial \varepsilon_{0}} \right)_{n_{0}} \zeta^{(\beta)}
+ \zeta^{(\theta)}, \\
&
\xi = \left(\frac{\partial P_{0}}{\partial n_{0}} \right)_{\varepsilon_{0}} \beta \xi^{(\alpha)} 
+ \left(\frac{\partial P_{0}}{\partial \varepsilon_{0}} \right)_{n_{0}} \xi^{(\beta)}
+ \xi^{(\theta)}, \\
&
- \chi = \left(\frac{\partial P_{0}}{\partial n_{0}} \right)_{\varepsilon_{0}} \beta \chi^{(\alpha)} 
+\left(\frac{\partial P_{0}}{\partial \varepsilon_{0}} \right)_{n_{0}} \chi^{(\beta)}
+ \chi^{(\theta)}, \\
&  
\kappa = \kappa^{(\alpha)} 
- \frac{P_{0}}{\varepsilon_{0} + P_{0}} \kappa^{(\beta)},\\
&
\lambda = - \lambda^{(\alpha)} 
+ \frac{P_{0}}{\varepsilon_{0} + P_{0}} \lambda^{(\beta)}.
\end{align} 
\end{subequations}
This implies that, in general, $\zeta \neq \zeta^{(\theta)}$ and $\kappa \neq \kappa^{(\alpha)}$, for example. This mapping between the coefficients was first derived via hydrodynamic frame transformations in \cite{kovtun:19first}, with the important difference that, in that reference, the relation is between the  Navier-Stokes coefficients in Landau matching conditions (only in this particular matching, $\zeta$ and $\kappa$ coincide, respectively, with the matching-invariant coefficients derived in Eqs.~\eqref{eq:matching-inv-NS}) and the BDNK matching-invariant coefficients \footnote{Some signs are different because of the difference in sign conventions of definitions \eqref{eq:def-coeffs-BDN-gn} and 2.4 of Ref.~\cite{kovtun:19first}.}
\begin{equation}
\begin{aligned}
& \zeta^{(i)} +  \left(\frac{\partial P_{0}}{\partial n_{0}} \right)_{\varepsilon_{0}} \xi^{(i)}
+ \left(\frac{\partial P_{0}}{\partial \varepsilon_{0}} \right)_{n_{0}} \chi^{(i)} = 
\sum_{n \geq 2} \sum_{m \geq 2}[\hat{S}^{-1}]_{nm} A_{m}^{(i)} \mathcal{H}_{n},
 \quad i = \alpha, \beta, \theta, \\
&  \kappa^{(i)} + \frac{n_{0}}{\varepsilon_{0} + P_{0}} \lambda^{(i)} = \sum_{n \geq 1} \sum_{m \geq 1}  [\hat{V}^{-1}]_{nm} B_{m}^{(i)}  \mathcal{J}_{n} \quad i = \alpha, \beta, 
\end{aligned}    
\end{equation}
where $\mathcal{H}_{n}$ and $\mathcal{J}_{n}$ were defined in Eq.~\eqref{eq:match-inv-NS-defs}. 

We would like to close this section with a brief comment on previous works where a derivation of BDNK theory from a microscopic description was investigated. As mentioned above, the first derivation of BDNK theory from kinetic theory was done in the original works \cite{bemfica:18causality,bemfica2019nonlinear}. After that, other approaches were pursued in \cite{Hoult:2021gnb,Biswas:2022cla}. Reference \cite{Hoult:2021gnb} discussed how BDNK may be derived from holography, using the fluid/gravity correspondence \cite{Bhattacharyya:2007vjd}. They also provided a derivation of BDNK from kinetic theory using ideas from the Hilbert series. Reference \cite{Biswas:2022cla} focused on the effects of a momentum-dependent relaxation time coefficient on the calculation of transport coefficients. 

\section{Transport coefficients in the relaxation time approximation}
\label{sec:RTA}

In order to provide some intuition on the constitutive relations derived in the previous sections, we calculate all transport coefficients using a simplified version of the linearized collision term: the relaxation time approximation (RTA). Here, since it is essential to consider unconventional matching conditions (Landau or Eckart matching conditions render the BDNK equations acausal), it is not possible to employ the relaxation time approximation proposed by Anderson and Witting \cite{andersonRTA:74}. In Ref.~\cite{rocha:21}, a novel RTA for the relativistic Boltzmann equation was proposed,
\begin{equation}
\label{eq:nRTA}
\begin{aligned}
f_{0\bf{p}}\hat{L}\phi_{\mathbf{p}} \approx - \frac{E_{\mathbf{p}}}{\tau_{R}} f_{\textbf{p}}^{\rm{eq}} \left\{ \phi_{\mathbf{p}} - 
\frac{\left( \phi_{\mathbf{p}} , \frac{E_{\mathbf{p}}}{\tau_{R}} \right)_{0}}{(1 , \frac{E_{\mathbf{p}}}{\tau_{R}})_{0}}  
-
\frac{\left( \phi_{\mathbf{p}} , \frac{E_{\mathbf{p}}}{\tau_{R}} \tilde{P}_{1} \right)_{0}}{\left( \tilde{P}_{1} , \frac{E_{\mathbf{p}}}{\tau_{R}} \tilde{P}_{1} \right)_{0}} \tilde{P}_{1} 
- 
\frac{\left( \phi_{\mathbf{p}} , \frac{E_{\mathbf{p}}}{\tau_{R}} p^{\langle \mu \rangle} \right)_{0}}{ \left( 1 , \frac{E_{\mathbf{p}}}{\tau_{R}}\right)_{1}} p_{\langle \mu \rangle}   \right\},
\end{aligned}
\end{equation}
where the first term on the right-hand side amounts to the traditional RTA. The remaining terms are inserted so that particle number, energy, and momentum are conserved regardless of the matching condition or energy dependence of the relaxation time employed.

The counter-terms, in their turn, denote a projector in the linear subspace of conserved quantities, which is spanned by the orthogonal basis $\{1, \Tilde{P}_{1}, p^{\langle \nu \rangle} \}$. The polynomial $\Tilde{P}_{1}$ is constructed so that it is orthogonal to $1$,
\begin{equation}
\label{eq:ort2}
\left( \tilde{P}_{1} , \frac{E_{\mathbf{p}}}{\tau_{R}} \right)_{0} = 0,
\end{equation}
where we defined the following scalar product
\begin{equation}
\label{eq:inner-prod-l}
\begin{aligned}
\left( \phi , \psi \right)_{\ell} 
= 
\frac{\ell !}{(2\ell+1)!!}
\int dP \left(\Delta_{\mu \nu}p^{\mu}p^{\nu}\right)^{\ell} \phi_{\mathbf{p}} \psi_{\mathbf{p}} f_{0\bf{p}}.
\end{aligned}    
\end{equation}
In general, we consider that the relaxation time depends on the microscopic energy through a power law, 
\begin{equation}
\label{eq:parametrization}
    \tau_{R} = t_{R} \left( \frac{E_{\mathbf{p}}}{T} \right)^{\gamma},
\end{equation}
which introduces the phenomenological parameter $\gamma$. In Ref.~\cite{rocha21-transient}, it was shown that this parameter significantly affects the transient evolution of the dissipative currents calculated using Israel-Stewart theory. Using the Ansatz \eqref{eq:nRTA}, the computation of transport coefficients is significantly simplified and explicitly calculations are carried out in the next subsection.

\subsection*{Transport coefficients: massless gas with constant relaxation time}
\label{sec:BDN-trsp-better-trun}

Now we apply  the relaxation time approximation, Eq.~\eqref{eq:nRTA}, to compute the matrix elements that appear in all perturbative schemes discussed so far, see Eqs.~\eqref{eq:mat-def-SVT-gn}. In the RTA, the scalar, vector, and tensor sector matrix  elements become 
\begin{equation}
\begin{aligned}
& 
S_{rn} \equiv \int dP 
P_{r}^{(0)}
\hat{L} \left[P_{n}^{(0)}\right] f_{0\bf{p}}
\approx
-\left( P_{r}^{(0)} , P_{n}^{(0)} \frac{E_{\mathbf{p}}}{\tau_{R}}\right)_{0} 
+
\frac{\left( P_{r}^{(0)} ,  \frac{E_{\mathbf{p}}}{\tau_{R}}\right)_{0}}
{\left( 1 ,  \frac{E_{\mathbf{p}}}{\tau_{R}}\right)_{0}}
\left( P_{n}^{(0)} ,  \frac{E_{\mathbf{p}}}{\tau_{R}}\right)_{0}
+
\frac{\left( P_{r}^{(0)} , \frac{E_{\mathbf{p}}}{\tau_{R}} \tilde{P}_{1} \right)_{0}}{\left( \tilde{P}_{1} , \frac{E_{\mathbf{p}}}{\tau_{R}} \tilde{P}_{1} \right)_{0}} \left( P_{n}^{(0)} , \frac{E_{\mathbf{p}}}{\tau_{R}} \tilde{P}_{1} \right)_{0} ,
\\
&
-\frac{1}{3}V_{rn}  \equiv -\frac{1}{3}\int dP 
P_{r}^{(1)}
p^{\langle \mu \rangle}
\hat{L} \left[P_{n}^{(1)} p_{\langle \mu \rangle} \right] f_{0\bf{p}} 
\approx
-\left( P_{r}^{(1)} , P_{n}^{(1)} \frac{E_{\mathbf{p}}}{\tau_{R}}\right)_{1} 
+
\frac{\left( P_{r}^{(1)} ,  \frac{E_{\mathbf{p}}}{\tau_{R}}\right)_{1}}{\left( 1 ,  \frac{E_{\mathbf{p}}}{\tau_{R}}\right)_{1}} \left( P_{n}^{(1)} ,  \frac{E_{\mathbf{p}}}{\tau_{R}}\right)_{1} , 
\\
&
\frac{1}{15}T_{rn} \equiv \frac{1}{15} \int dP 
P_{r}^{(2)}
p^{\langle \mu}p^{\nu \rangle}
\hat{L} \left[P_{n}^{(2)} p_{\langle \mu}p_{\nu \rangle} \right] f_{0\bf{p}} 
\approx
-\left( P_{r}^{(2)} , P_{n}^{(2)} \frac{E_{\mathbf{p}}}{\tau_{R}}\right)_{2}.
\end{aligned}    
\end{equation}

As shown in the last section, the computation of transport coefficients requires the inversion of matrices of infinite dimension. In practice this is performed by considering successive finite truncations of these matrices until the result converges. In this case, the choice of basis plays an important role. Indeed, the existence of convergence and its speed may depend on the choice of basis. In the present section, we use the set of functions
\begin{equation}
\label{eq:basis-x/(1+x)}
\begin{aligned}
& P_{m}^{(\ell)}(x) = \frac{x^{m-m_{\ell}}}{(1+x)^{N-n_{\ell}}}, \ m = 1, \cdots N 
\end{aligned}    
\end{equation}
as our basis in the $N$-th truncation step. The parameters $m_{\ell}$ and $n_{\ell}$ are judiciously chosen so that convergence is achieved. This basis set is inspired by Refs.~\cite{Arnold:2000dr,Arnold:2002zm}, where a similar set was used to perturbatively compute transport coefficients in gauge theories at high temperature using an effective kinetic theory approach. In the massless limit, the matrix elements then become 
\begin{equation}
\begin{aligned}
& 
S_{rn} =
- \frac{e^{\alpha}}{2 \pi^{2} t_{R} \beta^{3}} \left\{  F(r+n-2m_{0}+2-\gamma,2N-2n_{0}) - \frac{F(r-m_{0}+2-\gamma,2N-2n_{0})}{\Gamma(3-\gamma)} F(n-m_{0}+2-\gamma,2N-2n_{0})
\right.\\
&
\left.
-
\frac{[(3-\gamma)F(r-m_{0}+2-\gamma,2N-2n_{0})-F(r+3-\gamma,2N-2n_{0})]}{\Gamma(4-\gamma)} \right. \\
&
\left.
\times
[(3-\gamma)F(n-m_{0}+2-\gamma,2N-2n_{0})-F(n-m_{0}+3-\gamma,2N-2n_{0})]
\right\},
\\
&
V_{rn}  = \frac{e^{\alpha}}{2 \pi^{2} t_{R} \beta^{5}} \left[ F(r+n-2m_{1}+4-\gamma,2N-2n_{1}) - \frac{F(r-m_{1}+4-\gamma,2N-2n_{1})}{\Gamma(5-\gamma)} F(n+4-\gamma,2N-2n_{1}) \right],
\\
&
T_{rn} = - \frac{e^{\alpha}}{2 \pi^{2} t_{R} \beta^{7}} F(r+n-2m_{1}+6-\gamma,2N-2n_{1}),
\end{aligned}    
\end{equation}
where we defined
\begin{equation}
\begin{aligned}
& F(M,N) = \int_{0}^{\infty} dx \frac{x^{M}}{(1+x)^{N}} e^{-x}
=
\Gamma(M+1) U(N,N-M,1),
\end{aligned}    
\end{equation}
with $U(a,b,z)$ denoting the confluent hypergeometric function
$U(a,b,z) = [1/\Gamma(a)] \int_{0}^{\infty}2\Hat{\tau} \ e^{-zt} t^{a-1}(1+t)^{b-a-1}$, and $\Gamma(z)$ the Gamma function \cite{NIST:DLMF}. Moreover, the source term integrals in Eqs.~\eqref{eq:source-t-int-BDN} are
\begin{equation}
\begin{aligned}
& A^{(\alpha)}_{r} = \frac{e^{\alpha}}{2\pi^{2}\beta^{2}} F(r-m_{0}+2,N-n_{0})  
\quad
A^{(\beta)}_{r} = -\frac{e^{\alpha}}{2\pi^{2}\beta^{3}} F(r-m_{0}+3,N-n_{0})  \\
& B^{(\alpha)}_{r} = \frac{e^{\alpha}}{2\pi^{2}\beta^{4}} F(r-m_{1}+3,N-n_{1})
\quad
B^{(\beta)}_{r} = \frac{e^{\alpha}}{2\pi^{2}\beta^{5}} F(r-m_{1}+3,N-n_{1})
 \\
& C_{r} = \frac{e^{\alpha}}{2\pi^{2}\beta^{5}} F(r-m_{2}+5,N-n_{2}),
\end{aligned}    
\end{equation}
where we note that $A_{r}^{(\theta)} = (1/3)A_{r}^{(\beta)}$ in the massless limit. For the sake of simplicity, we take a constant relaxation time in the following calculations, i.e., we set the parameter $\gamma = 0$.

We compute the transport coefficients using two sets of
matching conditions in which the particle diffusion 4-current is set to zero. We shall refer to these types of frames as \textit{exotic Eckart matching conditions} \cite{rocha21-transient}. In the first type, we use Eqs.~\eqref{eq:matching_kinetic1} and \eqref{eq:matching_kinetic2} imposing $(q,z) = (1,0)$ and $s \neq 1,2$ so that $\nu^{\mu} \equiv 0$ and $\delta n \equiv 0$. The corresponding results for the transport coefficients of the BDNK equations can be seen in Table \ref{tab:coeffs1gn}, where we also indicate the values of $n_{\ell}$ and $m_{\ell}$ chosen. In the second type of matching conditions, we use $(q,z) = (2,0)$ and $s \neq 1,2$, so that $\nu^{\mu} \equiv 0$ and $\delta \varepsilon \equiv 0$. The corresponding results for the transport coefficients of the BDNK equations are listed in Table \ref{tab:coeffs2gn}. In these tables, we see that the normalized shear viscosity coefficient, $\eta/(P_{0} \tau_{R})$, which is matching independent, converges steadily to 0.8. All other transport coefficients seem to be exactly obtained at each order, as long as an appropriate choice of basis is employed. The reason for this behavior is clarified in Appendix \ref{sec:choice-basis}. It is also clear that $\xi^{(\alpha,\beta)}$ and $\chi^{(\alpha,\beta)}$ depend on the parameter $s$ employed to define the matching condition. The results for $\xi^{(\theta)}$ and $\chi^{(\theta)}$ are omitted due to the fact that, in the massless limit, $3 \xi^{(\theta)} = \xi^{(\beta)}$ and $3 \chi^{(\theta)} =  \chi^{(\beta)}$. Finally, we remark that the transport coefficients of Navier-Stokes theory can be obtained from the results in Tables \ref{tab:coeffs1gn} and \ref{tab:coeffs2gn} using relations \eqref{eq:rel-BDNK-NS}.
\begin{table}[!h]
    \centering
    \begin{tabular}{|c|c|c|c|c|c|c|c|}
    \hline
    Transp. coeff. / Trunc. ord. & 1 & 2 & 3 & 5 & 10\\
    \hline  
    $\eta/(P_{0} \tau_{R})$ ($m_{2}=0$, $n_{2}=1$)  & 0.428571 & 0.741457 & 0.795862 & 0.799938 & 0.7999998 \\
    \hline
     $\lambda^{(\alpha)}/(P_{0} \tau_{R})$ ($m_{1}=2$, $n_{1}=1$) & 1.33333 & 1.33333 & 1.33333 & 1.33333 & 1.33333\\
    \hline
    $\lambda^{(\beta)}/(P_{0} \tau_{R}) (m_{1} = 2 , n_{1} = 1)$ 
    & 4.00 & 4.00 & 4.00 & 4.00 & 4.00 \\
    \hline
     $\chi^{(\alpha)}/(P_{0} \tau_{R})$ ($m_{0}=-1$, $n_{0}=1$,$s=3$)  & 1.50 & 1.50 & 1.50 & 1.50 & 1.50 \\
    \hline
    $\chi^{(\beta)}/(P_{0} \tau_{R})$ ($m_{0}=-1$ , $n_{0} = 1$, $s=3$) & 7.50 & 7.50 & 7.50 & 7.50 & 7.50 \\
    \hline
    $\chi^{(\alpha)}/(P_{0} \tau_{R})$ ($m_{0} = -2$ , $n_{0} = 1$, $s=4$)  & 1.00 & 1.00 & 1.00 & 1.00 & 1.00\\
    \hline
    $\chi^{(\beta)}/(P_{0} \tau_{R})$ ($m_{0} = -2$  , $n_{0}=1$ , $s=4$)  & 6.00 & 6.00 & 6.00 & 6.00 & 6.00\\
    \hline

    \end{tabular}
    \caption{BDNK transport coefficients for Exotic Eckart frames with  $(q,z)=(1,0)$ and $\gamma=0$. The numbers 1, 2, 3, 5, and 10 on each column mean the first, second, third non-trivial truncation order, respectively.}
    \label{tab:coeffs1gn}
\end{table}

\begin{table}[!h]
    \centering
    \begin{tabular}{|c|c|c|c|c|c|c|c|}
    \hline
    Transp. coeff. / Trunc. ord. & 1 & 2 & 3 & 5 & 10\\
    \hline  
     $\xi^{(\alpha)}/(P_{0} \tau_{R})$ ($m_{0} = -1$  , $n_{0} = 1$, $s=3$)  & -1.00 & -1.00 & -1.00 & -1.00 & -1.00 \\
    \hline
    $\xi^{(\beta)}/(P_{0} \tau_{R})$ ($m_{0}=-1$, $n_{0} = 1$, $s=3$) & -5.00 & -5.00 & -5.00 & -5.00 & -5.00 \\
    \hline
    $\xi^{(\alpha)}/(P_{0} \tau_{R})$ ($m_{0}=-2$ , $n_{0}=1$, $s=4$)  & -0.50 & -0.50 & -0.50 & -0.50 & -0.50\\
    \hline
    $\xi^{(\beta)}/(P_{0} \tau_{R})$ ($m_{0}=-2$, $n_{0}=1$, $s=4$)  & -3.00 & -3.00 & -3.00 & -3.00 & -3.00\\
    \hline

    \end{tabular}
    \caption{BDNK  transport coefficients for Exotic Eckart frames with $(q,z)=(2,0)$ and $\gamma=0$. The numbers 1, 2, 3, 5, and 10 on each column mean the first, second, third non-trivial truncation order, respectively.}
    \label{tab:coeffs2gn}
\end{table}

The transport coefficients in Hilbert theory do not depend on any matching conditions and are listed in Table \ref{tab:coeffs-hilb}. As already mentioned, the shear viscosity coefficient is the same in both BDNK, Navier-Stokes, and Hilbert formulations. We also remind the reader that the coefficients $\zeta_{H}$, $\xi_{H}$, $\chi_{H}$ all vanish in the massless limit. Then, in this regime, the only new coefficients to be computed are $\kappa_{H}$ and $\lambda_{H}$, which converge to $n_0 \tau_R / 3$ and $-P_0 \tau_R$, respectively.    
\begin{table}[!h]
    \centering
    \begin{tabular}{|c|c|c|c|c|c|c|c|}
    \hline
    Transp. coeff. / Trunc. ord. & 1 & 2 & 3 & 5 & 10\\
    \hline
     $\kappa_{H}/(n_{0} \tau_{R})$ ($m_{1}=2$, $n_{1}=1$) & 0.33333 & 0.33333 & 0.33333 & 0.33333 & 0.33333\\
    \hline
     $\lambda_{H}/(P_{0} \tau_{R})$ ($m_{1}=2$, $n_{1}=1$,$s=3$)  & -1 & -1 & -1 & -1 & -1 \\ 
    \hline
    \end{tabular}
    \caption{Hilbert theory transport coefficients for $\gamma=0$. The numbers 1, 2, 3, 5, and 10 on each column mean the first, second, third non-trivial truncation order, respectively.}
    \label{tab:coeffs-hilb}
\end{table}

\section{Comparison between equations of motion: Exotic Eckart frames in Bjorken flow} 
\label{sec:comparison}

In this section we compare the solutions that emerge from each perturbative scheme discussed in the previous sections with solutions of Israel-Stewart theory and exact solutions of the Boltzmann equation in the relaxation time approximation. We assume that the system is composed of massless classical particles with a constant relaxation time. We shall further assume that the system undergoes a highly symmetric flow configuration -- Bjorken flow \cite{bjorken1983highly}. In this case, we have a longitudinally boost-invariant expanding fluid with a homogeneous transverse profile. In this setting, it is convenient to work with hyperbolic coordinates, $\tau = \sqrt{t^{2}-z^{2}}$ and $\eta = \tanh^{-1}(z/t)$. Then, the line element of Minkowski space reads $ds^{2} =  d\tau^{2} - dx^{2} - dy^{2} - \tau^{2} d \eta^{2}$ and the only non-vanishing Christoffel symbols are $\Gamma^{\tau}_{\eta \eta} = \tau$, $\Gamma^{\eta}_{\tau \eta}=\Gamma^{\eta}_{\eta \tau} = 1/\tau$. In this coordinate system, the fluid 4-velocity becomes trivial, $u^{\mu}=(1,0,0,0)$, and the fluid-dynamical equations simplify considerably. One further assumes that the system is invariant under reflections around the z--axis and, thus, any space-like vector such as $\nu^{\mu}$, $h^{\mu}$, $\nabla^{\mu} P_0$, and $\nabla^{\mu} \alpha$ vanishes identically. Finally, in Bjorken flow the shear tensor, the shear-stress tensor, and the expansion rate are expressed as 
\begin{equation}
\begin{aligned}
&
\sigma^{\mu}_{\ \nu} =  \text{diag}\left(0, -\frac{1}{3 \tau}, -\frac{1}{3\tau}, \frac{2}{3\tau} \right),\\
&
\pi^{\mu}_{\ \nu} = \text{diag}\left(0, - \frac{\pi}{2} , - \frac{\pi}{2}, \pi \right), \\
&
\theta = \frac{1}{\tau} . 
\end{aligned}
\end{equation}

\subsection{Hilbert equations of motion}

In Section \ref{sec:Hilb-expn} we derived the fluid-dynamical equations that emerge from the first order truncation of the Hilbert expansion. We found that 
the equilibrium fields $n_{0}$, $\varepsilon_{0}$, and $u^{\mu}$ satisfy the relativistic Euler equation \eqref{eq:euler-eqns}, while the dissipative currents obey the linear differential equations \eqref{eq:hydro-EoMs-hilb}, with the constrains given by Eqs.~\eqref{eq:constrains-hilbs-fin}. In the massless limit the coefficients $\xi_{H}$ and $\chi_{H}$ vanish and constraint \eqref{eq:constrains-hilbs-fin-sca} reduces to $\Pi_{(1)} = (1/3) \varepsilon_{(1)}$. Since all irreducible first rank tensors vanish in Bjorken flow, the remaining constraint \eqref{eq:constrains-hilbs-fin-vec} is trivially satisfied. With this information in mind and, using the notation $\delta \varepsilon \equiv \varepsilon_{(1)}$, the equations of motion obtained from the Hilbert series in Bjorken flow are 
\begin{subequations}
\label{eq:Hilb-EoM1-bjorken0}
\begin{align}
\label{eq:Hilb-EoM1-bjorken1}
&  \dot{n}_{0} 
 + 
 \frac{n_{0}}{\Hat{\tau}}
  = 0 , \\
&  \dot{\delta n} 
 + 
 \frac{\delta n}{\Hat{\tau}}
  = 0 , \\
&  \dot{\varepsilon}_{0} 
 + 
 \frac{4 \varepsilon_{0}}{3 \Hat{\tau}}
 = 0 , \\  
&  \dot{\delta \varepsilon} 
 + 
 \frac{4 \delta\varepsilon}{3 \Hat{\tau}}
 - \frac{16 \varepsilon_{0}}{45 \Hat{\tau}^{2}}  = 0,
 \label{eq:Hilb-EoM-eps-2}
\end{align}
\end{subequations}
where we defined the normalized time coordinate, $\Hat{\tau} = \tau/\tau_{R}$, and denoted $\dot{A} = dA/d\hat{\tau}$.
The above equations are solved by
\begin{equation}
\label{eq:hilbert-sol-bj}
\begin{aligned}
n_{0}(\tau) &= n_{0}(\tau_{0})\frac{\tau_{0}}{\tau}, \quad
\delta n(\tau) = \delta n(\tau_{0})\frac{\tau_{0}}{\tau}, \\
\varepsilon_{0}(\tau)  & = \varepsilon_{0}(\tau_{0}) \left(\frac{\tau_{0}}{\tau}\right)^{4/3} ,
\quad
\delta\varepsilon(\tau) = \left(\frac{\tau_{0}}{\tau}\right)^{4/3} \left[  \delta\varepsilon(\tau_{0}) + \frac{16}{45 \tau_{0}} \varepsilon_{0}(\tau_{0}) \right]  -  \frac{16 \varepsilon_{0}(\tau_{0}) \Hat{\tau}_{0}^{4/3}}{45 \Hat{\tau}^{7/3}}.
\end{aligned}    
\end{equation}
Hence, it can be seen that the ratio $\delta n/n_{0}$ is time-independent, whereas the ratio $\delta \varepsilon(\tau)/\varepsilon_{0}(\tau)$ becomes constant  asymptotically, with a transient component that decays as $1/\tau$. This is quite different than what happens to solutions of Navier-Stokes theory, which can also be solved analytically in this simplified scenario \cite{Denicol:2021}, 
\begin{equation}
\begin{aligned}
& \varepsilon_{NS}(\tau) = \varepsilon_{NS}(\tau_{0}) \left(\frac{\tau_{0}}{\tau}\right)^{4/3} \exp\left[-\frac{16}{45}\left(\frac{1}{\tau}-\frac{1}{\tau_{0}} \right)  \right].
\end{aligned}    
\end{equation}
This leads to a qualitative difference in the $1/\tau$ expansion for the normalized total energy. Indeed, for the Navier-Stokes solution, one finds the following terms when $\hat\tau\gg 1$
\begin{equation}
\begin{aligned}
& \frac{\varepsilon_{NS}(\tau)}{\varepsilon_{NS}(\tau_{0})}
=
\left(\frac{\tau_{0}}{\tau}\right)^{4/3} \exp\left(\frac{16}{45 \Hat{\tau}_{0}} \right) \left[1 - \frac{16}{45 \Hat{\tau}} + \cdots \right].
\end{aligned}    
\end{equation}
On the other hand, from Eq.~\eqref{eq:Hilb-EoM1-bjorken0}, we have for the Hilbert solution
\begin{equation}
\begin{aligned}
& \frac{\varepsilon_{0}(\tau)+\delta \varepsilon(\tau)}{\varepsilon_{0}(\tau_{0}) + \delta \varepsilon(\tau_{0})}
=
\left(\frac{\tau_{0}}{\tau}\right)^{4/3} \left[1 + \frac{16}{45} 
\frac{\varepsilon_{0}(\tau_{0})}{\varepsilon_{0}(\tau_{0}) + \delta \varepsilon(\tau_{0})} 
\left(
\frac{1}{\Hat{\tau_{0}}}
-
\frac{1}{\Hat{\tau}}\right)
 \right].
\end{aligned}    
\end{equation}
Hence, the $1/\hat\tau$ term of the Hilbert solution still displays a dependence on the initial condition, something that is not observed for the Navier-Stokes solution. This indicates that there are no attractor solutions for $\delta \varepsilon / \varepsilon$ in Hilbert theory. One may see this as a consequence of the infinite set of conservation laws that appear in this formalism. Finally, we note that for the Hilbert solution the series in square brackets ends at first order in $1/\hat\tau$, which is formally different than the Navier-Stokes solution. 

\subsection{Israel-Stewart equations of motion}
\label{sec:MIS-EoM}

In this subsection, we write the Israel-Stewart equations of motion for massless particles undergoing Bjorken flow with a constant relaxation time. These equations were recently derived using  general matching conditions in Ref.~\cite{rocha21-transient} and, for the sake of completeness, the derivation procedure is summarized in Appendix \ref{sec:MIS-gen-match}. In the following, we consider two sets of matching conditions.

\subsubsection{Exotic Eckart matching condition I: $\delta n = 0$, $\delta \varepsilon \neq 0$ ($q=1$, $s \neq 2$)} 

In this case, the continuity equation related to particle number conservation,
\begin{equation}
\label{eq:density-eom}
\begin{aligned}
\dot{n}_{0} + \frac{n_{0}}{\Hat\tau} = 0,
\end{aligned}    
\end{equation}
decouples from the rest of the  equations of motion. The remaining dynamical equations can be written as,
\begin{equation}
\label{eq:bjorken-eckart-matching-i}
\begin{aligned}
& 
\dot{\varepsilon_{0}}
+
\frac{4}{3 \Hat{\tau}}\varepsilon_{0}
- 
\delta \varepsilon
- 
\frac{\Gamma(s+4)}{20\Gamma(s+2)}\frac{\pi}{\Hat{\tau}}
=
0, \\
&
\dot{\delta \varepsilon}
+
\left(
1 + \frac{4}{3 \Hat{\tau}}\right) \delta \varepsilon
+
\left[\frac{\Gamma(s+4)}{20\Gamma(s+2)}-1\right]\frac{\pi}{\Hat{\tau}} = 0, \\
&
\dot{\pi} - \frac{16}{45 \Hat{\tau}} \left( \varepsilon_{0} + \delta \varepsilon \right)
+
\left( 1 + \frac{38}{21 \Hat{\tau} } \right) \pi =
0.
\end{aligned}
\end{equation}

\subsubsection{Exotic Eckart matching condition II: $\delta n \neq 0$, $\delta \varepsilon = 0$ ($q =2$, $s \neq 1$)}

In this case, the equation of motion related to particle number conservation does not decouple from the remaining equations of motion. The dynamical equations can be expressed in terms of the variables $n_{0}$, $\varepsilon_{0}$, $\delta n$, and $\pi$ as follows  
\begin{equation}
\label{eq:bjorken-eckart-matching-ii}
\begin{aligned}
& 
\dot{n_{0}}
+
\frac{n_{0}}{\Hat{\tau}}
- 
\delta n
- 
\left[ \frac{\Gamma(s+4)}{60\Gamma(s+2)} - \frac{1}{3\Gamma(s+2)} \right] \frac{(s-1)}{(s-2)} \frac{3n_{0}}{\varepsilon_{0}}\frac{\pi}{\Hat{\tau}}
=
0, \\
&
\dot{\delta n}
+
\left(
1 + \frac{1}{\Hat{\tau}}\right) \delta n
+
\left[ \frac{\Gamma(s+4)}{60\Gamma(s+2)} - \frac{1}{3\Gamma(s+2)} \right] \frac{(s-1)}{(s-2)} \frac{3n_{0}}{\varepsilon_{0}}\frac{\pi}{\Hat{\tau}}
 = 0, \\
& 
\dot{\varepsilon_{0}}
+
\frac{4}{3 \Hat{\tau}}\varepsilon_{0}
- 
\frac{\pi}{\Hat{\tau}}
=
0,\\
&
\dot{\pi} - \frac{16}{45 \Hat{\tau}} \left( \varepsilon_{0} + \delta \varepsilon \right)
+
\left( 1 + \frac{38}{21 \Hat{\tau} } \right) \pi =
0.
\end{aligned}
\end{equation}

\subsection{BDNK equations of motion}
\label{sec:BDN-EoM}

Now we proceed to write the BDNK equations of motion, derived in Sec.~\ref{sec:BDN-trsp-better-trun}, in Bjorken flow. As above, this is done assuming a gas composed of massless particles with a constant relaxation time and assuming two sets of matching conditions. First with $\delta n = 0$ and $\delta \varepsilon \neq 0$, then $\delta n = 0$ and $\delta \varepsilon \neq 0$. We also discuss the corresponding attractor solutions of the BDNK equations.

\subsubsection{Exotic Eckart matching condition I: $\delta n = 0$, $\delta \varepsilon \neq 0$ ($q=1$, $s \neq 2$)}
\label{sec:BDN-EoM-I} 

For this matching condition, the equation of motion for the particle density decouples and is given by \eqref{eq:density-eom}. The remaining equations of motion are,
\begin{subequations}
\begin{align}
\label{eq:BDN-EoM-eps-1}
&  \dot{\varepsilon}_{0} + \dot{\delta \varepsilon}  
 + 
 \frac{4}{3 \Hat{\tau}}\left(\varepsilon_{0}+\delta \varepsilon \right) - \frac{16 \varepsilon_{0}}{45 \Hat{\tau}^{2}}  = 0 , \\
&
 \label{eq:BDN-EoM-eps-2}
\dot{\varepsilon}_{0} 
+
\frac{4\varepsilon_{0}}{3 \Hat{\tau}} 
-
\delta \varepsilon
= 0.
\end{align}
\end{subequations}
The first equation of motion corresponds to the continuity equation related to energy conservation \eqref{eq:hydro-EoM-eps} while the second equation of motion corresponds to the constitutive relation \eqref{eq:hydro-EoM-eps}. The latter was re-written in terms of time derivatives of $\varepsilon_0$ using the equation of state and the equation of motion for particle density. It is important to notice that the equations of motion above do not depend on the matching parameter $s$, even though the transport coefficients themselves (cf.~Eq.~\eqref{eq:const-rels-eps-coefs}) do depend on $s$. This happens due to a fortuitous cancellation in the last two terms of Eq.~\eqref{eq:BDN-EoM-eps-2}, where we used $1 + (\chi^{(\alpha)}+\chi^{(\theta)})/(3 \chi^{(\alpha)} - \chi^{(\beta)}) = 4/3$ and 
$\varepsilon_{0}/(3 \chi^{(\alpha)} - \chi^{(\beta)}) = - 1$, respectively.

The coupled ordinary first order differential equations \eqref{eq:BDN-EoM-eps-1} and \eqref{eq:BDN-EoM-eps-2} can be solved analytically for $\varepsilon_0$ and $\delta \varepsilon$. This task can be performed by first converting these equations into a first order ordinary differential equation for $\Tilde{\delta \varepsilon} \equiv \delta \varepsilon/\varepsilon_{0}$,
\begin{equation}
\label{eq:BDN-eps-edo}
\begin{aligned}
& \dot{\Tilde{\delta \varepsilon}}
+
\Tilde{\delta \varepsilon}^{2}
+
\Tilde{\delta \varepsilon}
-
\frac{16}{45 \hat{\tau}^{2}} = 0,
\end{aligned}    
\end{equation}
which is a Ricatti differential equation \cite{zaitsev2002handbook} that can be solved with the Ansatz $\Tilde{\delta \varepsilon} \equiv \dot{y}/y$, leading to
\begin{equation}
\begin{aligned}
&  \ddot{y}
+
\dot{y}
-
\frac{16}{45 \Hat{\tau}^{2}} y = 0,
\end{aligned}    
\end{equation}
whose general solution is $y(\tau) = \tau^{1/2} e^{-\tau/2} \left[ a I_{\nu}\left(\tau/2 \right)
+
b K_{\nu}\left(\tau/2\right) \right]
$, where $I_{\nu}\left(x\right)$ and $K_{\nu}\left(x\right)$ are the modified Bessel functions \cite{gradshteyn2014table,NIST:DLMF} and, in the present case, $\nu = \sqrt{109/180} \approx 0.778175 $. Then, we have
\begin{equation}
\label{eq:BDN-anl-solu-eps-b}
\begin{aligned}
& \Tilde{\delta \varepsilon} (\Hat{\tau}) = \frac{(2-2\Hat{\tau})I_{\nu}\left(\frac{\Hat{\tau}}{2} \right) 
+ 
\Hat{\tau} I_{\nu+1}\left(\frac{\Hat{\tau}}{2} \right)
+
\Hat{\tau} I_{\nu-1}\left(\frac{\Hat{\tau}}{2} \right)
+
a
\left[(2-2\Hat{\tau})K_{\nu}\left(\frac{\Hat{\tau}}{2} \right) 
-
\Hat{\tau} K_{\nu-1}\left(\frac{\Hat{\tau}}{2} \right)
-
\Hat{\tau} K_{\nu+1}\left(\frac{\Hat{\tau}}{2} \right)
\right]}{ 4\Hat{\tau}\left[I_{\nu}\left(\frac{\Hat{\tau}}{2} \right) 
+
a K_{\nu}\left(\frac{\Hat{\tau}}{2} \right) \right]}.
\end{aligned}    
\end{equation}

\begin{figure}[!h]
    \centering
\begin{subfigure}{0.5\textwidth}
  \includegraphics[width=\linewidth]{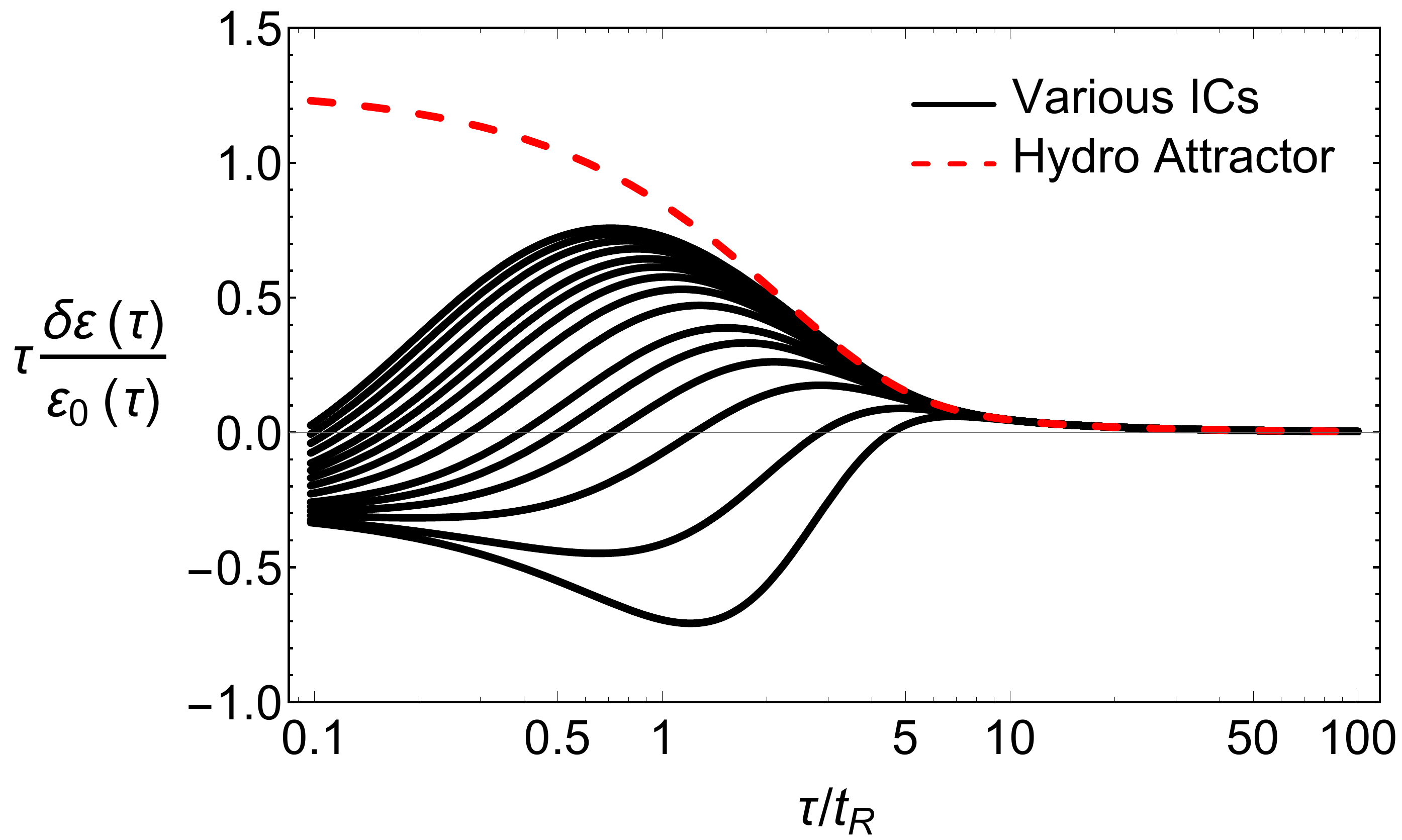}
  \caption{Late time attractor}
  \label{fig:attractor-BDN-1a}
\end{subfigure}\hfil
\begin{subfigure}{0.5\textwidth}
  \includegraphics[width=\linewidth]{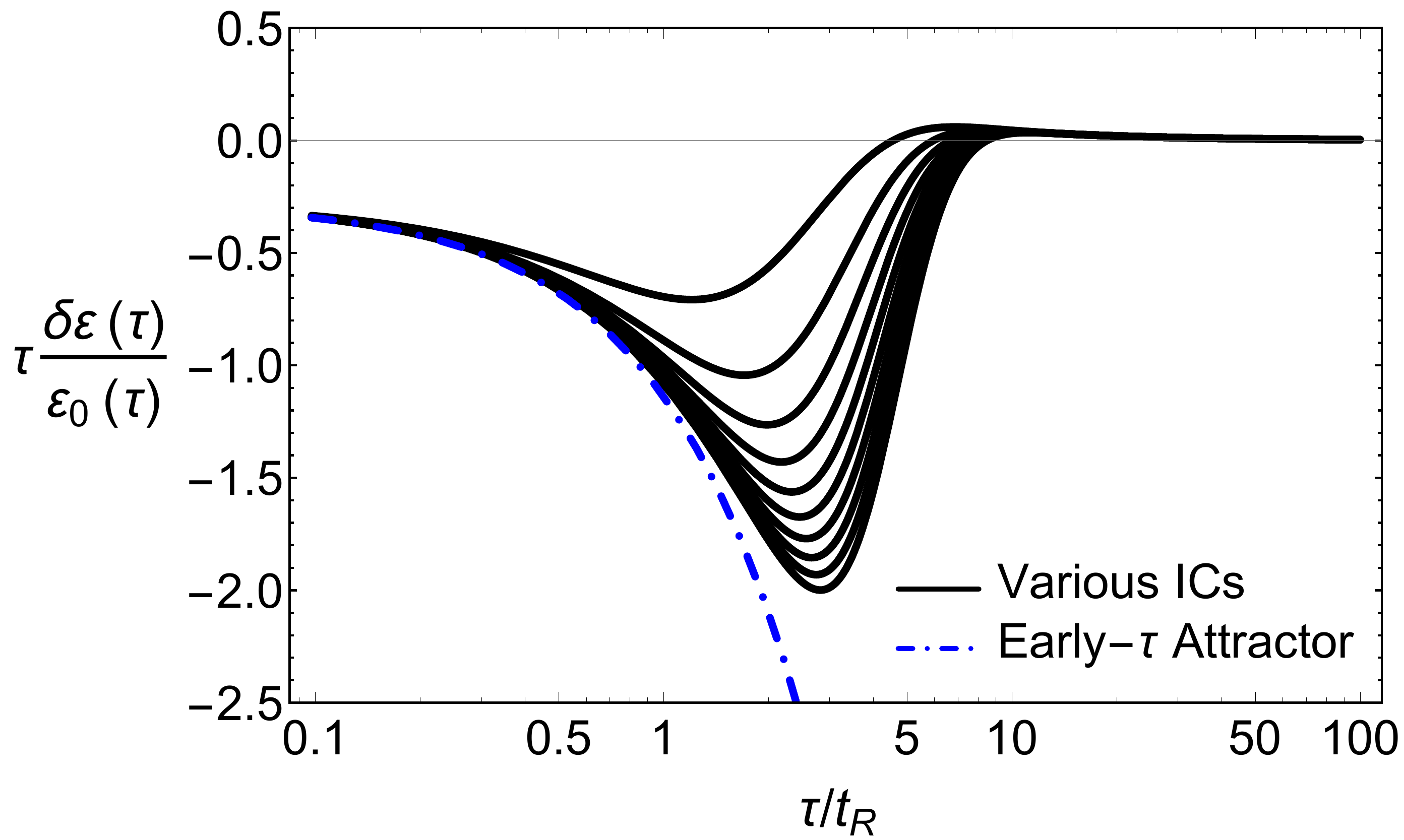}
  \caption{Early time attractor}
  \label{fig:attractor-BDN-1b}
\end{subfigure}\hfil
\caption{(Color online) Comparison between various solutions of BDNK theory,  Eq.~\eqref{eq:BDN-eps-edo}, and the attractor solutions. (a) The black curves represent solution \eqref{eq:BDN-anl-solu-eps-b} with $a=1,2,4,6,8,10,14,18,22,26,30,36,42,48,54$ and are shown in comparison with the hydrodynamic attractor \eqref{eq:BDN-atrac-hydro}. (b) The black curves represent the solutions of \eqref{eq:BDN-anl-solu-eps-b} with $a=1,1/2, 1/3, \cdots, 1/10$ shown in comparison with the early time attractor \eqref{eq:BDN-atrac-early}. 
 }
\label{fig:attractor-BDN-1}
\end{figure}

As $\Hat{\tau}$ goes to infinity, $K_{\nu}(\Hat{\tau}) \sim \sqrt{\pi/2} \ \Hat{\tau}^{-1/2} e^{-\Hat{\tau}}$ and $I_{\nu}(\Hat{\tau}) \sim (2 \pi)^{-1} \ \Hat{\tau}^{-1/2} e^{\Hat{\tau}} $ \cite{NIST:DLMF}. Hence, the terms of Eq.~\eqref{eq:BDN-anl-solu-eps-b} corresponding to the $I_{\nu}$'s dominate and any information about the initial condition is erased. The late time behavior of the solution becomes 
\begin{equation}
\label{eq:BDN-atrac-hydro}
\begin{aligned}
& \Tilde{\delta \varepsilon}_{\text{att}} = \frac{
(2-2\Hat{\tau})I_{\nu}\left(\frac{\Hat{\tau}}{2} \right) 
+ 
\Hat{\tau} I_{\nu-1}\left(\frac{\Hat{\tau}}{2} \right)
+
\Hat{\tau} I_{\nu+1}\left(\frac{\Hat{\tau}}{2} \right)}{ 4\Hat{\tau}
I_{\nu}\left(\frac{\Hat{\tau}}{2}  \right)}.
\end{aligned}
\end{equation}
We note that this also corresponds to the solution of \eqref{eq:BDN-eps-edo} with $a=0$ and, thus, it is referred to as an attractor solution. On the other hand, as $\Hat{\tau}$ approaches zero,  $I_{\nu}(\Hat{\tau}) \sim (1/2)\Gamma(\nu)^{-1}(\Hat{\tau}/2)^{\nu}$ and $K_{\nu}(\Hat{\tau}/2) \sim \Gamma(\nu + 1)(2\Hat{\tau})^{-\nu}$, the $K_{\nu}$ terms in Eqs.~\eqref{eq:BDN-anl-solu-eps-b} dominate and, once more, any information about any boundary condition is erased. Hence, the early-time solution is also universal and becomes
\begin{equation}
\label{eq:BDN-atrac-early}
\begin{aligned}
& \Tilde{\delta \varepsilon}_{\text{pb}} = \frac{
(2-2\Hat{\tau})K_{\nu}\left(\frac{\Hat{\tau}}{2} \right) 
- 
\Hat{\tau} K_{\nu-1}\left(\frac{\Hat{\tau}}{2} \right)
-
\Hat{\tau} K_{\nu+1}\left(\frac{\Hat{\tau}}{2} \right)}{ 4\Hat{\tau}
K_{\nu}\left(\frac{\tau}{2}  \right)}.
\end{aligned}
\end{equation}
This can be identified as the solution of \eqref{eq:BDN-eps-edo} with $a \to \infty$ and is referred to as a pullback attractor (see \cite{Behtash:2019txb}). Results for the evolution with various initial or boundary conditions, compared to the corresponding attractor solutions, can be seen in Fig.~\ref{fig:attractor-BDN-1}. There, we further normalize $\delta \varepsilon/\varepsilon_{0}$ with $1/ \tau$. 

\subsubsection{Exotic Eckart matching condition II: $\delta n \neq 0$, $\delta \varepsilon = 0$ ($q =2$, $s \neq 1$)} 
\label{sec:BDN-EoM-II}

\begin{figure}[!h]
    \centering
\begin{subfigure}{0.5\textwidth}
  \includegraphics[width=\linewidth]{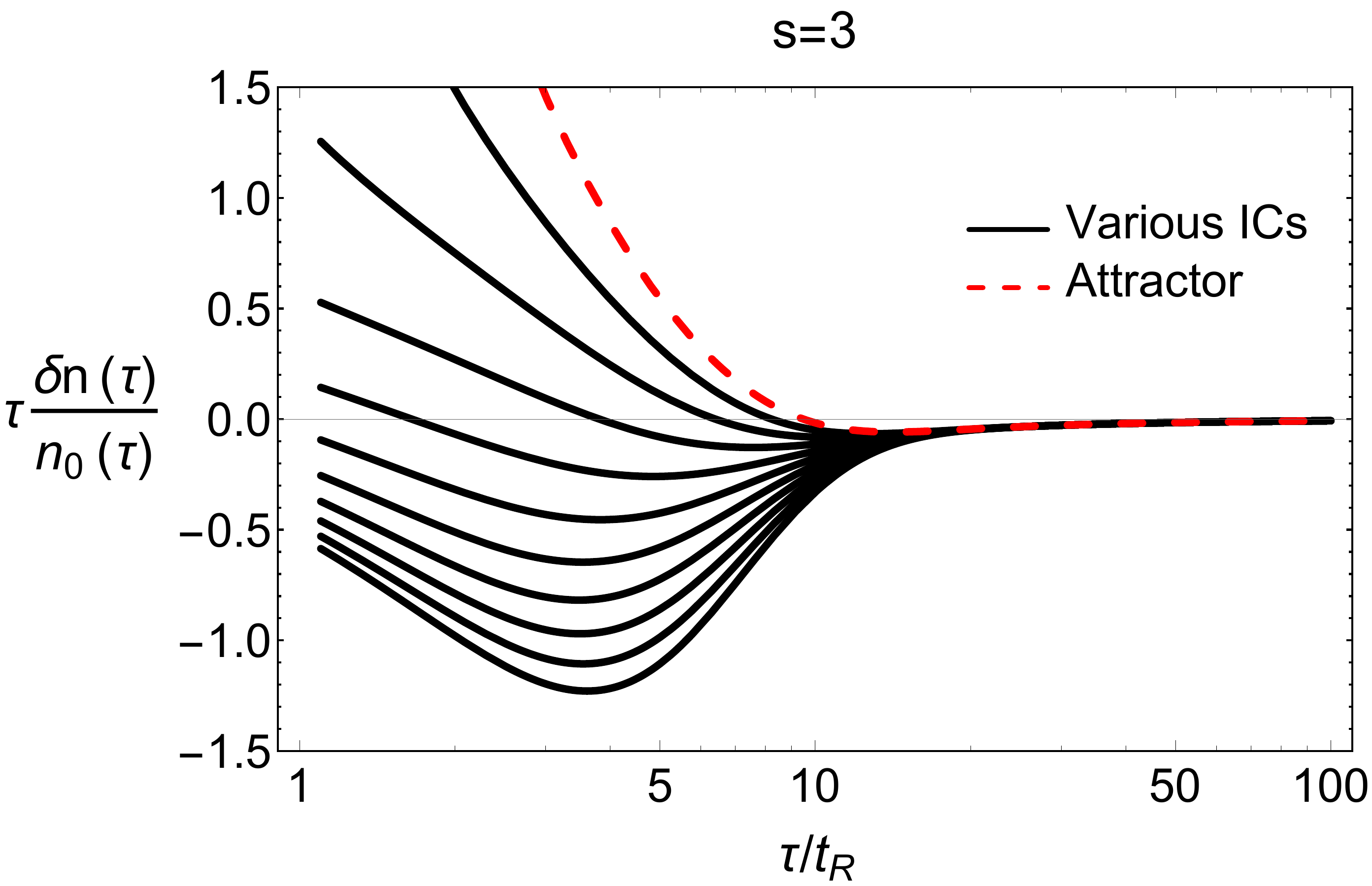}
  \caption{$s=3$}
  \label{fig:mu-q=0}
\end{subfigure}\hfil
\begin{subfigure}{0.5\textwidth}
  \includegraphics[width=\linewidth]{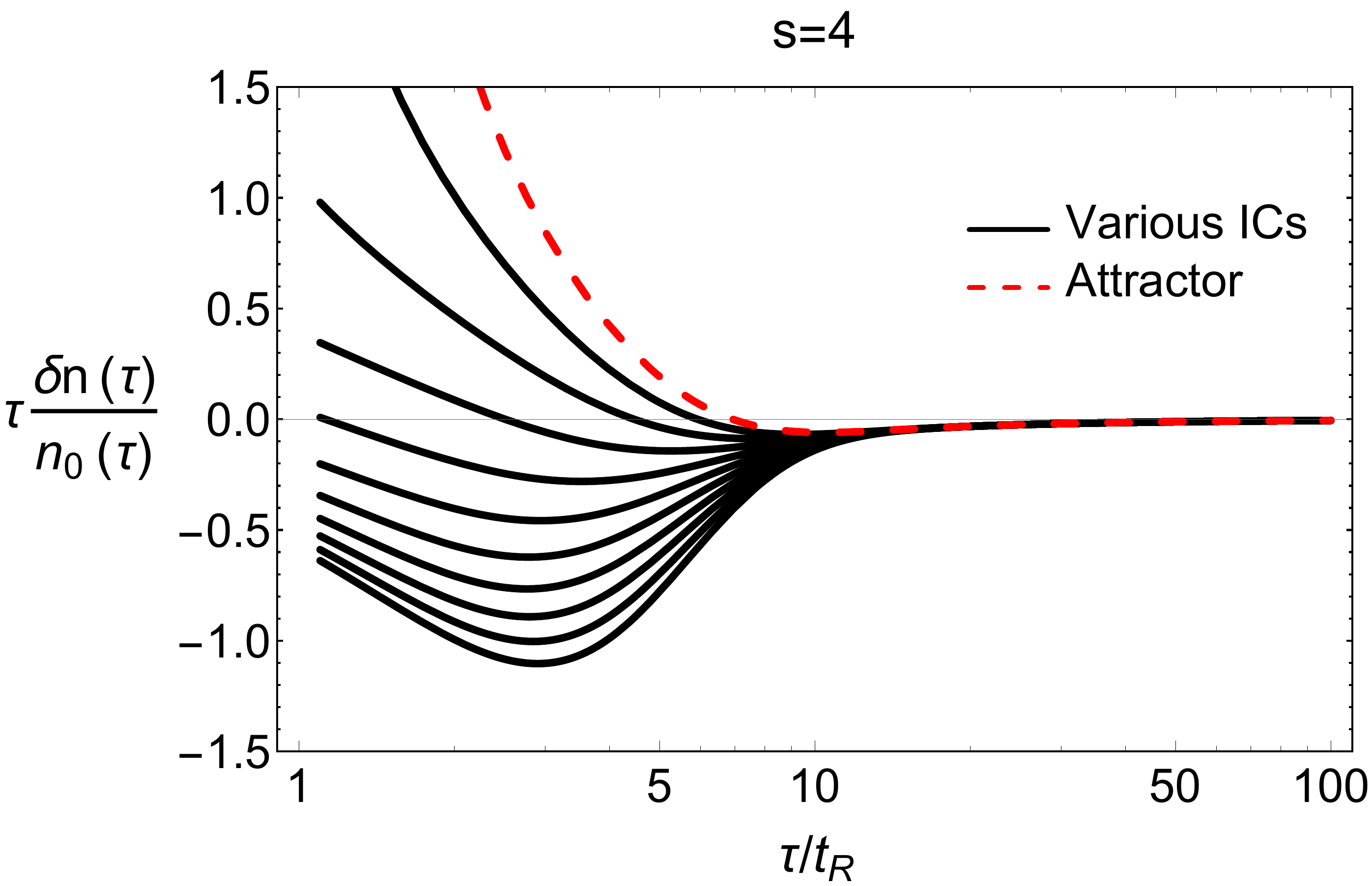}
  \caption{$s=4$}
  \label{fig:mu-q=1}
\end{subfigure}\hfil
\caption{(Color online) Comparison of hydrodynamic attractor solutions for different matching conditions, with matching parameter $s=3$ (a) and $s=4$ (b).}
\label{fig:attractors-type-ii-BDN}
\end{figure}

For the present matching conditions it is convenient to use $n_{0}$ and $\delta n$ as dynamical variables. The equations of motion are obtained from Eqs.~\eqref{eq:hydro-EoM-n0} and \eqref{eq:hydro-EoM-eps} assuming an equation of state of a gas of massless particles, $\varepsilon_{0} = 3 n_{0}/\beta = 3 e^{\alpha}/(\pi^{2} \beta^{4})$ , and constitutive relations \eqref{eq:const-rels-del-n}. Then, we have
\begin{subequations}
\begin{align}
\label{eq:EoM-BDN-n}
&  \dot{n}_{0}  
+
\dot{\delta n}
+ 
 \frac{1}{\Hat{\tau}} (n_{0}+\delta n)  = 0 , \\
&
 \label{eq:EoM-BDN-epsil}
\dot{n}_{0} 
+
\frac{n_{0}}{\Hat{\tau}} 
-
\frac{(s-2)}{(s-1)}\delta n 
-
\frac{16 n_{0}}{45 \Hat{\tau}^{2}}
= 0.
\end{align}
\end{subequations}
In contrast to the previous class of matching conditions, now the equations of motion depend explicitly on the matching parameter $s$. In a direct analogy with the previous case, we can derive an analytical solution for $\Tilde{\delta n} \equiv \delta n/n_{0}$ that obeys
\begin{equation}
\begin{aligned}
& \dot{\Tilde{\delta n}}
+
\frac{(s-2)}{(s-1)}
\Tilde{\delta n}^{2}
+
\frac{(s-2)}{(s-1)}
\Tilde{\delta n}
+
\frac{16}{45 \Hat{\tau}^{2}} \left( 
\Tilde{\delta n}
 + 1 \right) = 0,
\end{aligned}    
\end{equation}
which is also a Ricatti differential equation. The latter can be solved using the changes of variable $\Tilde{\delta n} \equiv \dot{y}/(A_{s} y)$ and $z \equiv \dot{y}+A_{s} y$, with $A_{s} \equiv (s-2)/(s-1)$, leading to the simple differential equation for $z$,
\begin{equation}
\begin{aligned}
& \dot{z}
+
\frac{16}{45 \Hat{\tau}^{2}} z = 0, 
\end{aligned}    
\end{equation}
which is solved by $z(\Hat{\tau}) = a e^{-A_{s} \Hat{\tau}} +  b e^{-A_{s}\Hat{\tau}} \int_{\Hat{\tau}_{0}}^{\Hat{\tau}} e^{A_{s} \Hat{\tau}+\frac{16}{45 \Hat{\tau}}} \, d\Hat{\tau}$ and, thus,
\begin{equation}
\begin{aligned}
& \Tilde{\delta n} = \frac{1}{A_{s}} \frac{ e^{A_{s}\Hat{\tau}+\frac{16}{45 \Hat{\tau}}}}{ \int_{\Hat{\tau}_{0}}^{\Hat{\tau}} e^{A_{s} \Hat{\tau}+\frac{16}{45 \Hat{\tau}}} \, d\Hat{\tau}+a'}-1,
\end{aligned}    
\label{solutions}
\end{equation}
where the only independent integration constant $a'= a/b$ has been chosen. The non-analytic behavior is evident and we can express the late-time attractor as
\begin{equation}
\begin{aligned}
& \Tilde{\delta n}_{\text{at}} = \frac{1}{A_{s}} \frac{  e^{A_{s}\Hat{\tau}+\frac{16}{45 \Hat{\tau}}}}{ \int_{\Hat{\tau}_{0}}^{\Hat{\tau}} e^{A_{s}\Hat{\tau}+\frac{16}{45 \Hat{\tau}}} \, d\Hat{\tau}}-1,
\end{aligned}    
\end{equation}
which corresponds to the solution with $a'=0$  displayed in Fig.~\ref{fig:attractors-type-ii-BDN}, compared with solutions \eqref{solutions} for various initial conditions, i.e., for several values of $a'$.


\subsection{Matching condition influence on evolution}

Now we are in position to compare the solutions of BDNK's and Israel-Stewart's equations of motion and the corresponding exact solutions from the Boltzmann equation in Bjorken flow. Here, we also assess the effect that matching conditions can have on such solutions. Due to the fact that any space-like 4-vector is identically zero in Bjorken flow, this analysis will be limited to the parameters $q$ and $s$ which define temperature and chemical potential (cf. \eqref{eq:matching_kinetic1} and \eqref{eq:matching_kinetic2}). In the present section, we shall only use type I and type II Exotic Eckart matching conditions.

\subsubsection{Exotic Eckart matching conditions I: $\delta n = 0$, $\delta \varepsilon \neq 0$ ($q=1$, $s \neq 2$)} 

\begin{figure}[!h]
    \centering
\begin{subfigure}{0.5\textwidth}
  \includegraphics[width=\linewidth]{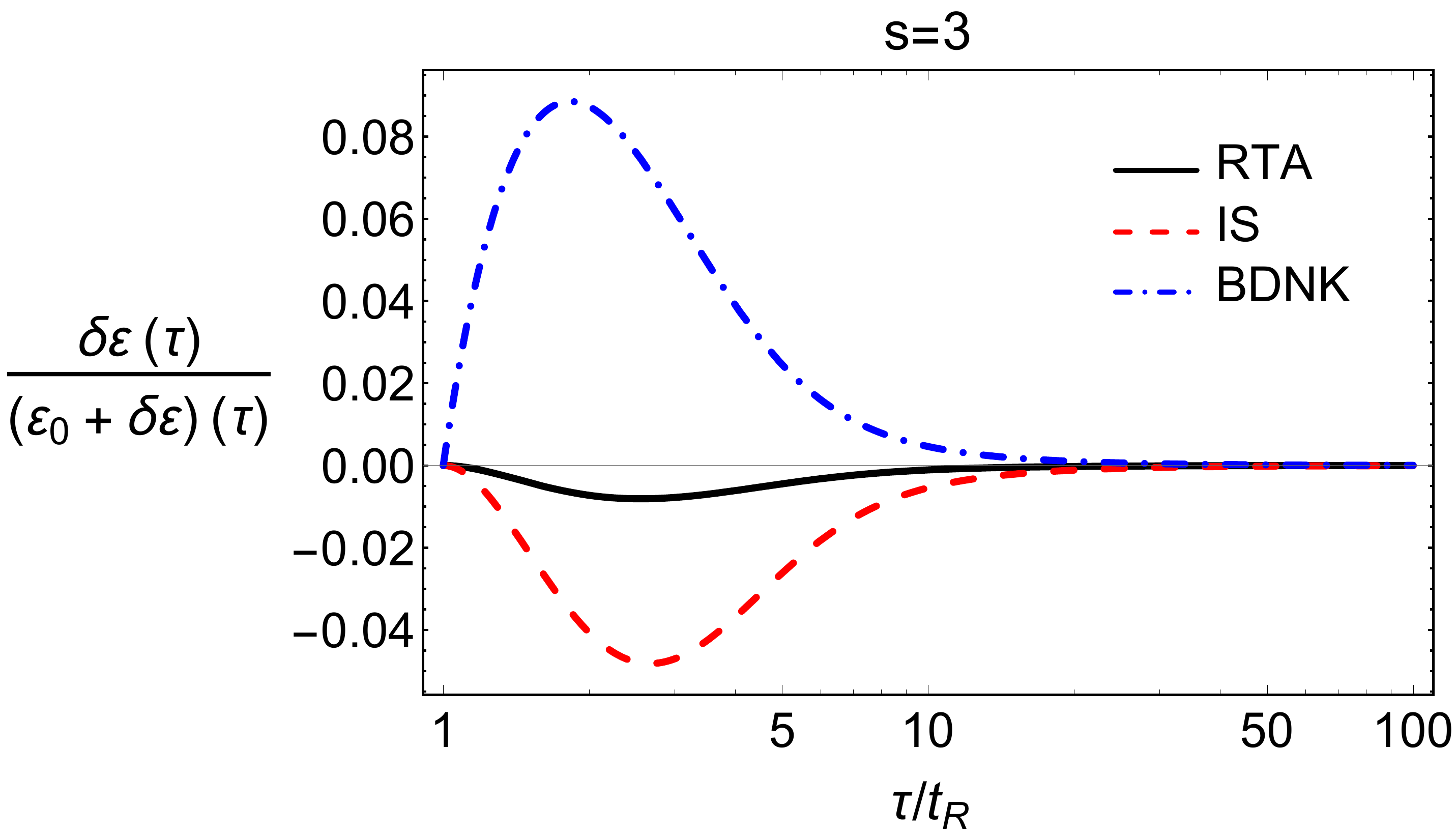}
  \label{fig:evol-deps-s=3-EQL}
\end{subfigure}\hfil
\begin{subfigure}{0.5\textwidth}
  \includegraphics[width=\linewidth]{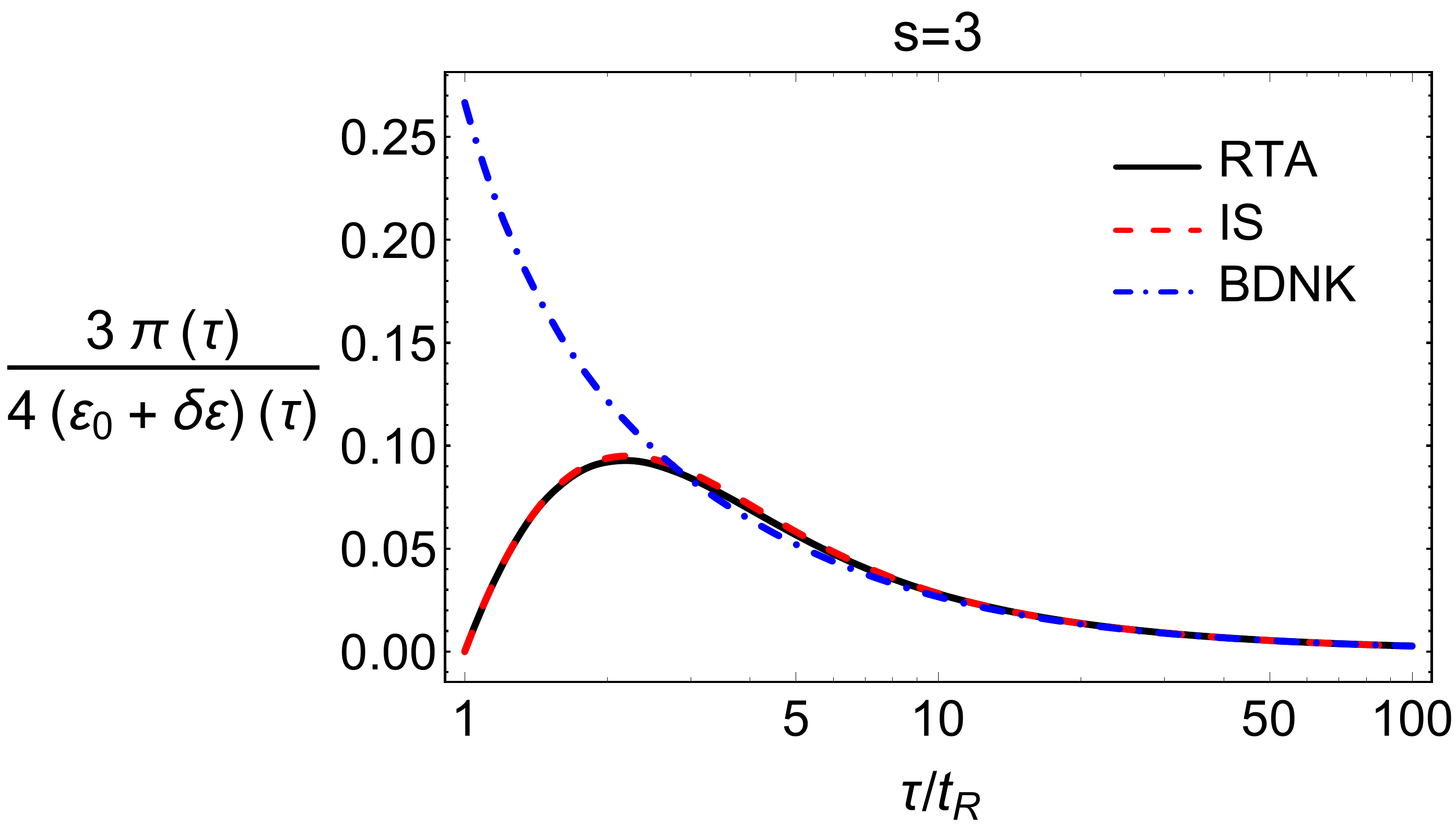}
  \label{fig:evol-pi-s=3-EQL}
\end{subfigure}\hfil
\caption{(Color online) Evolution of the non-equilibrium fraction of the energy density (right) and the normalized shear-stress tensor, $\pi/(\varepsilon_{0} + P_{0} + \delta \varepsilon + \Pi) = 3\pi/[4(\varepsilon_{0} + \delta \varepsilon)]$ (left), for the Boltzmann equation (RTA), Israel-Stewart (IS) and BDNK for type I exotic Eckart and $s=3$ and equilibrium initial conditions.}
\label{fig:evol-deps-pi-s=3-EQL}
\end{figure}

\begin{figure}[!h]
    \centering
\begin{subfigure}{0.5\textwidth}
  \includegraphics[width=\linewidth]{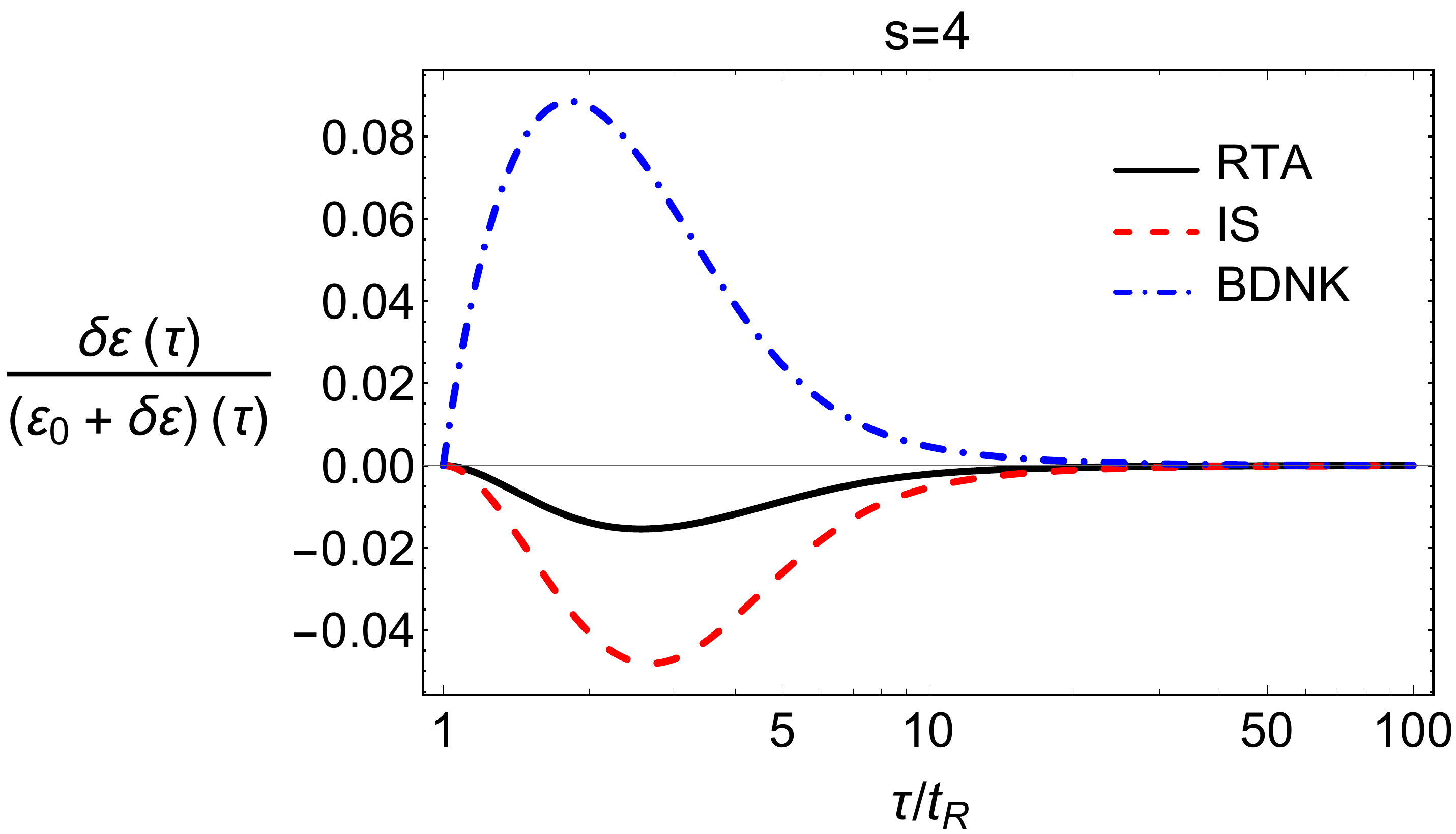}
  \label{fig:evol-deps-s=4-EQL}
\end{subfigure}\hfil
\begin{subfigure}{0.5\textwidth}
  \includegraphics[width=\linewidth]{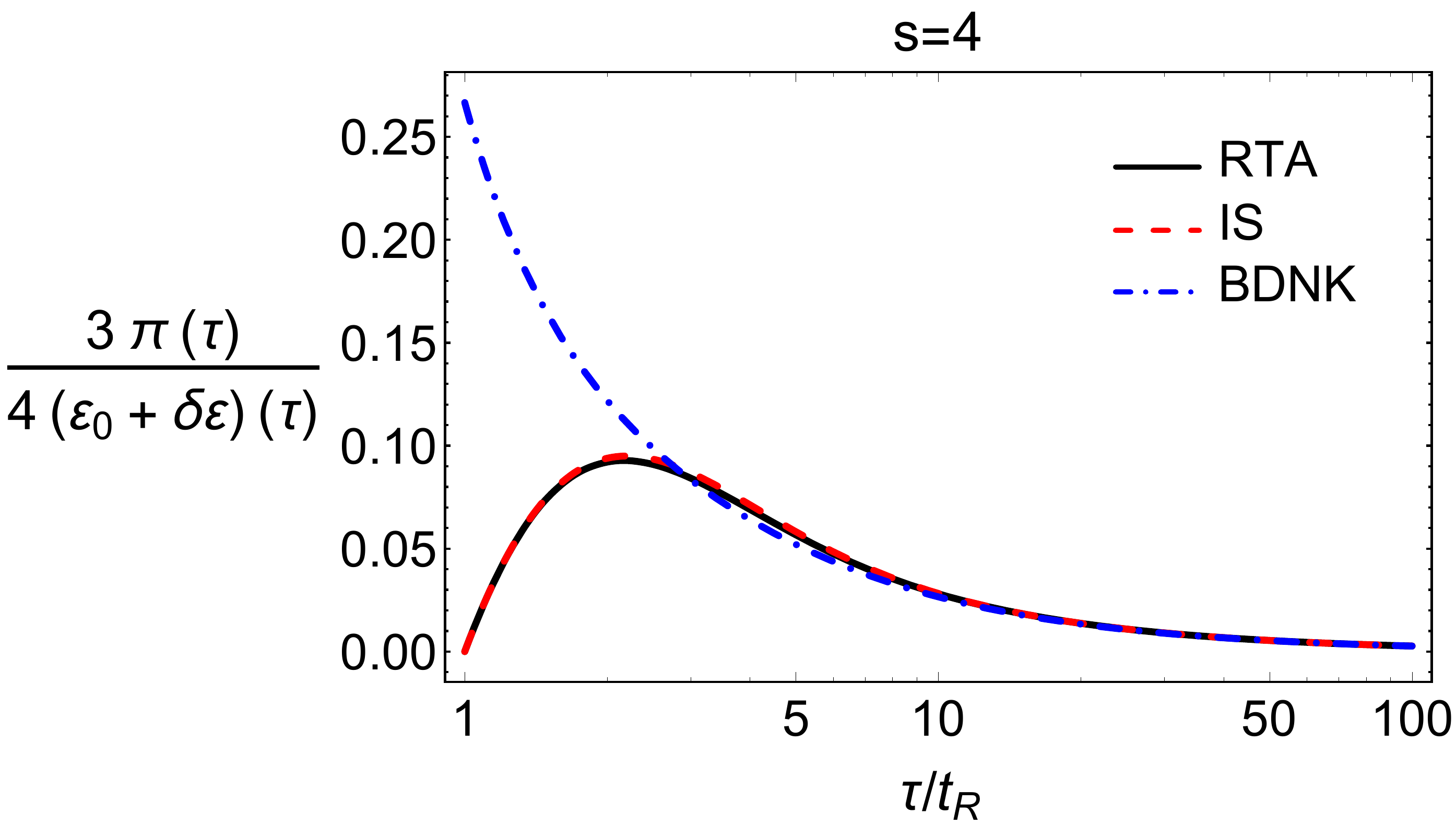}  
  \label{fig:evol-pi-s=4-EQL}
\end{subfigure}\hfil
\caption{(Color online) Evolution of the non-equilibrium fraction of the energy density (left) and the normalized shear-stress tensor, $\pi/(\varepsilon_{0} + P_{0} + \delta \varepsilon + \Pi) = 3\pi/[4(\varepsilon_{0} + \delta \varepsilon)]$ (right), for the Boltzmann equation (RTA), Israel-Stewart (IS), and BDNK  for type I exotic Eckart and $s=4$ and equilibrium initial conditions. }
\label{fig:evol-deps-pi-s=4-EQL}
\end{figure}

In this subsection we plot solutions of fluid-dynamical theories considering matching conditions in which $\delta n \equiv 0$, but $\delta \varepsilon \neq 0$ ($q =1$, $s \neq 2$). Unless stated otherwise, we consider initial conditions in which the dynamical variables are in local thermodynamic equilibrium (note that, in BDNK theory, the shear-stress tensor is not an independent dynamical variable and is determined by constitutive relations). Figures \ref{fig:evol-deps-pi-s=3-EQL} and \ref{fig:evol-deps-pi-s=4-EQL} portray the evolution of the (normalized) dissipative currents $\delta \varepsilon/(\varepsilon_0+\delta \varepsilon)$ and $3\pi/[4(\varepsilon_0+\delta \varepsilon)]$. In Fig.~\ref{fig:evol-deps-pi-s=3-EQL}, we employ $s=3$, while in Fig.~\ref{fig:evol-deps-pi-s=4-EQL} we display results for $s=4$. The corresponding solution obtained from the Boltzmann equation using the method of moments \cite{Denicol:2021} (see Appendix \ref{sec:BEq-EoM} for details) is  displayed in solid black lines, for the sake of comparison. 

We find that at late times ($\hat{\tau} \gtrsim 3$) all three solutions have approximately the same evolution for the shear-stress tensor. The difference observed among the solutions at early times comes mostly from the fact that we imposed equilibrium initial conditions for the solutions of Israel-Stewart theory and the Boltzmann equation, something that is not possible to implement for the shear-stress tensor in BDNK theory. On the other hand, the evolution of $\delta \varepsilon$ is very different in all three cases. The Israel-Stewart formalism yields a negative sign for this quantity, which is in qualitative agreement with the solution of the Boltzmann equation for this variable. However, Israel-Stewart theory clearly overpredicts the magnitude of this non-equilibrium correction. As already pointed out in Ref.~\cite{rocha21-transient}, in the Israel-Stewart formalism, this happens due to the dominance of the $\delta \varepsilon$-$\pi$ coupling coefficient (the last term in the second equation of \eqref{eq:bjorken-eckart-matching-i}, see also Table \ref{tab:couplings}), which yields a negative contribution to $\delta \varepsilon$, for $s=3 \text{ or } 4$. In contrast, the BDNK formalism yields a positive sign for $\delta \varepsilon$, due to the fact that it is driven by the shear term $16/(45 \Hat{\tau}^{2})$ in Eq.~\eqref{eq:BDN-EoM-eps-1}. The matching condition has a significant effect in solutions of the linearized Boltzmann equation and Israel-Stewart theory, increasing the non-equilibrium correction as one goes from $s=3$ to $s=4$. We have already demonstrated that, for this set of matching conditions, the solutions of BDNK theory for the normalized non-equilibrium energy density have no dependence on $s$.

\begin{figure}[!h]
    \centering
\begin{subfigure}{0.5\textwidth}
  \includegraphics[width=\linewidth]{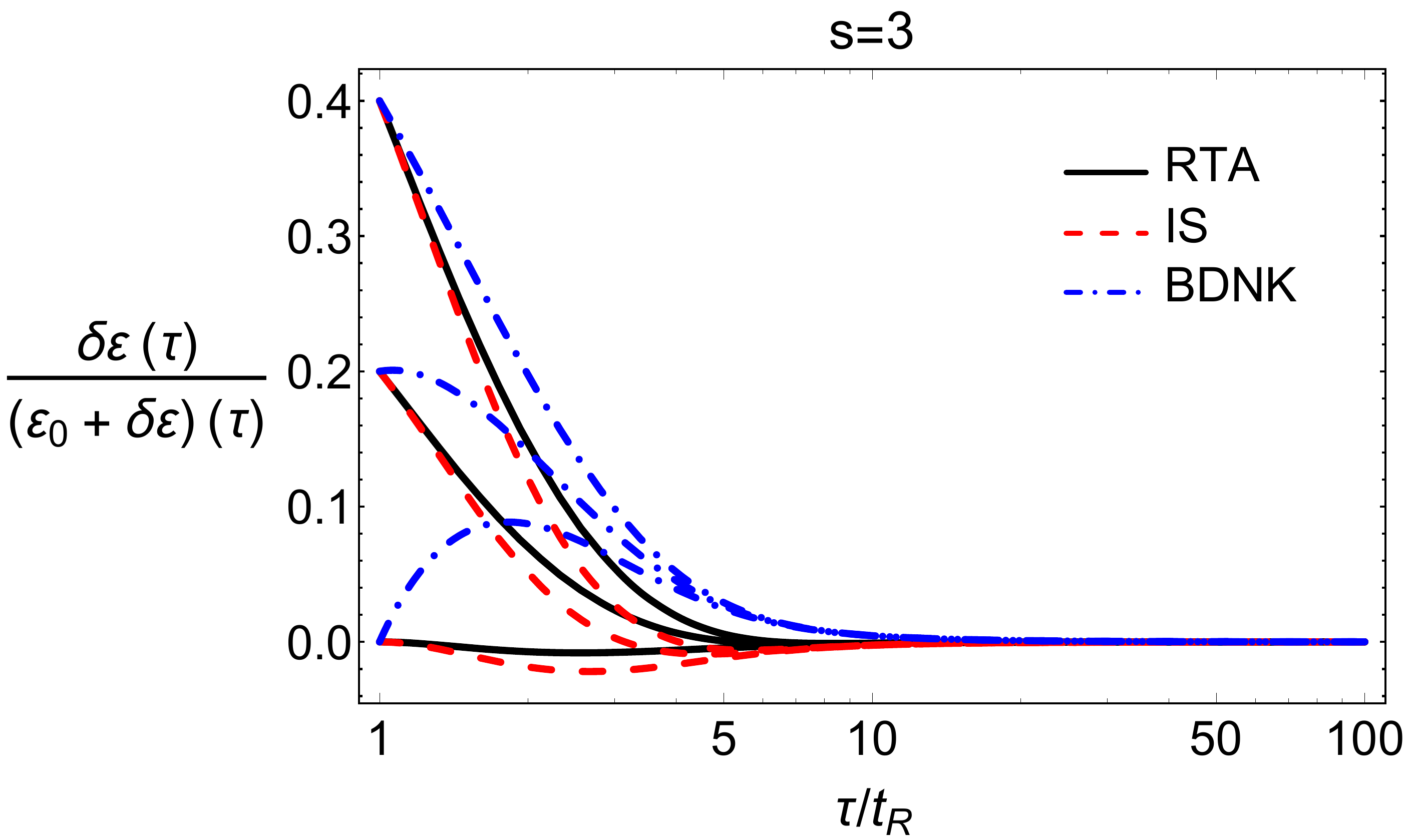}
  \label{fig:atr-deps-pi-s=3-BDN}
\end{subfigure}\hfil
\begin{subfigure}{0.5\textwidth}
  \includegraphics[width=\linewidth]{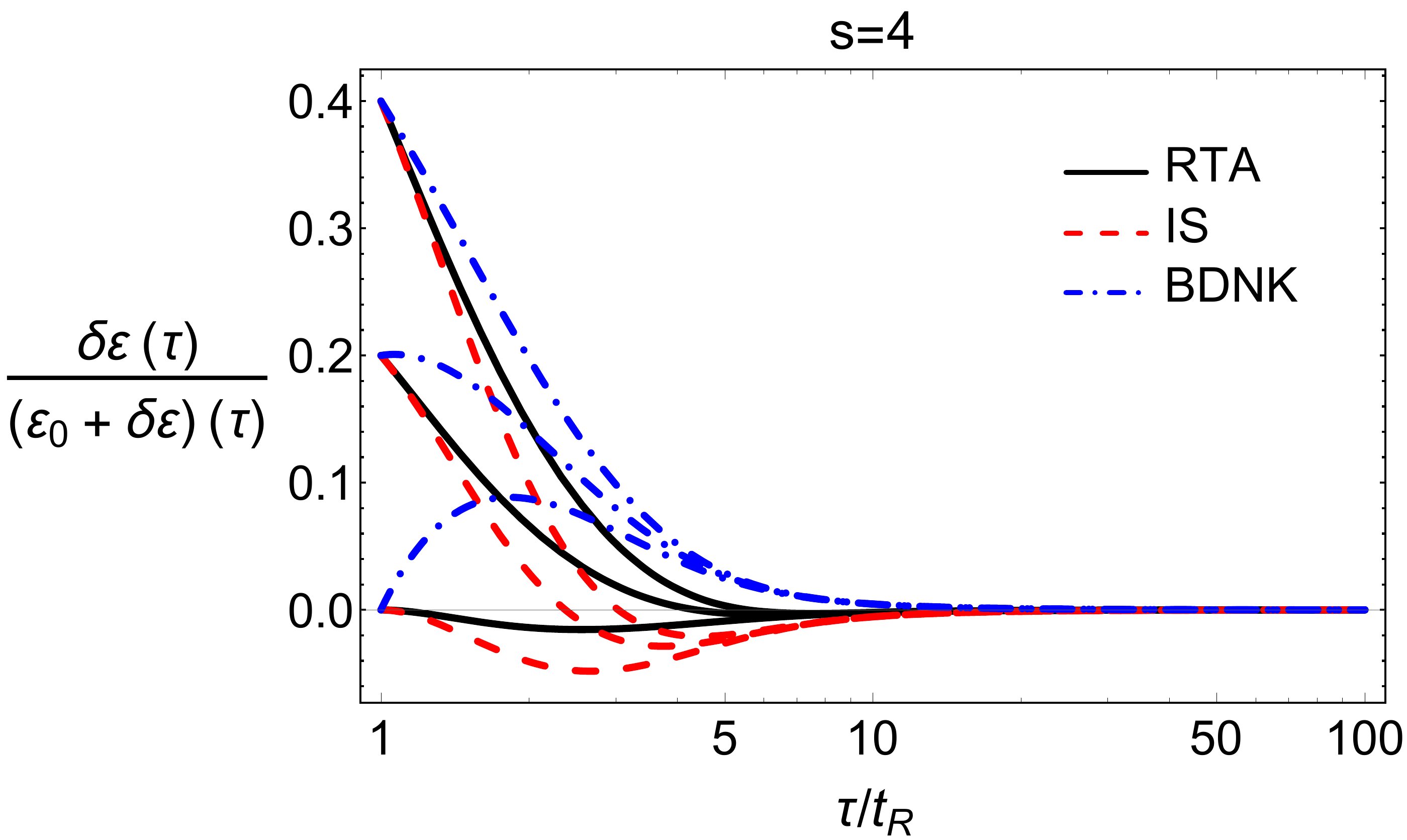}
   \label{fig:atr-deps-pi-s=4-BDN}
\end{subfigure}\hfil
\caption{(Color online) Comparison between the time evolution (for various initial conditions) of the non-equilibrium correction for the energy density found by solving the RTA Boltzmann equation, Israel-Stewart (IS), and BDNK using type I exotic Eckart for $s=3$ (left) and $s=4$ (right). The initial conditions are so that $\delta \varepsilon(\Hat{\tau}_{0})/[(\varepsilon_{0} + \delta \varepsilon)(\Hat{\tau}_{0})] =0$, $0.2$, $0.4$, and, when applicable, $\pi(\Hat{\tau}_{0}) = 0$. }
\label{fig:atr-deps-pi}
\end{figure}

We now consider solutions of Israel-Stewart, BDNK, and the Boltzmann equation for several initial values of $\delta \varepsilon$. The remaining dynamical variables of each theory are still set to their respective local equilibrium values. The idea is to visualize the attractor dynamics that each formalism displays. The results are shown in Fig.~\ref{fig:atr-deps-pi}, where we considered simulations with $\Hat{\tau}_{0}\delta \varepsilon(\Hat{\tau}_{0})/\varepsilon(\Hat{\tau}_{0}) =0$, $0.2$, and $0.4$. As expected, we see that the solutions of each theory display universal behavior at late times ($\hat{\tau} \gtrsim 5$), indicating the existence of late-time attractor solutions. We note that such attractor solutions were already explicitly derived for BDNK theory in Sec.~\ref{sec:BDN-EoM}. We also note that the late time solutions of Israel-Stewart theory and the Boltzmann equation for $\delta \varepsilon$ are qualitatively similar, in particular when it comes to the sign of the energy density non-equilibrium correction. The quantitative agreement between these solutions is not very good and worsens as one increases $s$. The BDNK formalism, on the other hand, clearly has a different attractor solution, which always displays a positive energy density non-equilibrium correction.

\subsubsection{Exotic Eckart matching conditions II: $\delta n  \neq 0$, $\delta \varepsilon \equiv 0$ ($q=2$, $s \neq 1$)} 

\begin{figure}[!h]
    \centering
\begin{subfigure}{0.5\textwidth}
  \includegraphics[width=\linewidth]{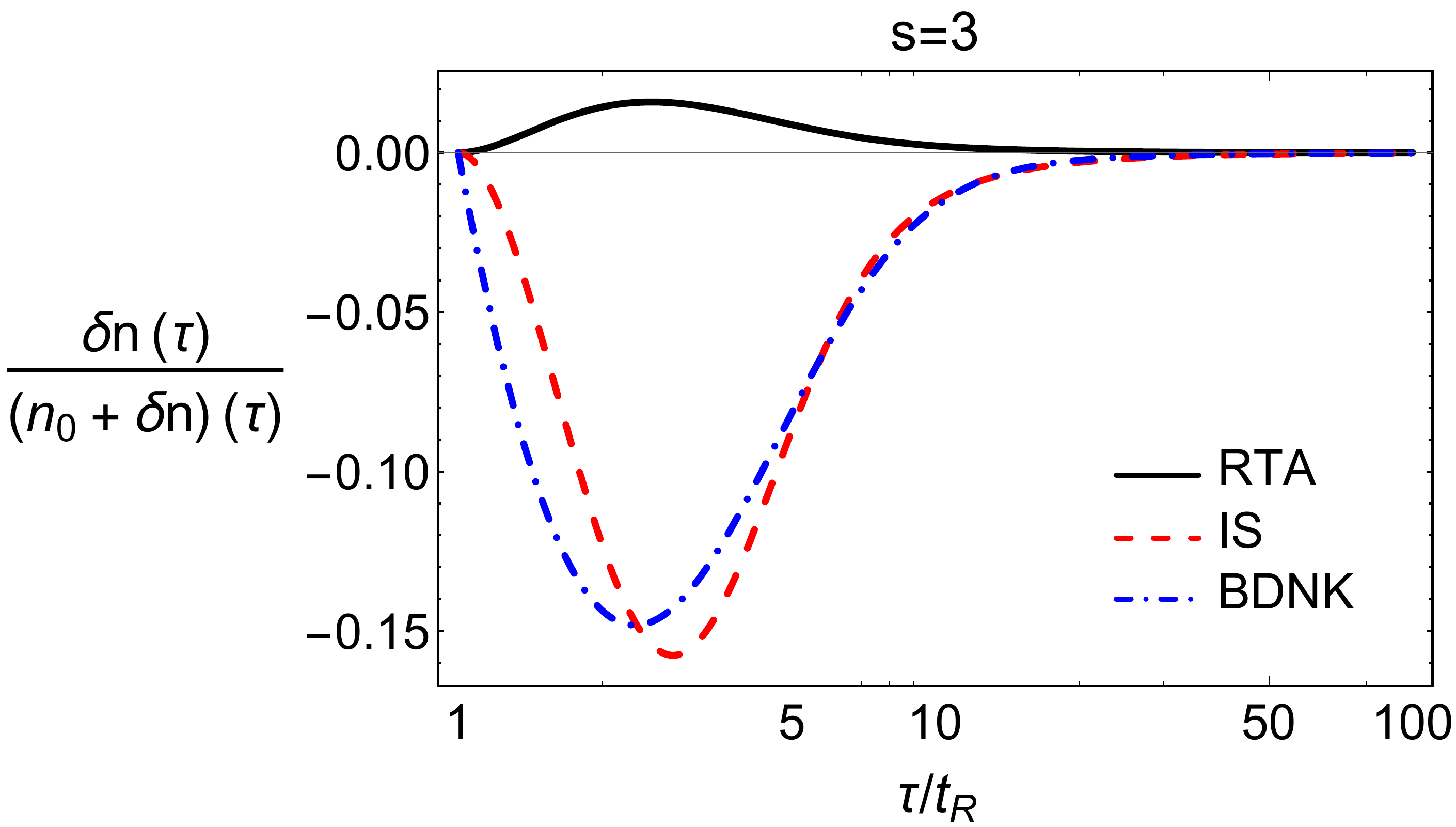}
  \label{fig:evol-dn-s=3-EQL}
\end{subfigure}\hfil
\begin{subfigure}{0.45\textwidth}
  \includegraphics[width=\linewidth]{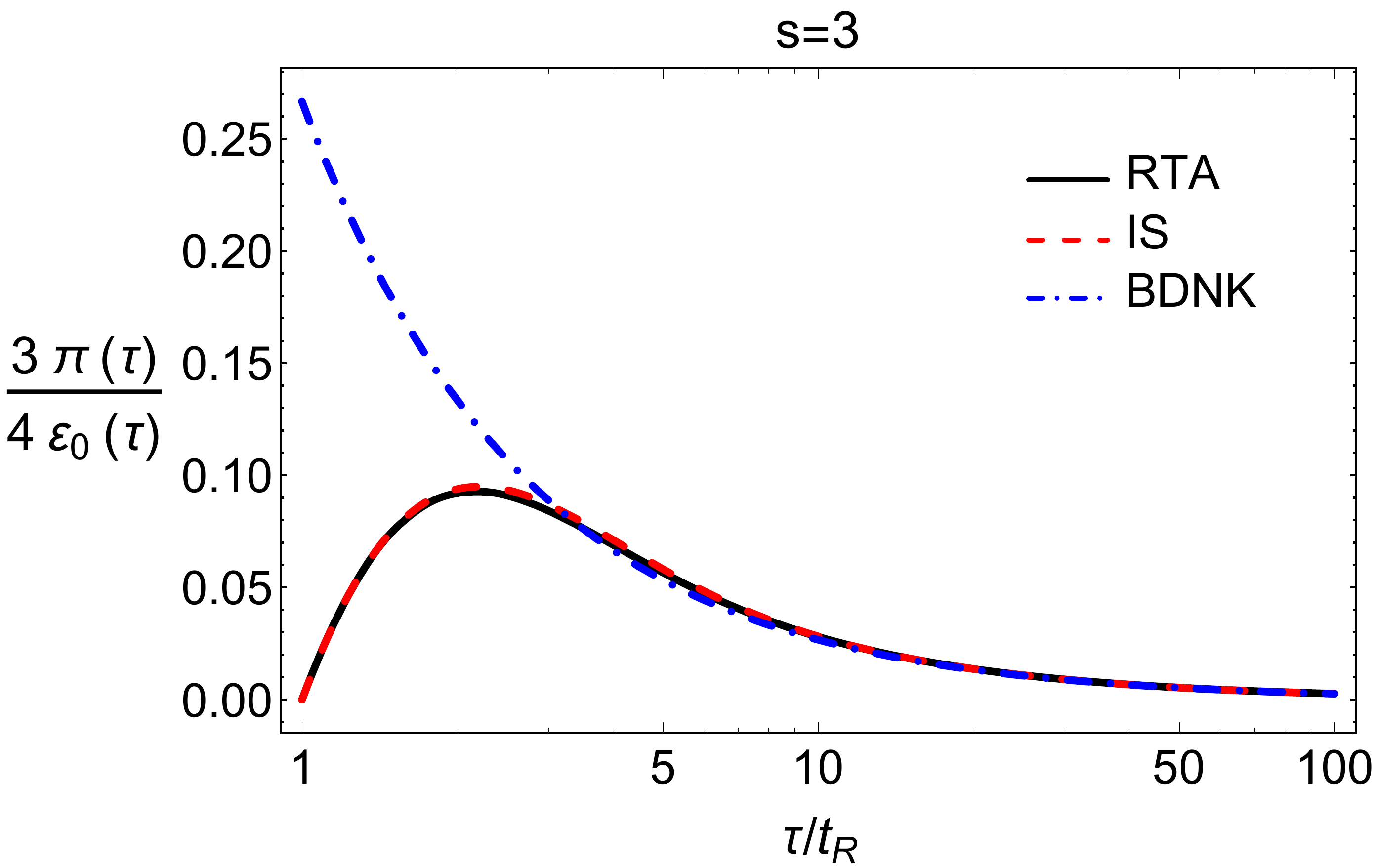}
  \label{fig:evol-pi-s=3-EQL1}
\end{subfigure}\hfil
\caption{(Color online) Evolution of the non-equilibrium fraction of the particle density (left) and the normalized shear-stress tensor, $\pi/(\varepsilon_{0} + P_{0}) = 3\pi/(4 \varepsilon_{0})$ (right), found by solving the RTA Boltzmann equation, Israel-Stewart (IS), and BDNK  for type II exotic Eckart with $s=3$ and equilibrium initial conditions.}
\label{fig:evol-dn-pi-s=3-EQL}
\end{figure}

\begin{figure}[!h]
    \centering
\begin{subfigure}{0.5\textwidth}
  \includegraphics[width=\linewidth]{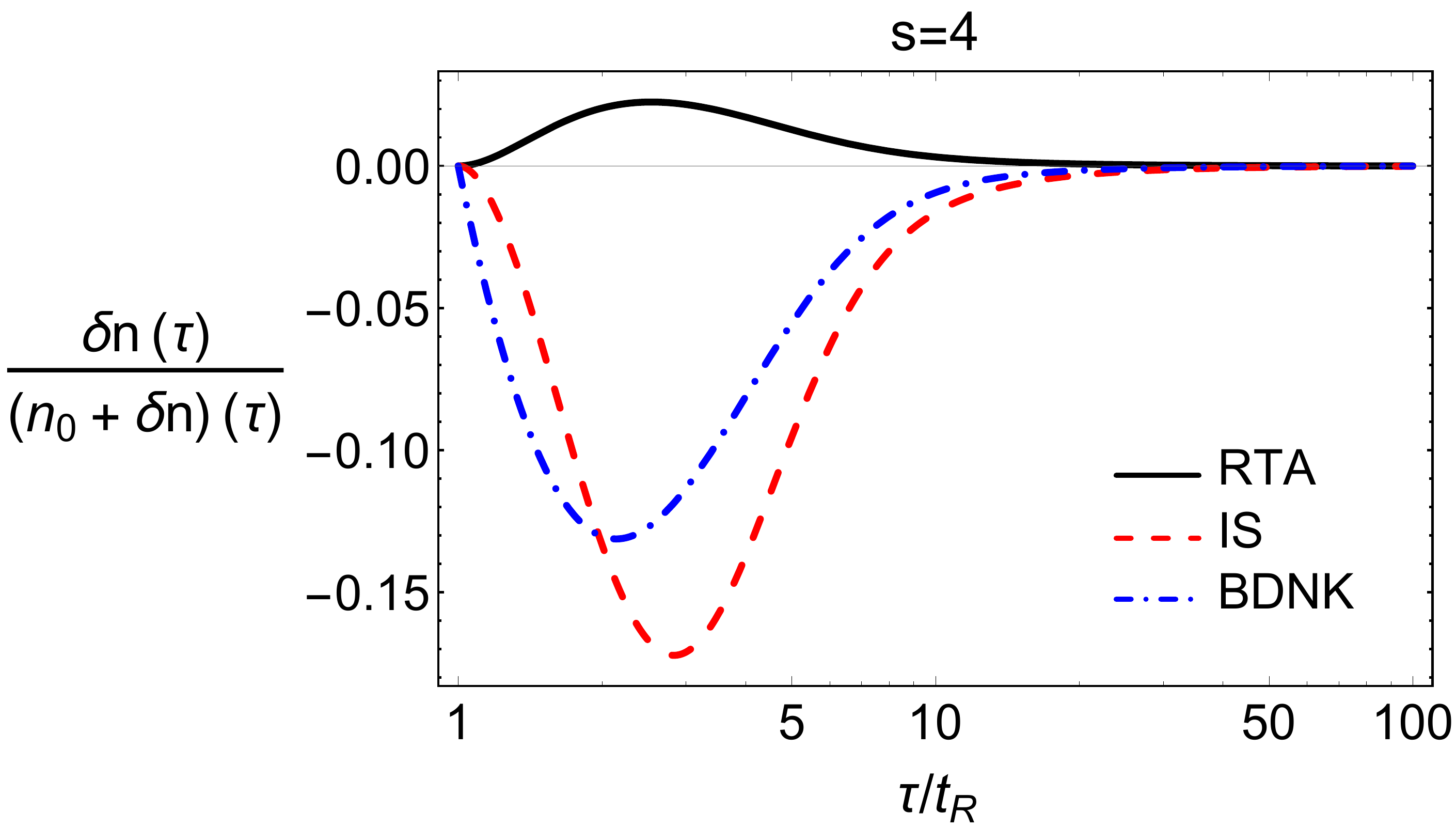}
  \label{fig:evol-dn-s=4-EQL}
\end{subfigure}\hfil
\begin{subfigure}{0.45\textwidth}
  \includegraphics[width=\linewidth]{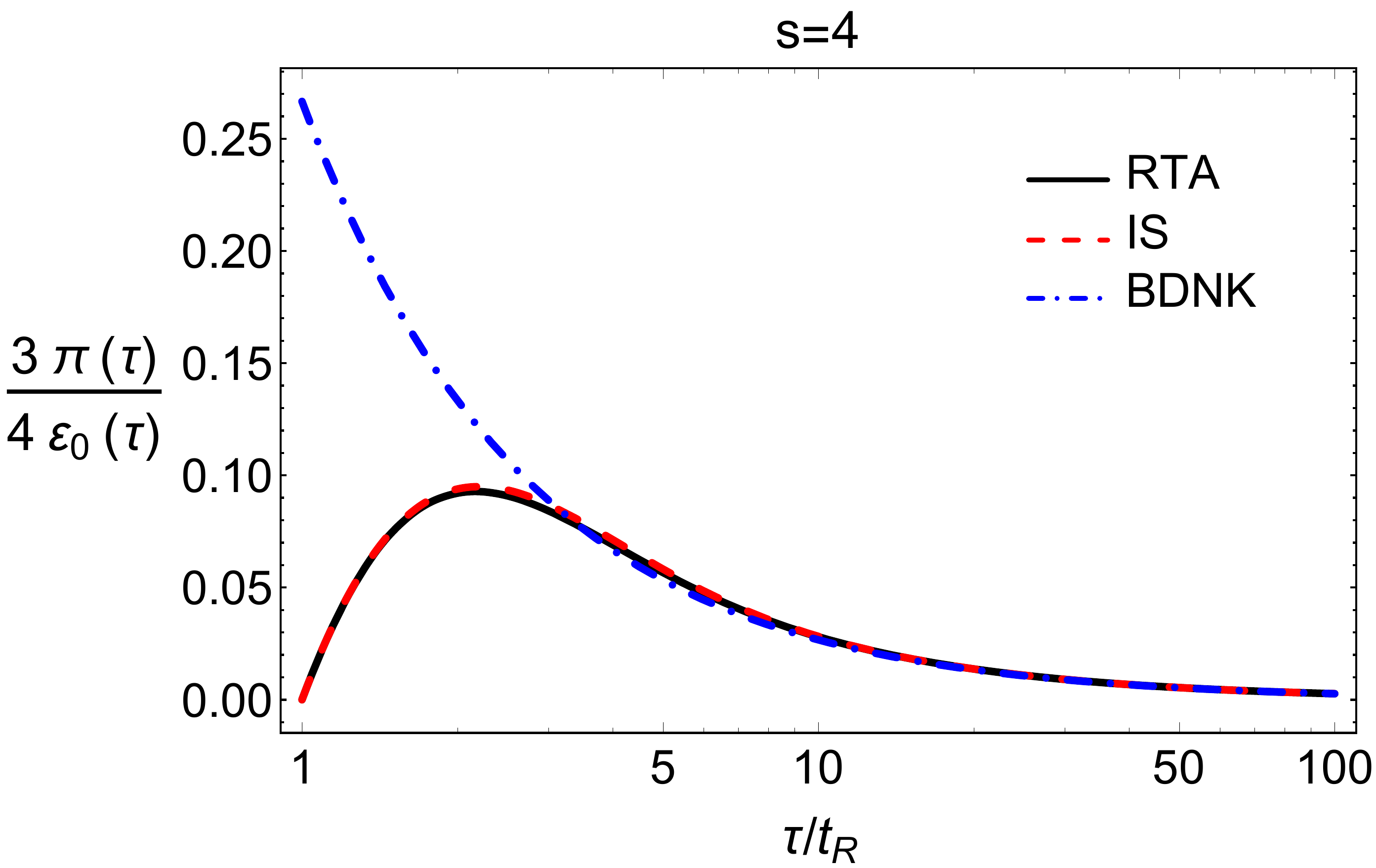}
  \label{fig:evol-pi-s=4-EQL}
\end{subfigure}\hfil
\caption{(Color online) Evolution of the non-equilibrium fraction of the particle density (left) and the normalized shear-stress tensor, $\pi/(\varepsilon_{0} + P_{0}) = 3\pi/(4 \varepsilon_{0})$ (right), found by solving the RTA  Boltzmann equation, Israel-Stewart (IS), and BDNK  for type II exotic Eckart with $s=4$ and equilibrium initial conditions.}
\label{fig:evol-dn-pi-s=4-EQL}
\end{figure}

Now we investigate the solutions of fluid-dynamical theories for the matching conditions where $\delta n \neq 0$ but $\delta \varepsilon \equiv 0$ ($q=2$, $s \neq 1$). In Figs.~\ref{fig:evol-dn-pi-s=3-EQL} and \ref{fig:evol-dn-pi-s=4-EQL}, we depict the evolution of the dissipative currents assuming local equilibrium initial conditions for the corresponding dynamical variables of each framework. As it happened in the last section, for late times ($\hat{\tau} \gtrsim 3$) the shear-stress tensor evolution coincides for all three formalisms. However, for the evolution of the non-equilibrium component of the particle number density, $\delta n$, we see that the three formulations display rather different solutions.    

In the present case, solutions of both Israel-Stewart and BDNK theories are qualitatively similar, displaying negative values for $\delta n$. In Israel-Stewart theory, this happens because of the $\delta n$-$\pi$ coupling term in the second equation displayed in Eq.~\eqref{eq:bjorken-eckart-matching-ii}. In contrast to the previous matching condition, the evolution of $\delta n$ in the BDNK formalism also yields negative values due to the dominance of the $n_{0}/\tau$ term in Eq.~\eqref{eq:EoM-BDN-n}. Finally, we note that solutions of the Boltzmann equation for $\delta n$ differ significantly and always display positive values for this quantity.

We now consider solutions of Israel-Stewart, BDNK, and the RTA Boltzmann equation for several initial values of $\delta n$. The remaining dynamical variables of each theory are still set to their respective local equilibrium values. As in the previous subsection, the goal is to gain some intuition on the attractor dynamics that each formalism may display. The results are shown in Fig.~\ref{fig:atr-dn-pi-s=3,4-BDN}, where we considered simulations with $\delta n(\Hat{\tau}_{0})/[(n + \delta n)(\Hat{\tau}_{0})] =0$, $0.2$, and $0.4$. We see that the solutions of each theory display universal behavior at late times ($\hat{\tau} \gtrsim 10$), indicating the existence of late-time attractor solutions. Again, we note that such attractor solutions were already explicitly derived for BDNK theory in Sec.~\ref{sec:BDN-EoM}. For these matching conditions, one sees that the late time solutions of Israel-Stewart theory and BDNK theory for $\delta n$ are qualitatively similar, even though the quantitative agreement between these solutions is not very good. Both fluid-dynamical frameworks appear to be unable to describe the solutions for $\delta n$ found in  the  Boltzmann equation in the relaxation time approximation.

\begin{figure}[!h]
    \centering
\begin{subfigure}{0.5\textwidth}
  \includegraphics[width=\linewidth]{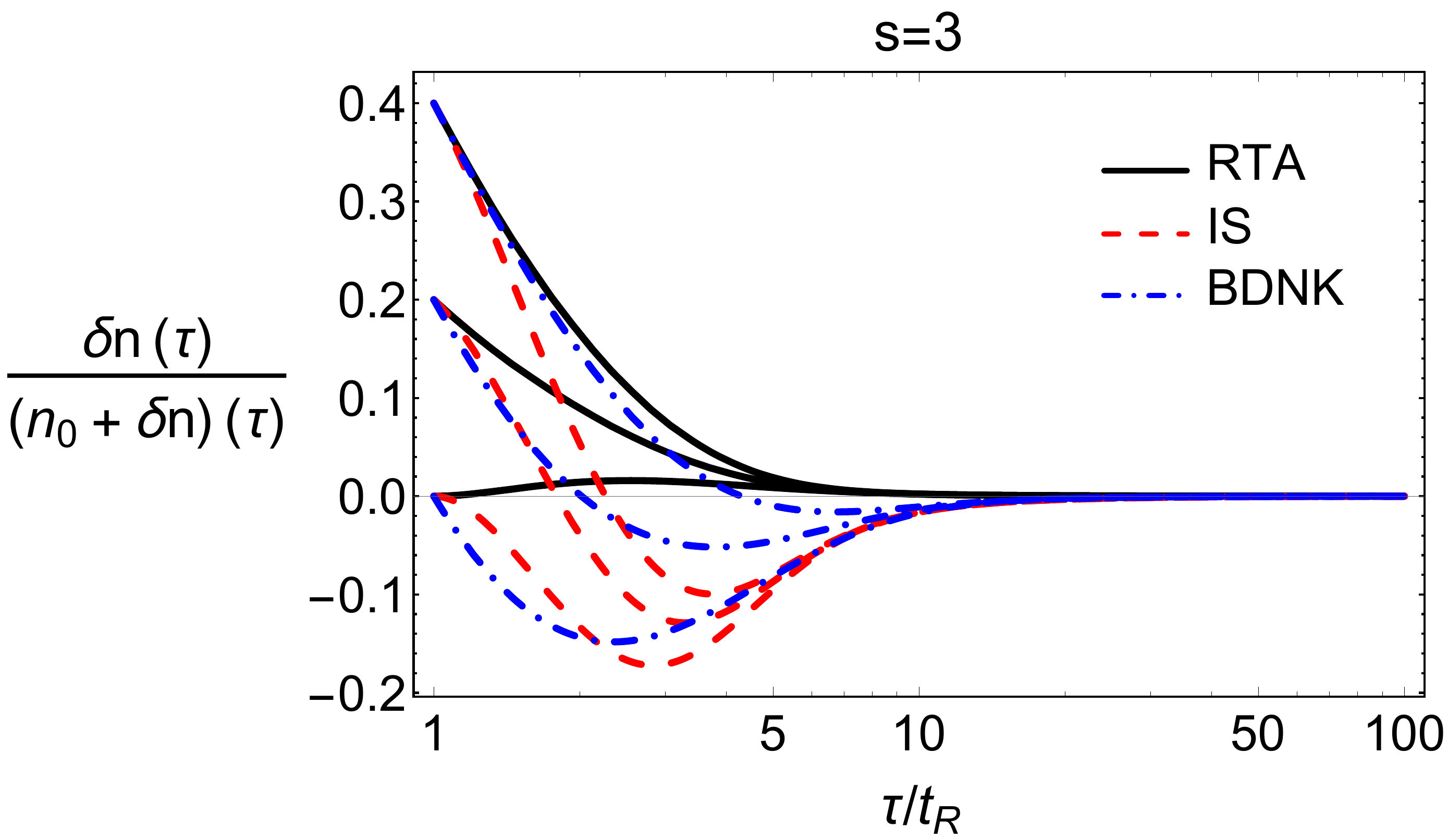}
  \label{fig:atr-dn-pi-s=3-BDN}
\end{subfigure}\hfil
\begin{subfigure}{0.5\textwidth}
  \includegraphics[width=\linewidth]{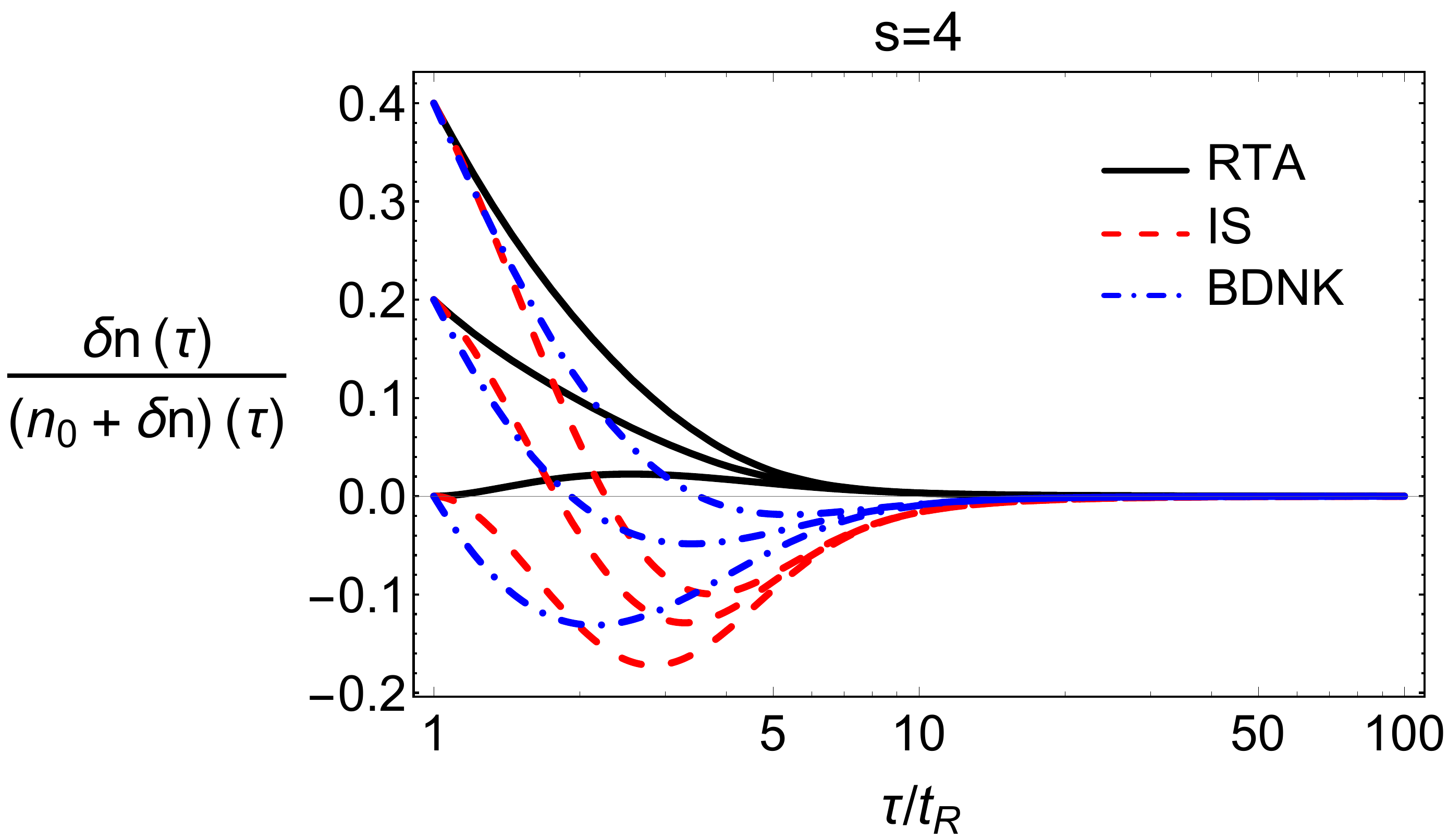}
  \label{fig:atr-dn-pi-s=4-BDN}
\end{subfigure}\hfil
\caption{(Color online) Comparison between time evolution of the non-equilibrium correction to the particle density found by solving the RTA Boltzmann equation,  Israel-Stewart (IS), and BDNK for type II exotic Eckart with $s=3$ (left) and $s=4$ (right) and various initial conditions. The initial conditions are such that $\delta n(\Hat{\tau}_{0})/[(n + \delta n)(\Hat{\tau}_{0})] =0.0, 0.2, 0.4$ and $\pi(\hat{\tau}_{0}) = 0$, when applicable.}
\label{fig:atr-dn-pi-s=3,4-BDN}
\end{figure}


\newpage

\section{Final remarks and discussion}
\label{sec:concl}

In this paper we discussed how relativistic fluid-dynamical theories can be derived from the Boltzmann equation by imposing different perturbative schemes. First, we showed how the traditional Chapman-Enskog and Hilbert expansions are used to obtain macroscopic solutions of the Boltzmann equation for arbitrary matching conditions. We then introduced a novel perturbative scheme to microscopically derive the BDNK equations of fluid dynamics from the Boltzmann equation, also considering arbitrary matching conditions. The main difference between our approach and the traditional Chapman-Enskog expansion is to construct a perturbative scheme using \emph{moments} of the Boltzmann equation (for a given basis) instead of the Boltzmann equation itself. With this prescription, the compatibility conditions for the inversion of the linearized collision operator are avoided and one is not required to replace time-like derivatives of the fluid-dynamical variables by space-like ones -- a feature of the traditional gradient expansion that is not present and imposed in the BDNK equations. 

We obtained microscopic expressions for all the transport coefficients of relativistic Navier-Stokes theory, Hilbert theory, and the BDNK equations. As far as the authors know, this is the first time such full  expressions are obtained for arbitrary matching conditions. We remark that the transport coefficients were calculated assuming a system composed of classical particles that only interact through binary collisions. However, we note that our approach can be generalized to compute the transport coefficients in more realistic systems of phenomenological interest, such as QCD effective kinetic theory \cite{Arnold:2002zm}. We then explicitly calculated these transport coefficients imposing the relaxation time approximation for the linearized collision term and assuming that the particles are massless. This was done using the transport coefficients calculated within the relaxation time approximation, considering relaxation times that are energy independent. In this setting, we were able to obtain analytical solutions of Hilbert theory and the BDNK equations. We further argued that there are no attractor solutions of Hilbert theory for the normalized nonequilibrium energy density fraction. On the hand, we showed that the BDNK equations display attractor solutions for this quantity.

We then investigated and compared the solutions of these fluid-dynamical frameworks with exact solutions of the Boltzmann equation for a gas of massless classical particles undergoing Bjorken flow. One of our goals was to compare solutions obtained with different matching conditions and understand the effect of the latter on such rapidly expanding systems. Investigating the solutions of BDNK theory, we found that the attractor structure is largely affected by the matching conditions. Indeed, for matching conditions such that $\delta n \equiv 0$ and $\delta \varepsilon \neq 0$, we found that both the late-time (hydrodynamic) and early-time attractors can be analytically obtained. This class of matching conditions is physically motivated by causality \cite{bemfica2019nonlinear}, and we have seen that the evolution does not depend on the parameter $s$, and neither do the attractors, which is a surprising feature. As for matching conditions such that $\delta n \neq 0$ and $\delta \varepsilon \equiv 0$, there is only a hydrodynamic attractor and it depends on the matching parameter $s$. 

For the sake of completeness, we compare the evolution and the attractor structure obtained by solving the BDNK equations of motion with those found by solving Israel-Stewart theory under general matching conditions and also the full moment equations of the RTA Boltzmann equation. The dynamics of the latter two equations of motion is also affected by matching conditions, with the 19-moments truncation explicitly depending on the parameter $s$ and the slight change of variables of the moment equations in the RTA. At sufficiently late times, we found that the shear tensor (which in Bjorken flow is reduced to only one independent function $\pi(\tau)$) evolves similarly in BDNK, Israel-Stewart, and the RTA moment equations. In contrast, the time evolution profiles displayed by $\delta \varepsilon$ and $\delta n$ are very different among the three formulations.

The attractors are also quite different in the three approaches. This is in contrast to other results obtained using Landau matching conditions \cite{Heller:2015dha,Romatschke:2017vte,Blaizot:2021cdv,Soloviev:2021lhs,Chattopadhyay:2021ive,Almaalol:2020rnu}, where the attractors of the hydrodynamic theory match the one from kinetic theory. This suggests that the truncation method employed in the alternative matching conditions for the IS formalism has to be improved beyond the moments approximation implemented in Ref.~\cite{rocha21-transient}. Moreover, the attractor mismatch may evidence that other implementations of the alternative matching conditions may be more adequate to perform the analysis.  

In future works, it would be relevant to analyze the effects from assuming momentum-dependent relaxation times into account. Furthermore, an obvious next step is the calculation of BDNK transport coefficients for massive particles and also other type of interactions.

\section*{Acknowledgements}

G.S.R. is financially funded by Conselho Nacional de Desenvolvimento Científico e Tecnológico (CNPq), process No. 142548/2019-7. G.S.D. also acknowledges CNPq as well as Fundação Carlos Chagas Filho de Amparo à Pesquisa do Estado do Rio de Janeiro (FAPERJ), process No. E-26/202.747/2018. J.N. is partially supported by the U.S.
Department of Energy, Office of Science, Office for Nuclear Physics under Award No. DE-SC0021301.

\appendix

\section{Moments method and Israel-Stewart theory under general matching conditions }
\label{sec:MIS-gen-match}

In this Appendix we summarize the truncation procedure to obtain Israel-Stewart \cite{israel1979annals,israel1979jm} equations of motion under general matching conditions used in Sec.~\ref{sec:MIS-EoM} and proposed in  Ref.~\cite{rocha21-transient}. In the Israel-Stewart formalism, non-equilibrium dissipative currents are considered as fields with independent dynamics. Thus, for general matching conditions, none of the hydrodynamic fields $(n_{0}, \delta n, \varepsilon_{0}, \delta \varepsilon, \Pi , u^{\mu}, \nu^{\mu}, h^{\mu}, \pi^{\mu \nu})$ are zero and 19 equations of motion are required to have a closed system in terms of these variables. Five equations are given by the conservation laws \eqref{eq:basic-hydro-EoM}. The remaining equations of motion are obtained by truncating the exact equations for the irreducible moments \eqref{eq:irreducible_moments} of the Boltzmann equation, which can be seen in Refs.~\cite{denicol2012derivation,rocha21-transient}.

To truncate this infinite system of partial differential equations, in Ref.~\cite{rocha21-transient} we followed a generalization of the ideas put forward by Grad \cite{grad:1949kinetic,struchtrup2005macroscopic} in the non-relativistic case, and later on by Israel and Stewart \cite{israel1979annals} under general matching conditions. Then, we consider the expansion of the deviation function $\phi_{\bf p} = (f_{\bf p} - f_{0\mathbf{p}})/f_{0\mathbf{p}}$ in an irreducible and orthogonal basis.
This expansion is truncated by consistently constraining the deviation function with the definitions of the hydrodynamic fields \eqref{eq:def_kinetic} and the matching conditions \eqref{eq:matching_kinetic1} and \eqref{eq:matching_kinetic2}. Effectively, this implies that all non-hydrodynamic moments $\rho_{r}^{\mu_{1} \cdots \mu_{\ell}}$ vanish for $\ell \geq 3$ and that the moments coefficients $\rho_{r}$, $\rho_{r}^{\mu}$ and $\rho_{r}^{\mu \nu}$ are linear combinations of the hydrodynamic fields $(\delta n, \delta \varepsilon, \Pi, \nu^{\mu}, h^{\mu}, \pi^{\mu \nu})$. This leads to a closed system of coupled relaxation equations for the dissipative currents that depend on the parameters $q$, $s$, and $z$. These equations greatly simplify in the massless limit and in the case of Bjorken flow.

\subsection*{Bjorken flow}

For Bjorken flow, in the massless limit, and using exotic Eckart matching conditions such that $\delta n \equiv 0$ and $\rho_{s} \equiv 0$, the equations of motion, which include the conservation laws and the relaxation equations, can be cast in the form
\begin{equation}
\label{eq:bjorken-eckart-matching-i-apn}
\begin{aligned}
& 
\left(\begin{array}{ccc}
\dot{\varepsilon_{0}} \\
\dot{\delta \varepsilon} \\
\dot{\pi} 
\end{array}\right)
+
\left(\begin{array}{cccc}
\frac{4}{3 \tau} &  - \frac{1}{\tau_{\delta \varepsilon}}  & -\frac{1}{ \tau} \left( \frac{\lambda_{\delta \varepsilon \pi}}{\tau_{\delta \varepsilon}} + 1 \right)  \\
0 & \frac{4}{3 \tau} + \frac{1}{\tau_{\delta \varepsilon}}   & \frac{1}{\tau} \frac{\lambda_{\delta \varepsilon \pi}}{\tau_{\delta \varepsilon}}  \\
-\frac{16}{45 \tau} & -\frac{16}{45 \tau} & \frac{38}{21 \tau} + \frac{1}{\tau_{\pi}}
\end{array}\right)
\left(\begin{array}{ccc}
\varepsilon_{0} \\
\delta \varepsilon \\
\pi 
\end{array}\right)
=
\left(\begin{array}{ccc}
0 \\
0 \\
0
\end{array}\right),
\end{aligned}
\end{equation}
where $\tau_{\delta \varepsilon}$ and $\tau_{\pi}$ denote the relaxation times associated with $\delta \varepsilon$ and $\pi^{\mu \nu}$, respectively. For a constant relaxation time, $\tau_{\delta \varepsilon}= \tau_{\pi} = \tau_{R}$. We also have the matching-dependent coupling constant $\lambda_{\delta \varepsilon \pi}$, which is expressed as
\begin{equation}
\begin{aligned}
& 
\frac{\lambda_{\delta \varepsilon \pi}}{\tau_{\delta \varepsilon}} = 
\frac{\Gamma (s+4)}{20 \Gamma (s+2)} - 1.
\end{aligned}    
\end{equation}
whereas for exotic Eckart matching conditions such that $\delta \varepsilon \equiv 0$ and $\rho_{s} \equiv 0$, the equations of motion are 
\begin{equation}
\label{eq:bjorken-eckart-matching-ii-apn}
\begin{aligned}
& 
\left(\begin{array}{ccc}
\dot{n_{0}} \\
\dot{\delta n} \\
\dot{\varepsilon_{0}} \\
\dot{\pi} 
\end{array}\right)
+
\left(\begin{array}{cccc}
\frac{1}{\tau} &  -\frac{1}{\tau_{\delta n}} & 0 & -\frac{\lambda_{\delta n \pi}}{\tau}    \\
 0 & \frac{1}{\tau} + \frac{1}{\tau_{\delta n}} & 0 &  \frac{\lambda_{\delta n \pi}}{\tau}  \\
0 & 0 & \frac{4}{3 \tau} & -\frac{1}{\tau} \\ 
0 & 0 & -\frac{16}{45 \tau}  & \frac{38}{21 \tau} + \frac{1}{\tau_{\pi}}\end{array}\right)
\left(\begin{array}{ccc}
n_{0} \\
\delta n \\
\varepsilon_{0} \\
\pi 
\end{array}\right)
=
\left(\begin{array}{ccc}
0 \\
0 \\
0 \\
0
\end{array}\right),
\end{aligned}
\end{equation}
where we have the relaxation time $\tau_{\delta n}$ related to the dissipative current $\delta n$. For a constant relaxation time $\tau_{\delta n} = \tau_{R}$. We also have the matching-dependent coupling 
\begin{equation}
\begin{aligned}
& 
\frac{\lambda_{\delta n \pi}}{\tau_{\delta n}} = \frac{s-1}{(s-2) \Gamma (s+2)} \left( \frac{\Gamma (s+4)}{60} -\frac{1}{3} \right).
\end{aligned}    
\end{equation}

In this particular background, the dynamics of $n_{0} + \delta n$ and $\varepsilon_{0} + \delta \varepsilon$ are matching invariant. Indeed, summing the two first rows of Eq.~\eqref{eq:bjorken-eckart-matching-i} and \eqref{eq:bjorken-eckart-matching-ii}, the result does not depend on the matching-dependent coefficients $\tau_{\delta n}$, $\tau_{\delta \varepsilon}$, $\lambda_{\delta n \pi}$, and $\lambda_{\delta \varepsilon \pi}$. Values for the couplings can be seen in Table \ref{tab:couplings} for the values of $s$ used in the main text.
\begin{table}[!h]
    \centering
    \begin{tabular}{|c|c|c|c|}
    \hline
      &  $s=3$ & $s=4$\\
    \hline
    $\lambda_{\delta \varepsilon \pi }[\tau_{\delta \varepsilon}]$ & $1/2$ & $11/10$\\
    \hline
    $\lambda _{\delta n \pi}[\tau_{\delta n}]$  & 35/36 &  251/240 \\
    \hline     
    \end{tabular}
    \caption{Couplings in units of $\tau_{\delta \varepsilon}$ and $\tau_{\delta \varepsilon}$ for $s=3,4$ exotic Eckart matching conditions.}
    \label{tab:couplings}
\end{table}

\section{Basis with zero modes}
\label{sec:zero-mode-CE}

In this Appendix, we show the procedure that can be performed in order to compute transport coefficients in the Chapman-Enskog expansion in a basis that contains zero modes. We note that the methods outlined here can be  extended to the modified expansion used to derive BDNK hydrodynamics. To this end, we use the basis
\begin{equation}
\begin{aligned}
& 
P^{(\ell)}_{n} = E_{\bf p}^{n}, \ n = 0, 1, \cdots .
\end{aligned}    
\end{equation}
Once again, since the linearized collision term $\hat{L}$ is a linear operator, the particular solution $\phi_{\bf p}^{\rm{part}}$ must have the general form \eqref{eq:part-sol-phi-CE},
\begin{equation}
\label{eq:part-sol-phi-CE-E^n}
\begin{aligned}
\phi_{\bf p}^{\rm{part}} = \mathcal{S}_{\bf p} \theta +  \mathcal{V}_{\bf p} p^{\langle \mu \rangle} \nabla_{\mu} \alpha 
+
\mathcal{T}_{\bf p} p^{\langle \mu}p^{\nu \rangle} \sigma_{\mu \nu}, 
\end{aligned}    
\end{equation}
The next step is to replace the particular solution Eq.~\eqref{eq:part-sol-phi-CE} into Eq.~\eqref{eq:eq-int-tensors-CE}, and then we obtain again Eq.~\eqref{eq:int-SVT}, which is a coupled integral equation for $\mathcal{S}$, $\mathcal{V}$, and $\mathcal{T}$. We proceed by multiplying this equation by $E_{\bf p}^{r}$, $E_{\bf p}^{r}  p^{\langle \mu \rangle}$ and $E_{\bf p}^{r}p^{\langle \mu}p^{\nu \rangle}$ and integrating in over momentum. Then, after expanding $\mathcal{S}_{\bf p}$, $\mathcal{V}_{\bf p}$, and $\mathcal{T}_{\bf p}$ as polynomials in $E_{\bf p}$ such that 
\begin{equation}
\begin{aligned}
&
\mathcal{S}_{\bf p} =  \sum_{n \geq 0} s_{n} E_{\bf p}^{n}, \ \
\mathcal{V}_{\bf p} =  \sum_{n \geq 0} v_{n} E_{\bf p}^{n}, \ \
\mathcal{T}_{\bf p} = \sum_{n \geq 0} t_{n} E_{\bf p}^{n},
\end{aligned}    
\end{equation}
we have the following systems of equations, which are analogous to Eqs.~\eqref{eq:sys-SVT} 
\begin{subequations}
\label{eq:sys-SVT-E^n}
\begin{align}
&
\label{eq:sys-SVTa-E^n}
\sum_{n} S_{rn}^{'} s_{n} = A_{r}^{'}, \\
&
\label{eq:sys-SVTb-E^n}
\sum_{n} V_{rn}^{'} v_{n} = B_{r}^{'},\\
&
\label{eq:sys-SVTc-E^n}
\sum_{n} T_{rn}^{'} t_{n} = C_{r}^{'},
\end{align}    
\end{subequations}
where 
\begin{equation}
\label{eq:mat-def-SVT}
\begin{aligned}
& 
S_{rn}^{'} \equiv \int dP E_{\bf p}^{r}\hat{L}\left[E_{\bf p}^{n}\right] f_{0\bf{p}}, 
\quad 
A_{r}^{'} = \int dP E_{\bf p}^{r}\left( \mathcal{A}_{\bf p} -  \frac{\beta}{3} \Delta^{\lambda \sigma}p_{\lambda} p_{\sigma} \right)f_{0\bf{p}},
\\
&
V_{rn}^{'} \equiv \int dP  E_{\bf p}^{r} p^{\langle \mu \rangle} \hat{L} \left[E_{\bf p}^{n} p_{\langle \mu \rangle} \right] f_{0\bf{p}}, 
\quad  
B_{r}^{'} = \int dP \left(\Delta^{\mu \nu}p_{\mu} p_{\nu} \right) E_{\bf p}^{r} \left(1 - \frac{n_{0} E_{\mathbf{p}}}{\varepsilon_{0}+ P_{0}} \right) f_{0\bf{p}}, \\\
&
T_{rn}^{'} \equiv \int dP 
E_{\bf p}^{r}
p^{\langle \mu}p^{\nu \rangle}
\hat{L} \left[E_{\bf p}^{n} p_{\langle \mu}p_{\nu \rangle} \right] f_{0\bf{p}}, 
\quad 
C_{r}^{'} = - \beta \int dP \left(\Delta^{\mu \nu}p_{\mu} p_{\nu} \right)^{2} E_{\bf p}^{r}f_{0\bf{p}}. 
\end{aligned}    
\end{equation} 
Equations \eqref{eq:sys-SVTc-E^n} can be schematically inverted as 
\begin{equation}
\begin{aligned}
& t_{n} = \sum_{m} [T^{'-1}]_{nm} C^{'}_{m}.
\end{aligned}    
\end{equation}
As for equations \eqref{eq:sys-SVTa-E^n} and \eqref{eq:sys-SVTb-E^n}, the inversion process is not so simple due to the presence of zero modes. In fact, these systems of equations have the matrix forms   
\begin{equation}
\begin{aligned}
\left(
\begin{array}{cccccc}
    0 &  0 & 0 & 0 & 0 & \cdots \\
    0 & 0 & 0 & 0 & 0 & \cdots \\ 
    0 & 0 & S_{22} & S_{23} & S_{24} & \cdots  \\
    0 & 0 & S_{32} & S_{33} & S_{34} & \cdots  \\
    \vdots & \vdots  & \vdots & \vdots & \vdots  & \vdots 
\end{array}
\right)
\left(\begin{array}{c}
    s_{0}  \\
    s_{1}  \\
    s_{2}  \\
    s_{3}  \\
    \vdots  \\
\end{array} \right)
&=
\left(\begin{array}{c}
    0  \\
    0  \\
    A_{2}  \\
    A_{3}  \\
    \vdots  \\
\end{array} \right)
, \\
\left(
\begin{array}{ccccc}
    0 &  0 & 0 & 0 & \cdots \\
    0 & V_{11} & V_{12} & V_{13} & \cdots \\
    0 & V_{21} & V_{22} & V_{23} & \cdots \\
    \vdots & \vdots  & \vdots & \vdots & \vdots 
\end{array}
\right)
\left(\begin{array}{c}
    v_{0}  \\
    v_{1}  \\
    v_{2}  \\
    \vdots  \\
\end{array} \right)
& =
\left(\begin{array}{c}
    0  \\
    B_{1}  \\
    B_{2}  \\
    \vdots  \\
\end{array} \right),
\end{aligned}    
\end{equation}
where the vanishing of the first (the first and the second) line(s) of the matrix $V$ ($S$) stems from the self-adjoint property \eqref{eq:selfadjoint}. Additionally, $A_{0} = A_{1} = B_{0} = 0$ due to property \eqref{eq:orthg-srcs}. 

Hence, to solve the related linear equations, it is necessary to remove linear sub-spaces corresponding to the zero-modes. Then, denoting the submatrices of $S'$ and $V'$ as $\hat{S}$ and $\hat{V}$, respectively, we have as a result of the schematic inversion 
\begin{equation}
\begin{aligned}
& v_{n} = \sum_{m \geq 1}  [\hat{V}^{-1}]_{nm} B_{m}
, \ \ n = 1, 2, \cdots ,\\
& s_{n} = \sum_{m \geq 2} [\hat{S}^{-1}]_{nm} A_{m}, \ \
n = 2, 3, \cdots .
\end{aligned}    
\end{equation}

The coefficients related to the zero modes, $s_{0}$, $s_{1}$, and $v_{0}$ cannot be obtained. Nevertheless, this is not a problem since they can be incorporated into the homogeneous solution by a  redefinition of the $a$ and $b^{\mu}$ coefficients in $\phi^{\rm{hom}}_{\bf p}$. They are then obtained from the matching conditions \eqref{eq:matching_kinetic1} and \eqref{eq:matching_kinetic2}, which when substituted in Eq.~\eqref{eq:phi-hom-phi-part} lead to the conditions
\begin{equation}
\begin{aligned}
& I_{q,0} a + I_{q+1,0} b_{\mu} u^{\mu} 
=
- \left\langle E_{\bf p}^{q} \mathcal{S}_{\bf p} \right\rangle \theta ,
\\
& I_{s,0} a + I_{s+1,0} b_{\mu} u^{\mu} 
=
- \left\langle E_{\bf p}^{s} \mathcal{S}_{\bf p} \right\rangle \theta ,
\\
& I_{z+2,1} b_{\langle \mu \rangle} 
=
\frac{1}{3} \left\langle \left(\Delta^{\mu \nu}p_{\mu} p_{\nu} \right) E_{\bf p}^{z} \mathcal{V}_{\bf p} \right\rangle_{0} \nabla_{\mu} \alpha,
\end{aligned}    
\end{equation}
where it was used that $p^{\mu} = E_{\bf p} u^{\mu} + p^{\langle \mu \rangle}$. These are solved with 

\begin{equation}
\label{eq:a,b-mu-E^n}
\begin{aligned}
&
a = \frac{I_{q+1,0} \langle E_{\bf p}^{s} \mathcal{S}_{\bf p} \rangle_{0} - \langle E_{\bf p}^{q} \mathcal{S}_{\bf p} \rangle_{0} I_{s+1,0}}{G_{s+1,q}}
\theta,
\\
&
b^{\mu} u_{\mu} = \frac{\langle E_{\bf p}^{q} \mathcal{S}_{\bf p}\rangle_{0} I_{s,0} - I_{q,0} \langle E_{\bf p}^{s} \mathcal{S}_{\bf p} \rangle_{0}}{G_{s+1,q}} \theta, \\
&
b_{\langle \mu \rangle} 
=
\frac{1}{3}\frac{\left\langle \left(\Delta^{\mu \nu}p_{\mu} p_{\nu} \right) E_{\bf p}^{z} \mathcal{V}_{\bf p} \right\rangle_{0}}{I_{z+2,1}} \nabla_{\mu} \alpha. 
\end{aligned}    
\end{equation}
Finally, we have as the solution for the first order Chapman-Enskog deviation function
\begin{equation}
\begin{aligned}
&
\phi_{\bf p} = 
\tilde{\mathcal{S}}_{\bf p} 
\theta 
+
\tilde{\mathcal{V}}_{\bf p} 
p^{\langle \mu \rangle} \nabla_{\mu} \alpha 
+
\mathcal{T}_{\bf p} p^{\langle \mu}p^{\nu \rangle} \sigma_{\mu \nu}. 
\end{aligned}    
\end{equation}
with 
\begin{equation}
\label{eq:til-SVT-sol-E^n}
\begin{aligned}
& 
\tilde{\mathcal{S}_{\bf p}} = \sum_{n \geq 2} \sum_{m \geq 2} [\hat{S}^{-1}]_{nm} A_{m} \left( E_{\bf p}^{n}
+
\frac{I_{q+1,0} I_{s+n,0} - I_{q+n,0} I_{s,0}}{G_{s+1,q}}
+
\frac{I_{q+n,0} I_{s,0} - I_{q,0} I_{s+n,0} }{G_{s+1,q}} E_{\bf p} 
\right),
\\
&
\tilde{\mathcal{V}}_{\bf p} 
=
\sum_{n \geq 1} \sum_{m \geq 1}  [\hat{V}^{-1}]_{nm} B_{m} \left( E_{\bf p}^{n} 
- 
\frac{I_{z+n+2,1}}{I_{z+2,1}} \right), \\
&
\mathcal{T}_{\bf p} =
\sum_{n \geq 0} \sum_{m \geq 0} [T^{-1}]_{nm} C_{m} E_{\bf p}^{n}.
\end{aligned}    
\end{equation}
From this solution,  constitutive relations for the non-equilibrium corrections can be obtained. Indeed, definitions \eqref{eq:def_kinetic} yield  
\begin{equation}
\label{eq:def-coeffs-E^n}
\begin{aligned}
\Pi & = - \zeta \theta,  \
\delta n =  - \xi \theta, \
 \delta \varepsilon = \chi \theta,\\
\nu^{\mu} & = \kappa \nabla^{\mu} \alpha, \
h^{\mu} = - \lambda \nabla^{\mu} \alpha, \\
\pi^{\mu \nu} & = 2 \eta \sigma^{\mu \nu},
\end{aligned}
\end{equation}
with transport coefficients given by 
\begin{equation}
\label{eq:coeffs-chap-ensk-E^n}
\begin{aligned}
&
\zeta = \sum_{n \geq 2} \sum_{m \geq 2}[\hat{S}^{-1}]_{nm} A_{m}
H_{n}^{(\zeta)}, 
\ \
\xi = \sum_{n \geq 2} 
 \sum_{m \geq 2} [\hat{S}^{-1}]_{nm} A_{m} H_{n}^{(\xi)},
\ \
\chi = - \sum_{n \geq 2} 
 \sum_{m \geq 2} [\hat{S}^{-1}]_{nm} A_{m}
H_{n}^{(\chi)}, \\
&
\kappa = \sum_{n \geq 1} \sum_{m \geq 1}  [\hat{V}^{-1}]_{nm} B_{m} J_{n}^{(\kappa)}, 
\ \
\lambda = \sum_{n \geq 1} \sum_{m \geq 1}  [\hat{V}^{-1}]_{nm} B_{m} J_{n}^{(\lambda)},\\
&
\eta =  \sum_{n \geq 0} \sum_{m \geq 0} [T^{-1}]_{nm} C_{m} I_{n+4,2} ,
\end{aligned}    
\end{equation}
where
\begin{equation}
\label{eq:H,J-expn-E^n}
\begin{aligned}
& H_{n}^{(\zeta)}  =  I_{n+2,1}
-
I_{q+n,0}
\frac{I_{2,1}
 I_{s+1,0}-
I_{s,0} I_{3,1} }{G_{s+1,q}}
+
I_{s+n,0}
\frac{I_{2,1} I_{q+1,0}
-
I_{q,0} I_{3,1}}{G_{s+1,q}} ,  \\
&
H_{n}^{(\xi)}  = I_{n+1,0}
-
I_{q+n,0}
\frac{G_{s+1,1}}{G_{s+1,q}}
+
I_{s+n,0}
\frac{G_{q+1,1}}{G_{s+1,q}},  
\\
&
H_{n}^{(\chi)} = - I_{n+2,0}
+
I_{q+n,0}
\frac{G_{s+1,2}}{G_{s+1,q}}
-
I_{s+n,0}\frac{G_{q+1,2}}{G_{s+1,q}} ,\\
&
J_{n}^{(\kappa)}  =  -I_{n+2,1}
+
\frac{I_{2,1}}{I_{z+2,1}} 
I_{z+n+2,1},
\\
&
J_{n}^{(\lambda)}  =  I_{n+3,1}
-
\frac{I_{3,1}}{I_{z+2,1}}  I_{z+n+2,1},
\end{aligned}   
\end{equation}
where the conclusions of the end of Section \ref{sec:Chap-Ensk-1} are also valid here. The inversion procedure described here also has consequences for the perturbative procedure outlined in Section \ref{sec:Chap-Ensk-2}. In that case, the zero modes \emph{must} be explicitly excluded from the expansion of the functions $\mathcal{S}^{(\alpha,\beta,\theta)}$, $\mathcal{V}^{(\alpha,\beta)}$, and $\mathcal{T}$.

\section{Choice of basis for the computation of transport coefficients}
\label{sec:choice-basis}

In this Appendix we show the reason for the specific choice of the parameters $m_{\ell}$ and $n_{\ell}$ which define the bases used to compute the transport coefficients in Tables \ref{tab:coeffs1gn} and \ref{tab:coeffs2gn}. To this end, we digress to the exact derivation of transport coefficients using the basis constructed using powers of energy. In the particular case of  RTA, Eq.~\eqref{eq:nRTA}, we employ the basis so that $P_{n}^{(0)} = E_{\bf p}^{n+2}$, $P_{0}^{(1)} = E_{\bf p}^{-1}$, $P_{1}^{(1)} = E_{\bf p}$, $P_{2}^{(1)} = E_{\bf p}^{-1}$, $\ldots$, and $P_{n}^{(2)} = E_{\bf p}^{n-1}$ so we have  
\begin{subequations}
\label{eq:sys-RTA-l's}
\begin{align}
\label{eq:sys-RTA-l=0}
&  D\alpha I_{r+1,0} - D\beta I_{r+2,0} + \beta I_{r+2,1} \theta
=
- \frac{1}{\tau_{R} } \left\{ \rho_{r+1} 
+
\frac{\Gamma(r+3)}{\beta^{r}\Gamma(3)} \left[
(r-1)\delta n
-
\beta \frac{r}{3} \delta \varepsilon \right] 
\right\},
 \ \ \ r = 2, 3, 4, \cdots,
\\
& \label{eq:sys-RTA-l=1}
-\nabla^{\mu} \alpha I_{r+2,1} 
 +  
 \left( \nabla^{\mu}\beta + \beta D u^{\mu} \right) I_{r+3,1}
= - \frac{1}{ \tau_{R}}\left[ \rho^{\alpha}_{r+1}
-
\frac{1}{\beta^{r}} \frac{\Gamma(r+5)}{\Gamma(5)}h^{\alpha}
\right], \ \ \ r = 1, 2, 3, \cdots ,
\\
&
\label{eq:sys-RTA-l=2}
-2 \beta I_{r+4,2} \sigma^{\alpha \beta}
=
- \frac{1}{\tau_{R} } \rho^{\alpha \beta}_{r+1}, \ \ r = 0, 1, 2, \cdots, 
\end{align}
\end{subequations}
where we defined the irreducible moments 
\begin{equation}
\label{eq:irreducible_moments}
\begin{aligned}
\rho^{\mu_{1} \cdots \mu_{\ell}}_{r} = \int dP  E_{\mathbf{p}}^{r} p^{\langle \mu_{1}} \cdots p^{\mu_{\ell} \rangle} \phi_{\mathbf{p}} f_{0\bf p}.
\end{aligned}    
\end{equation}
The hydrodynamic fields correspond to particular instances of these general moments such that $\delta n = \rho_{1}$, $\delta \varepsilon = \rho_{2}$, $\nu^{\mu} = \rho_{0}^{\mu}$, $h^{\mu} = \rho_{1}^{\mu}$, and $ \pi^{\mu \nu} = \rho_{0}^{\mu \nu}$. The particular case of a constant relaxation time allows us to express the collision integrals solely in terms of a few integer moments. Thus, to obtain the constitutive relations, one can simply use the appropriate value of $r$ in Eqs.~\eqref{eq:sys-RTA-l's} and use the information contained in the matching conditions. In terms of the irreducible moments \eqref{eq:irreducible_moments}, the matching conditions of the type \eqref{eq:matching_kinetic2} are 
\begin{subequations}
\begin{align}
&
\rho_{q} = \rho_{s} = 0, \ \ \ q \neq s, \\
&
\rho_{z}^{\mu} = 0.
\end{align}    
\end{subequations}

We start with the coefficients related to scalar hydrodynamic fields. In this case we use \textit{type I exotic Eckart}, $q=1$, $s \neq 2$, to obtain the transport coefficients, we take $r=s-1$ in Eq.~\eqref{eq:sys-RTA-l=0}, then we have    
\begin{equation}
\label{eq:const-rels-eps}
\begin{aligned}
& \delta \varepsilon 
= \chi^{(\alpha)} D\alpha - \chi^{(\beta)} \left( \frac{D\beta }{\beta} - \frac{1}{3} \theta \right),
\end{aligned}    
\end{equation}
with
\begin{equation}
\label{eq:const-rels-eps-coefs}
\begin{aligned}
&  \chi^{(\alpha)} = \frac{\tau_{R}}{(s-1)} \varepsilon_{0}, \\
&  \chi^{(\beta)} =  \frac{(s+2) \tau_{R}}{(s-1)} \varepsilon_{0},
\end{aligned}    
\end{equation}
which is consistent with the results of Table \ref{tab:coeffs1gn}. Otherwise, if we use \textit{type II exotic Eckart matching conditions}, we have $q=2$, $s \neq 1$. Again, we take $r=s-1$ in Eq.~\eqref{eq:sys-RTA-l=0} to obtain  
\begin{equation}
\label{eq:const-rels-del-n}
\begin{aligned}
& \delta n 
= \xi^{(\alpha)} D\alpha - \xi^{(\beta)} \left( \frac{D\beta }{\beta} - \frac{1}{3} \theta \right),
\end{aligned}    
\end{equation}
with
\begin{equation}
\label{eq:const-rels-del-n-coefs}
\begin{aligned}
&  \xi^{(\alpha)} = -  \tau_{R}   
\frac{n_{0}}{(s-2)},
 \\
&  \xi^{(\beta)} =  -  \tau_{R}   
  \frac{n_{0}(s+2)}{(s-2)},
\end{aligned}    
\end{equation}
which is consistent with the results of Table \ref{tab:coeffs2gn}. Now it is possible to explain the behavior seen in Tables \ref{tab:coeffs1gn} and  \ref{tab:coeffs2gn}: we obtain the exact values from the matrix inversion procedure because the power which was used in Eq.~\eqref{eq:sys-RTA-l=0} to obtain the transport coefficients, $E_{\bf p}^{s-1}$, can be exactly expanded as the finite sum of basis elements. For instance, for $s=3$ we have chosen the basis \eqref{eq:basis-x/(1+x)} with $n_{0} = -1$ $m_{0} = 1$, and in this case 
\begin{equation}
\begin{aligned}
&x^{2} = 1 \frac{x^{2}}{1+x} + 1 \frac{x^{3}}{1+x} + 0 \frac{x^{4}}{1+x}, \\
& x^{2} = 1 \frac{x^{2}}{(1+x)^{3}} + 3 \frac{x^{3}}{(1+x)^{3}} + 3 \frac{x^{4}}{(1+x)^{3}} + 1 \frac{x^{5}}{(1+x)^{3}},
\end{aligned}    
\end{equation}
for the second and fourth order truncation orders, respectively.

Similarly, for the vector dissipative currents constitutive relations, the transport coefficients depend explicitly on the parameter $z$ used to define the velocity 4-vector $u^{\mu}$. Indeed, if we use 
\textit{Eckart matching conditions}, $\nu^{\mu} = 0$ ($z=0$), then, we choose $r=-1$ in Eq.~\eqref{eq:sys-RTA-l=1} and we readily obtain the constitutive relation for the energy diffusion vector
\begin{equation}
\begin{aligned}
h^{\mu} 
= \lambda^{(\alpha)} \nabla^{\mu} \alpha - \lambda^{(\beta)} \left( \frac{1}{\beta} \nabla^{\mu} \beta +   D u^{\mu}\right)
\end{aligned}    
\end{equation}
with
\begin{equation}
\label{eq:lambdas-E^n}
\begin{aligned}
&
\lambda^{(\alpha)} = \frac{4}{9} \tau_{R}  \varepsilon_{0}, \quad
\lambda^{(\beta)} = \frac{4}{3} \tau_{R} \varepsilon_{0}. 
\end{aligned}    
\end{equation}
In the case where \textit{Landau matching conditions} are used, $h^{\mu} = 0$ ($z=1$), then we choose again $r=-1$ in Eq.~\eqref{eq:sys-RTA-l=1} and we readily obtain the constitutive relation for the particle diffusion vector
\begin{equation}
\label{eq:kappas-E^n}
\begin{aligned}
\nu^{\mu} 
= \kappa^{(\alpha)} \nabla^{\mu} \alpha - \kappa
^{(\beta)} \left( \frac{1}{\beta} \nabla^{\mu} \beta +   D u^{\mu}\right),
\end{aligned}    
\end{equation}
where
\begin{equation}
\begin{aligned}
&
\kappa^{(\alpha)} = \frac{n_{0}}{3}\tau_{R},  
\quad
\kappa^{(\beta)} = n_{0} \tau_{R}. 
\end{aligned}    
\end{equation}
Once again, the power of $E_{\bf p}$ which was used to obtain the transport coefficients from  Eq.~\eqref{eq:sys-RTA-l=1}, $E^{-1}_{\bf p}$ can be exactly expanded as the finite sum of basis elements. Indeed, 
\begin{equation}
\begin{aligned}
&\frac{1}{x} = 1 \frac{1}{x(1+x)} + 1 \frac{1}{1+x} + 0 \frac{x}{1+x}, \\
& \frac{1}{x} = 1 \frac{1}{x(1+x)^{3}} + 3 \frac{1}{x(1+x)^{3}} + 3 \frac{x}{x(1+x)^{3}} + 1 \frac{x^{2}}{x(1+x)^{3}},
\end{aligned}    
\end{equation}
for the second and fourth order truncation orders, respectively.

The constitutive relations for the rank-two tensors can be readily obtained, as they do not depend on matching conditions and the RTA collision term is diagonal. In this case, to obtain the constitutive relation for the shear-stress tensor, we integrate Eq.~\eqref{eq:nRTA} with $E_{\mathbf{p}}^{-1} p^{\langle \mu} p^{\nu \rangle}$. Then, we have the familiar constitutive relation
\begin{equation}
\label{eq:BDN-shear}
\begin{aligned}
& \pi^{\mu \nu} = 2 \eta \sigma^{\mu \nu},
\end{aligned}    
\end{equation}
with
\begin{equation}
\label{eq:shear-RTA}
\begin{aligned}
& \eta = \beta \tau_{R} I_{2,2},
\end{aligned}    
\end{equation}
which coincides with the expression for the shear viscosity within the Chapman-Enskog expansion in RTA \cite{rocha:21} and  is also consistent with the results of Table \ref{tab:coeffs1gn}.

\section{Boltzmann's moment equations of motion in Bjorken flow}
\label{sec:BEq-EoM}

In this Appendix we discuss the set of moment equations of the Boltzmann equation within the relaxation time approximation \eqref{eq:nRTA} in Bjorken flow  \cite{Denicol:2021,Denicol:2019lio}. In curved spacetime, the on-shell Boltzmann equation is expressed as \cite{Denicol:2014xca,Denicol:2014tha,Bazow:2015dha} 
\begin{equation}
\begin{aligned}
& p^{\mu} \partial_{\mu}f_{\bf p} 
+
\Gamma^{\alpha}_{\mu i} p_{\alpha} p^{\mu} \frac{\partial f_{\bf p}}{\partial p_{i}} 
=
C[f_{\bf p}].
\end{aligned}    
\end{equation}
In Bjorken flow, the only non-vanishing components of the Christoffel symbols are $\Gamma^{\tau}_{\eta \eta} = \tau$ $\Gamma^{\eta}_{\tau \eta} = 1/\tau$, hence, 
\begin{equation}
\begin{aligned}
& p^{\tau} \partial_{\tau}f_{\bf p} 
=
C[f_{\bf p}].
\end{aligned}    
\end{equation}

The Boltzmann equation can be re-expressed in terms of an infinite set of coupled differential equations for the irreducible moments of $f_{\bf p}$. The underlying symmetries of Bjorken background imply that $f_{\bf p} = f(\tau, p_{\eta}, p^{\tau})$ and also that it is possible to expand $f$ in terms of Legendre polynomials and powers of $p^{\tau}$. This motivates the use of the moments
\begin{equation}
\label{eq:BE-moments-Bjorken}
\begin{aligned}
& \rho_{n,m} = \int dP \left(p^{\tau}\right)^{n+1} P_{2m}(\cos \Theta) f_{\textbf{p}},
\end{aligned}    
\end{equation}
to describe the dynamics. In the equation above, $P_{2m}(\cos \Theta)$ denotes the Legendre polynomial \cite{gradshteyn2014table} in the variable $\cos \Theta \equiv p_{\eta}/(p^{\tau} \tau)$. We further notice that parity symmetry implies that moments constructed from $P_{n}(\cos \Theta)$ with odd $n$ are zero. If the distribution is that of particles in local equilibrium \eqref{eq:f_0}, then the moments reduce to
\begin{equation}
\label{eq:EQL-BE-moments}
\begin{aligned}
& \rho_{n,m}^{(0)} = \int dP \left(p^{\tau}\right)^{n+1} P_{2m}(\cos \Theta) f_{0\bf{p}} = e^{\alpha} \frac{\Gamma(n+3)}{2 \pi^{2}\beta^{n+3}} \delta_{m,0}.
\end{aligned}    
\end{equation}
The relevant hydrodynamic variables can be expressed in terms of the moments in Eqs.~\eqref{eq:BE-moments-Bjorken} and \eqref{eq:EQL-BE-moments} 
\begin{equation}
\begin{aligned}
&  n_{0} = \rho_{0,0}^{(0)} \quad \delta n = \rho_{0,0} - \rho_{0,0}^{(0)}, \\
& \varepsilon_{0} = \rho_{1,0}^{(0)} \quad \delta \varepsilon = \rho_{1,0} - \rho_{1,0}^{(0)}, \\
&
\pi = -\frac{2}{3} \rho_{1,1} ,
\end{aligned}    
\end{equation}
where the latter identification can be performed from the fact that $\pi \equiv \pi^{\eta}_{\ \eta}$ can be written as
\begin{equation}
\begin{aligned}
& \pi = \int dP \left( p^{\tau} \right)^{2} \left[ \left( \frac{p_{\eta}}{\tau p^{\tau}} \right)^{2} - \frac{1}{3} \right] f_{\bf p}.
\end{aligned}    
\end{equation}
The matching conditions \eqref{eq:matching_kinetic1} and \eqref{eq:matching_kinetic2} are expressed as \cite{Denicol:2016bjh}
\begin{equation}
\begin{aligned}
& \rho_{q-1,0} \equiv \rho_{q-1,0}^{(0)}, \\
& \rho_{s-1,0} \equiv \rho_{s-1,0}^{(0)}.
\end{aligned}    
\end{equation}
Then, after some algebraic steps, one obtains that the equations of motion for the moments of \eqref{eq:BE-moments-Bjorken} in RTA for constant relaxation time, $\gamma = 0$, are 
\begin{equation}
\label{eq:EoMs-BEq}
\begin{aligned}
& 
\dot{\rho}_{n,m} + \frac{\rho_{n,m}}{\Hat{\tau}}
+
\frac{2m(2m-1)(n+2m+1)}{16m^{2}-1} \frac{\rho_{n,m-1}}{\Hat{\tau}}
\\
&
+
\frac{2m(2m+1)+(8m^{2}+4m-1)n}{(4m-1)(4m+3)} \frac{\rho_{n,m}}{\Hat{\tau}}
+
\frac{(n-2m)(2m+1)(2m+2)}{(4m+1)(4m+3)} \frac{\rho_{n,m+1}}{\Hat{\tau}}
\\
&
=
-
(\rho_{n,m} - \rho_{n,m}^{(0)})
+
 \frac{\Gamma(n+3)}{\Gamma(3)\beta^{n}} (1-n)\delta n \ \delta_{m,0}
+
\beta n  \frac{\Gamma(n+3)}{\beta^{n}\Gamma(4)} 
\delta \varepsilon  \ \delta_{m,0}, \quad n=0,1,\cdots, \quad m=0,1,\cdots, 
\end{aligned}    
\end{equation}
where $\delta_{n,m}$ denotes the Kronecker delta. The equations above form an infinite system of first-order coupled differential equations. The symmetry assumptions lead to the fact that moments with different values of $n$ decouple, and only trios of consecutive moments couple for the Legendre index $m$. For computation purposes this tower is truncated at some high moment $m = M_{trun}$. In the computations of the main text we chose $M_{trun} = 25$ (we checked that our results are robust with respect to variations of this quantity). It should also be noticed that the counter-terms which are characteristic of the novel RTA formulation only have an effect for scalar ($m=0$) moments.


\bibliographystyle{apsrev4-1}
\bibliography{liography}

\end{document}